\documentclass{article}

\usepackage{arxiv}

\usepackage[utf8]{inputenc} % allow utf-8 input
\usepackage[T1]{fontenc}    % use 8-bit T1 fonts
\usepackage{hyperref}       % hyperlinks
\usepackage{url}            % simple URL typesetting
\usepackage{booktabs}       % professional-quality tables
\usepackage{amsfonts}       % blackboard math symbols
\usepackage{nicefrac}       % compact symbols for 1/2, etc.
\usepackage{microtype}      % microtypography
\usepackage{lipsum}		% Can be removed after putting your text content
\usepackage{graphicx}
\usepackage{doi}
\usepackage{amsthm,amsfonts,amssymb,amsmath,graphicx} 
\newtheorem{theorem}{Theorem}
\newtheorem{proposition}{Proposition}
\newtheorem{example}{Example}
\newtheorem{claim}{Claim}
\newenvironment{customproof}[1][Proof]{%
  \begin{proof}[#1]%
}{%
  \end{proof}%
}

\newtheorem{Remark}{Remark}
\newtheorem{lemma}[theorem]{Lemma}

\usepackage{natbib}
 \bibpunct[, ]{(}{)}{,}{a}{}{,}%

\usepackage{bbm}
\usepackage{caption}
\usepackage{multirow}
\usepackage{mathtools}
\usepackage{comment}
\usepackage{caption}
\usepackage{graphicx}
\usepackage{booktabs}
\usepackage{soul}
\usepackage{subfigure}
\usepackage{xurl}
\usepackage{adjustbox}
\usepackage{soul}
\usepackage{floatrow}
\usepackage{enumitem}
\usepackage{dsfont}

\usepackage{floatrow}
\newcommand{\Halmos}{\ensuremath{\Box}}
\newfloatcommand{capbtabbox}{table}[][\FBwidth]

\newcommand{\setrounds}{\mathcal{R}}
\newcommand{\nrounds}{\mathsf{r}}
\newcommand{\indexround}{r}

\newcommand{\setteams}{\mathcal{T}}
\newcommand{\nteams}{\mathsf{t}}
\newcommand{\indexteam}{t}

\newcommand{\setgames}{\mathcal{G}}
\newcommand{\ngames}{\mathsf{g}}
\newcommand{\indexgame}{g}

\newcommand{\nentries}{\mathsf{e}}
\newcommand{\nsimulations}{\mathsf{w}}
\newcommand{\indexentry}{E}
\newcommand{\newindexentry}[1]{\indexentry^{#1}}
\newcommand{\indexentryi}{e}
\newcommand{\setentries}{\mathcal{E}}

\newcommand{\Pteam}{\mathbf{P}^{\text{team}}}
\newcommand{\Pgame}{\mathbf{P}^{\text{game}}}
\newcommand{\Pround}{\mathbf{P}^{\text{round}}}

\newcommand{\Pteami}{P^{\text{team}}}
\newcommand{\Pgamei}{P^{\text{game}}}
\newcommand{\Proundi}{P^{\text{round}}}

\newcommand{\tourney}{\mathtt{M}}
\newcommand{\indexbracket}{B}
\newcommand{\setbrackets}
{\mathcal{B}}
\newcommand{\indexoutcome}
{O}

\newcommand{\indexsimulation}
{W}
\newcommand{\indexsimulationi}
{w}
\newcommand{\newsimulation}[1]{\indexsimulation^{#1}}
\newcommand{\setsimulations}
{\mathcal{W}}

\newcommand{\Propplus}{\texttt{PROP}\textsuperscript{+}\xspace }
\newcommand{\Prop}{\texttt{PROP}\xspace }
\newcommand{\SIP}{\texttt{SIP}\xspace }
\newcommand{\GreedySAA}{\texttt{G-SAA}\xspace }
\newcommand{\SAA}{\texttt{SAA}\xspace }

\newcommand{\SIM}{\texttt{Sim}\xspace }

\newcommand{\completebrackets}{\mathcal{B}}

\newcommand{\paren}[1]{\left( #1 \right)}

\newcommand{\angles}[1]{\left\langle #1 \right\rangle}

\newcommand{\expectation}[1]{\mathbb{E}\paren{#1}}

\usepackage[ruled,vlined]{algorithm2e}

\usepackage{hyperref}
\hypersetup{
    colorlinks=true,
    linkcolor=blue,
    filecolor=magenta,      
    urlcolor=cyan,
    citecolor=blue,
    hypertexnames=false
    }

\title{The Madness of Multiple Entries in \textit{March Madness}}

\author{ \href{https://orcid.org/0000-0001-7939-7714}{\includegraphics[scale=0.06]{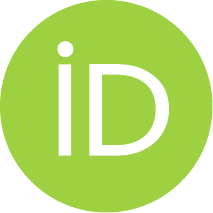}\hspace{1mm}Jeff Decary} \\
	Department of Operations and Information Management\\
	University of Connecticut
        Storrs, CT 06268\\
	\texttt{Jeff.Sylvestre-Decary@uconn.edu} \\
	\And
	\href{https://orcid.org/0000-0002-5566-5224}{\includegraphics[scale=0.06]{orcid.pdf}\hspace{1mm}David Bergman} \\
	Department of Operations and Information Management\\
	University of Connecticut
        Storrs, CT 06268\\
	\texttt{David.Bergman@uconn.edu} \\
	\And
	\href{https://orcid.org/0000-0002-1439-5205}{\includegraphics[scale=0.06]{orcid.pdf}\hspace{1mm}Carlos Cardonha} \\
	Department of Operations and Information Management\\
	University of Connecticut
        Storrs, CT 06268\\
	\texttt{Carlos.Cardonha@uconn.edu} \\
        \And
	Jason Imbrogno \\
	Department of Finance, Economics, and Data Analytics\\
        University of North Alabama\\
        Florence, AL 35632\\
	\texttt{jimbrogno@una.edu} \\
        \And
	\href{https://orcid.org/0000-0001-9269-633X}{\includegraphics[scale=0.06]{orcid.pdf}\hspace{1mm}Andrea Lodi} \\
	Jacobs Technion-Cornell Institute
        Cornell Tech\\
        New York, NY 10044\\
	\texttt{andrea.lodi@cornell.edu} \\
}

\renewcommand{\shorttitle}{The Madness of Multiple Entries in \textit{March Madness}}

\hypersetup{
pdftitle={The Madness of Multiple Entries in \textit{March Madness}},
pdfsubject={March Madness},
pdfauthor={Jeff Decary, David Bergman, Carlos Cardonha, Jason Imbrogno, Andrea Lodi},
pdfkeywords={March Madness, Mixed-Integer Linear Programming, Simulation Optimization, Sports Betting},
}

\begin{document}
\maketitle

\begin{abstract}
This paper explores multi-entry strategies for betting pools related to single-elimination tournaments. In such betting pools, participants select winners of games, and their respective score is a weighted sum of the number of correct selections. Most betting pools have a top-heavy payoff structure, so the paper focuses on strategies that maximize the expected score of the best-performing entry. There is no known closed-formula expression for the estimation of this metric, so the paper investigates the challenges associated with the estimation and the optimization of multi-entry solutions. We present an exact dynamic programming approach for calculating the maximum expected score of any given fixed solution, which is exponential in the number of entries. We explore the structural properties of the problem to develop several solution techniques. In particular, by extracting insights from the solutions produced by one of our algorithms, we design a simple yet effective problem-specific heuristic that was the best-performing technique in our experiments, which were based on real-world data extracted from recent \textit{March Madness} tournaments. In particular, our results show that the best 100-entry solution identified by our heuristic had a 2.2\% likelihood of winning a \$1 million prize in a real-world betting pool.
%\emph{DraftKings} contest against top sports bettors.
\end{abstract}

% keywords can be removed

\section{Introduction}
\label{sec:introduction}

\textit{March Madness} refers to the annual National Collegiate Athletic Association (NCAA) Division I men's basketball tournament, featuring 68 teams. In a preliminary round, eight teams play in four separate games, with the winning teams joining sixty others in a single-elimination tournament, where the champion emerges after winning six consecutive games.  Colloquially, the beginning of the 64-team tournament is often referred to as the start of the tournament.

\begin{figure}[ht!]
    \centering
    \includegraphics[scale=0.5]{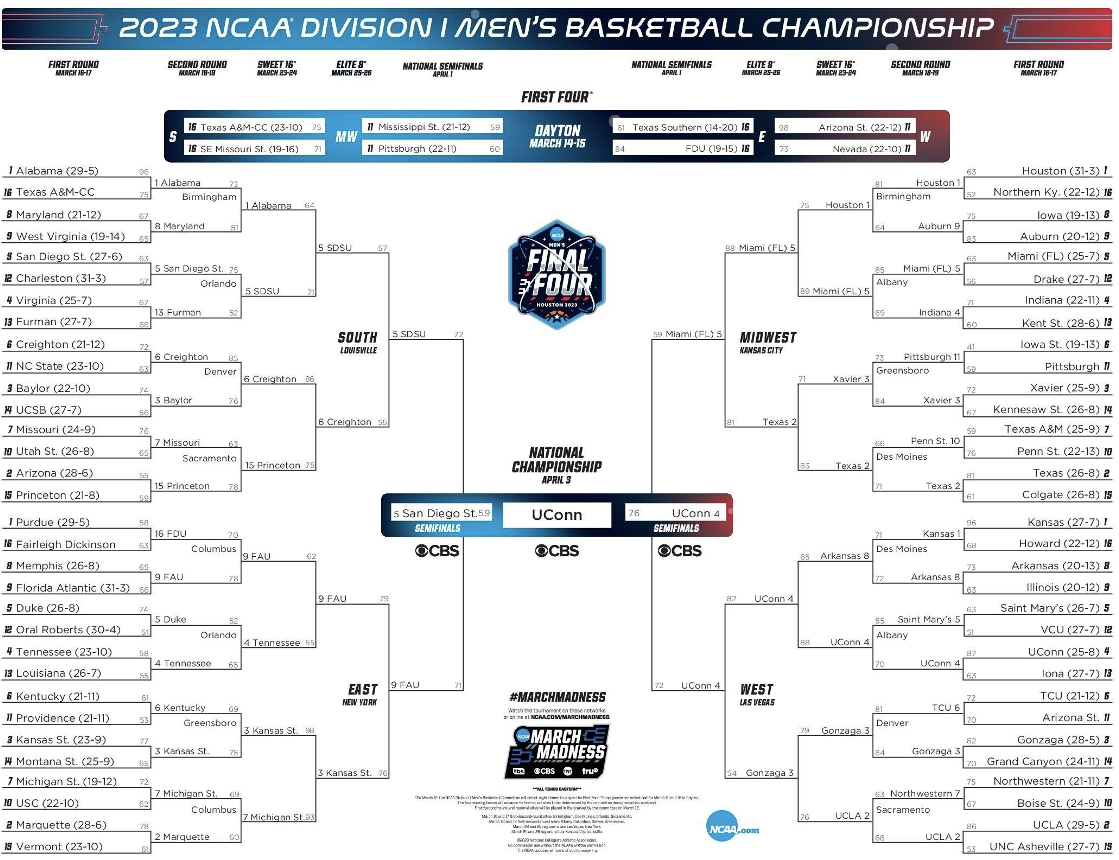}
    \caption{2023 NCAA Men's Tournament Bracket}
    \label{fig:2019_MM}
\end{figure}

The bracket of the 64-team tournament is divided into four ``regions," designated by the locations where the first two rounds are played. Each region has 16 teams seeded from 1 to 16, where seed 1 is the strongest team in the region and seed 16 is the weakest. Game arrangements ensure that, in the first round, the strongest team plays against the weakest, the second-strongest plays against the second-weakest, and so on. In the first round within each region, seed number $n \in \{1, 2, 3, ..., 8\}$ plays against seed number $17 - n$. If there are no upsets (i.e., a higher-seeded team beating a lower-seeded one) in the first round, the second-round match-ups have seed numbers $n\in \{1, 2, 3, 4\}$ playing against seed numbers $9 - n$. The teams are not ``reseeded" after the start, so the ``strongest versus weakest'' arrangement of match-ups is only guaranteed in the first round. The four regional bracket winners advance to the tournament stage known as the \textit{Final Four}, where two semifinal games and the championship game determine the winning team.  To claim the championship, the \textit{March Madness} winner must secure six consecutive victories from the start of the tournament. Figure \ref{fig:2019_MM}, courtesy of the NCAA,\footnote{NCAA: %\url{https://rb.gy/5y2lg7}
\url{www.ncaa.com/news/basketball-men/article/2023-04-18/2023-ncaa-bracket-scores-stats-march-madness-mens-tournament} 
(Accessed: February 3, 2024)}
%\cite{MM2023Bracket}, 
illustrates the bracket and results of the 2023 tournament, won by the University of Connecticut.
%of 2023 is shown in Figure \ref{fig:2019_MM}, courtesy of \cite{MM2023Bracket}. 

Betting pools are often organized based on \textit{March Madness}. Arguably, 
%According to \cite{NCAAhistory}, 
the first \textit{March Madness} betting pool ran in 1977 out of a Staten Island bar.\footnote{Smithsonian Magazine: 
\url{https://www.smithsonianmag.com/history/when-did-filling-out-march-madness-bracket-become-popular-180950162/}
%\url{https://shorturl.at/tyIM2}
(Accessed: February 3, 2024)
} The game's popularity has consistently grown, as evidenced by a survey conducted by \textit{Morning Consult} and produced by the American Gaming Association (AGA) %\citeyearpar{MMAgamingA} 
in 2019.\footnote{American Gaming Association: %\url{https://rb.gy/xp2ao3}
\url{https://www.americangaming.org/new/americans-will-wager-8-5-billion-on-march-madness/} 
(Accessed: February 3, 2024)} The survey revealed that over 40 million people in the U.S. spent \$8.5 billion to fill out 149 million \textit{March Madness} brackets that year. More recently, the AGA 
%\citeyearpar{MMAgamingA2} 
reported that for \textit{March Madness} 2023, 68 million Americans were planning to wager \$15.5 billion.\footnote{American Gaming Association: 
\url{https://www.americangaming.org/resources/2023-march-madness-wagering-estimates/}
%\url{https://shorturl.at/ivAD9}
(Accessed: February 3, 2024)}

In a \textit{March Madness} betting pool, participants must predict the winners for all tournament games before the start of the first game in the 64-team tournament, and entries cannot be modified afterward.
% Each entry of a \textit{March Madness} betting pool must specify a team for all games in the tournament, where the participant entering the pool seeks to specify the winner of those games. Entries must be registered before the tournament starts and cannot be modified after the beginning of the first game. 
A typical scoring system used in \textit{March Madness} betting pools awards points for each correct selection, with increasing weights in later rounds. In the most common, standard system, each of the 32 first-round games is worth 1 point, the 16 second-round games are worth 2 points, the 8 third-round games are worth 4 points, and so on. With 32 available points in each round, an entry can score up to 192 points. Participants are ranked based on their accumulated points throughout the tournament. 

There are three primary reasons why \textit{March Madness} provides such an exciting platform for competition. (1) US college athletics is a massive industry with a market size of \$19 billion \citep{StatistaCollegeSports,Chung2017}. (2) Betting is deeply ingrained in American culture, so this game of chance solicits interest. (3) Lastly, and perhaps most importantly, the selection of winners throughout the tournament is notoriously difficult. Warren Buffett famously offered \$1 billion in 2014 to anyone filling out a perfect bracket.\footnote{Business Insider: 
\url{https://www.businessinsider.com/warren-buffett-billion-dollar-bracket-2014-1}
%\url{https://shorturl.at/wxzD5}
(Accessed: February 3, 2024)
}
%\citep{Warren1}. 
This offer has since been changed to \$1 million a year for life to any of his employees for correctly picking all 48 games in the first two rounds;\footnote{
CNBC:
\url{https://www.cnbc.com/2019/03/21/inside-warren-buffetts-multimillion-dollar-march-madness-challenge.html}
%\url{https://t.ly/4sGdJ}
(Accessed: February 3, 2024)
}
%\citep{Warren2}; 
no one has conquered this to date. 

This article delves into the challenge of selecting winning brackets for \textit{March Madness}, focusing on scenarios where a participant can enter multiple brackets. The techniques developed here apply to betting pools in various single-elimination tournament settings. Examples include the advanced rounds of tournaments such as the Federation Internationale de Football Association (FIFA) World Cup and almost all competitions organized by the Association of Tennis Professionals (ATP). 

\subsection{Sports Betting Industry}

\begin{figure}[ht!]
    \centering
    \includegraphics[width=\textwidth]{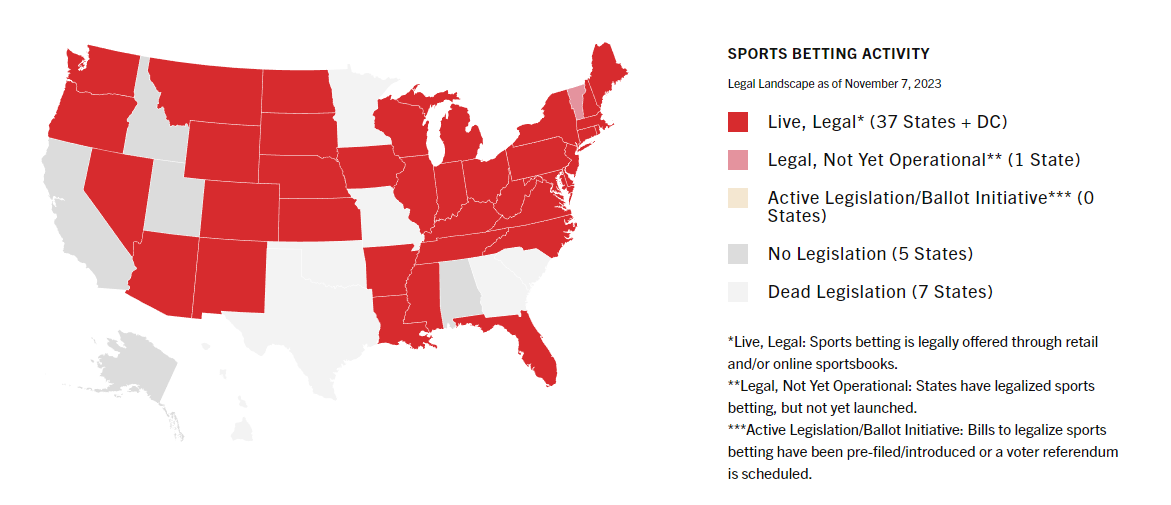}
    \caption{Sport Betting Legalization Status as of November 7, 2023}
    \label{fig:AGAMAP}
\end{figure}

The sports betting industry is experiencing substantial growth in the U.S., propelled by favorable recent changes in regulations. A key milestone was the Supreme Court ruling in \textit{Murphy v. National Collegiate Athletic Association} \citeyearpar{Supreme18} that lifted the federal government's ban on state-sponsored sports gambling. This decision has resulted in 37 states legalizing sports betting to date (see Figure~\ref{fig:AGAMAP}, courtesy of the AGA\footnote{American Gaming Association: 
%\url{https://shorturl.at/fkAW3}
\url{https://www.americangaming.org/research/state-gaming-map/}
(Accessed: February 3, 2024)
}).
In this newly legalized and regulated environment, the AGA reported that people in the U.S. wagered \$93.2 billion on sports, generating \$7.5 billion in revenue for sportsbooks.\footnote{American Gaming Association: %\url{https://rb.gy/v6famq} 
\url{https://www.americangaming.org/new/2022-commercial-gaming-revenue-tops-60b-breaking-annual-record-for-second-consecutive-year/}
(Accessed: February 3, 2024)} %\citeyearpar{MMAgamingA1}. 
\href{https://www.fanduel.com/}{\emph{FanDuel}} and \href{https://www.draftkings.com/}{\emph{DraftKings}} are the two major companies offering legalized sports betting, holding 44.6\% and 27.3\% of the gross gaming revenue for 2023, respectively.\footnote{Wall Street Journal: %\url{https://rb.gy/tgz8kl}
\url{https://www.wsj.com/business/media/disney-agonized-about-sports-betting-now-its-going-all-in-70b31f3b}
(Accessed: February 3, 2024) }% \citep{WSJ_Disney}.

\textit{Betting pools} represent a significant component of the sports gambling market. In betting pools, all participants pay an entry fee to enroll in the competition and become eligible to compete for prizes.  These competitions require participants to choose outcomes of several sporting events, and the participants who score more points (generally by making more weighted correct picks) stand to win monetary rewards.  Often, a participant can enter a limited number of multiple entries by repeatedly paying the entry fee.  The pool operator typically retains a portion of the funds as a management fee.

\subsection{The Difficulty in Selecting Entries}

There are $2^{63}=9,223,372,036,854,775,808$ possible outcomes for a \textit{March Madness} tournament. Naturally, some outcomes are more likely to occur than others, but empirical evidence shows that participants in \textit{March Madness} betting pools tend to over-pick the favorites \citep{metrick1996march,clair2007optimal,wright2016method,Morewedge2018}. This strategy is affected by \textit{upsets}, which refers to the victory of the underdog (i.e., higher-seeded team) in a game.
Table~\ref{tab:upsetTable} shows the average number of upsets per round for all \textit{March Madness} tournaments spanning from 1985 to 2023. On average, upsets happen in 20\% to 25\% of the games.
\begin{table}[ht!]
\scriptsize
\begin{tabular}{|c|c|c|c|c|c|c||c|}
\hline
        & \begin{tabular}[c]{@{}c@{}}Round 1\\ (32 games)\end{tabular} & \begin{tabular}[c]{@{}c@{}}Round 2\\ (16 games)\end{tabular} & \begin{tabular}[c]{@{}c@{}}Round 3\\ (8 games)\end{tabular} & \begin{tabular}[c]{@{}c@{}}Round 4\\ (4 games)\end{tabular} & \begin{tabular}[c]{@{}c@{}}Round 5\\ (2 games)\end{tabular} & \begin{tabular}[c]{@{}c@{}}Round 6\\ (1 game)\end{tabular} & \begin{tabular}[c]{@{}c@{}}Total\\ (63 games)\end{tabular} \\ \hline
Average & 8.29                                                         & 4.68                                                         & 2.37                                                        & 1.76                                                        & 0.50                                                        & 0.18                                                       & 17.79                                                      \\ \hline
\end{tabular}
\caption{Upsets Per Round During \textit{March Madness}}
\label{tab:upsetTable}
\end{table}
Consequently, it is estimated that 66.7\% of the selections made by an average participant's entry are correct.\footnote{NCAA: 
%\url{https://shorturl.at/qtNUX}
\url{https://www.ncaa.com/news/basketball-men/bracketiq/2023-03-16/perfect-ncaa-bracket-absurd-odds-march-madness-dream}
(Accessed: February 3, 2024)
}

\subsection{Relevant Literature on Sports Betting Pools}

As popular interest in college sports and sports betting has grown, so has the academic interest. For example, fundamental questions about random knockout tournaments have been studied since the 1950s \citep{david1959tournaments,horen1985comparing,knuth1987random}, and there is a vast literature on sports scheduling \citep{NemTri98,ribeiro2012sports}.

The existing literature on \textit{March Madness} focuses on developing prediction models for game outcomes and on strategies for single-entry pools \citep{caudill2003predicting,kvam2006logistic,hoegh2015nearest,gumm2015machine,Palley2019}. Regarding single-entry strategies, \citet{kaplan2001march} present an exact dynamic programming algorithm that identifies an entry with the highest expected score. \cite{clair2007optimal} demonstrate that the optimal strategy for a participant in a single entry pool varies with the number of participants in a betting pool. They show that participants in large betting contests tend to over-pick specific teams, thus diminishing the expected payoff from those selections.

While prior research has predominantly concentrated on single-entry betting strategies for \textit{March Madness}, recent betting data from 2019 reveals that, on average, each participant entered 3.7 entries \citep{MMAgamingA}.The rationale behind having multiple entries is to enable participants to manage uncertainty effectively and enhance their likelihood of selecting a successful entry by diversifying their picks. This article provides the first systematic study for picking entries composing a multi-entry bet for a \textit{March Madness} betting pool. 
 
The study of multi-entry strategies for sports betting competitions has gained popularity in the last few years. \citet{bergman2017surviving} introduce a mathematical programming-based multi-entry strategy for a National Football League (NFL) survivor pool in which participants sequentially select a team every week, and they can only remain in the competition if the selected team wins in that week's round. \citet{hunter2016picking} present an iterative multi-entry strategy for daily fantasy sports (DFS) contests that selects entries with the highest individual expected score in each step and then forces diversification of new entries to be selected by incorporating constraints that limit the number of identical players with the previously selected entries. \citet{bergman2022optimizing} present an exact approach for selecting two entries for DFS contests under the assumption that the scores are normally distributed. \citet{haugh2018play} introduce a model to simulate the opponents' selection in DFS (as in~\cite{clair2007optimal}) and identify a score to beat to be \emph{in-the-money}; the authors use an iterative method similar to~\cite{hunter2016picking} to extend their formulation to the multi-entry problem.  Finally, \citet{liu2023picking} reformulate the maximization problem of the expected score of the best-performing entry to a semi-definite program and solve it using a distributionally robust optimization approach.

\subsection{Maximizing the Expectation of the Maximum-Scoring Entry}

Numerous challenges are associated with identifying an optimal multi-entry betting strategy for a participant controlling multiple entries in a \textit{March Madness} pool. 
One prominent challenge arises from the highly top-heavy payout structure characteristic of typical \textit{March Madness} betting pools, i.e., a disproportionately large portion of the total prize pool is awarded to the top performers. For example, in 2023, \emph{DraftKings} ran a high-rollers competition that required \$100 per entry to play, where each participant could enter up to 100 entries. With 12,605 total entries, the participant with the sole winning entry received an 80\% share of the payout, totaling \$1 million. Given this payout distribution, a natural objective function for a multi-entry bettor is to select a collection containing one that will perform exceptionally well. Therefore, we investigate strategies for maximizing the expectation of the maximum-scoring entry.  
\subsection{Contributions}

The paper focuses on identifying multi-entry solutions for single-elimination tournaments, focusing on \textit{March Madness} tournaments. The main contributions of this paper are the following:

\begin{itemize}
    \item We formally define the problem of selecting multiple entries in a \textit{March Madness} betting pool with a top-heavy payout structure;
    \item We explore the structural properties of the problem, unveiling its difficulty;
    \item We present an exact method for evaluating the expectation of the maximum scoring entry, which is exponential in the number of entries;
    \item We introduce four algorithms to identify high-quality multi-entry solutions. These procedures are based on a mathematical programming formulation of the single-entry problem;
    \item We analyze the solutions of our best-performing mathematical programming strategy to develop a heuristic that provides a higher integration level in the selection of entries (numerical experiments on real-world data show that this heuristic outperforms the other strategies);
    \item We apply our methodology to a real-world betting pool, exhibiting that our approach compares favorably against some of the best sports bettors in the world.
\end{itemize}

\subsection{Organization}
The paper is organized as follows. Section~\ref{sec:problem} introduces the notation used throughout the manuscript and describes the problem. 
Section~\ref{sec:EMS_Calculation} investigates the exact calculation and the estimation of the expected maximum score of a fixed set of entries. Section~\ref{sec:StructuralProperties} investigates the structural properties of the problem. 
Section~\ref{sec:Optimization} presents our optimization algorithms. Section~\ref{sec:ExperimentalResults} presents a thorough evaluation of our estimation procedures along with a comparison of our algorithms.  Section~\ref{sec:DKContest} presents results comparing the solutions produced by our best-performing algorithm with other submitted entries on a recent real-world \textit{March Madness} tournament. Finally, Section~\ref{sec:Conclusion} concludes the paper with directions for future work.

\section{Notation and Problem Description}
\label{sec:problem}

This section introduces the notation used throughout the paper and formally defines the problem we aim to solve.

\subsection{Tournament}

A \textit{single-elimination tournament} $\tourney \coloneqq \angles{\setteams,\setrounds,\setgames}$ is defined by a set of teams~$\setteams$ and a set of games~$\setgames$ organized in rounds $\setrounds$.   Specifically:
\begin{itemize}
    \item $\setteams \coloneqq \{1, \ldots, \nteams\}$, with cardinality $\nteams$ and elements indexed by $t$;
    \item $\setrounds \coloneqq \{1, \ldots, \nrounds\}$, with cardinality $\nrounds$ and elements indexed by $\indexround$; and
    \item $\setgames \coloneqq \{1, \ldots, \ngames\}$, with cardinality $\ngames$ and elements indexed by %which we index with 
    $\indexgame$. 
\end{itemize}
We sometimes refer to a team, round, or game by its index for ease of notation.

Each game consists of a match-up between two teams.  Every team plays a single game in round $\indexround = 1$, so there are $\frac{\nteams}{2}$ games in round 1. Exactly one team wins each game. Any team that loses a game is eliminated from the tournament, so half of the teams proceed to the subsequent round. Therefore, each round~$\indexround$ has~$\frac{\nteams}{2^\indexround}$ games, and we have~$\ngames=\nteams-1$ and~$\nrounds = \log_2{\nteams}$. In particular, a  \textit{March Madness} tournament has $\nteams = 64$ teams, $\ngames = 63$ games, and $\nrounds=6$ rounds.

We define a layered acyclic graph~$D = (\setgames,A)$ over the set of games to describe the tournament's structure. Each \textit{node} (\textit{game}) $\indexgame \in \setgames$ is in the \emph{layer} (\emph{round}) $\indexround(\indexgame) \in \setrounds$.\footnote{In this section, we use the terms \textit{nodes} and \textit{games}, as well as \textit{rounds} and \textit{layers}, interchangeably.} Each \textit{arc} in~$A$ represents a progression step in the tournament bracket, i.e., arc~$(\indexgame,\indexgame')$ indicates that the winner of game~$\indexgame$ plays game~$\indexgame'$ next, where $\indexround(\indexgame') = \indexround(\indexgame) + 1$; we say that game~$\indexgame$ leads to game~$\indexgame'$ if~$(\indexgame,\indexgame')$ is in~$A$.
%There are $\frac{\nteams}{2}$ games in round 1, each with 
Nodes in the first layer have no incoming arcs and one outgoing arc with head in the second layer (i.e., the winners of the first-round games play in the second round).  The rest of the graph is defined iteratively. Namely, each node~$\indexgame$ in layer~$\indexround(\indexgame) \in \{2, \ldots, \nrounds-1\}$ has two incoming arcs with tails in layer~$\indexround(\indexgame)-1$ and one outgoing arc with head in layer~$\indexround(\indexgame)+1$. The final layer~$\nrounds$ has a singleton node with two incoming arcs and no outgoing arc, representing the final game. 

We denote the set of teams that might play game $\indexgame \in \setgames$ by $\setteams(\indexgame)$. The construction of~$\setteams(\indexgame)$ is also iteratively defined by round.  For round 1, each team $\indexteam \in \setteams$ is assigned to a unique game,
so we have~$|\setteams(\indexgame)| = 2$ if~$\indexround(\indexgame) = 1$. If~$\indexround(\indexgame) > 1$,  we have  
\[ 
\setteams(\indexgame) \coloneqq \bigcup\limits_{\indexgame' \in \Gamma^{-}(\indexgame)} \setteams(\indexgame'),
\] where %(recalling that, by standard graph theory notation, 
$\Gamma^{-}(\indexgame)$ is the set composed of the two games in round~$\indexround(\indexgame) - 1$ with outgoing arcs directed at $\indexgame$. Moreover, for  $\indexround(\indexgame) > 1$, we denote by $\gamma^{-}(\indexgame,\indexteam)$ the  game $\indexgame'$ in round $\indexround(\indexgame)-1$ in which $t \in \setteams(\indexgame')$, i.e., team~$\indexteam$ has to win game~$\gamma^{-}(\indexgame,\indexteam)$ to play game~$\indexgame$.  We also let $\delta^-(\indexgame,\indexteam)$ be the game in round $r(\indexgame)-1$ that has head $\indexgame$ for which team $\indexteam$ cannot play, i.e., $\indexteam \notin \setteams(\delta^-(\indexgame,\indexteam))$. 
Each team has a unique path to the final game, so the sets~$\setteams(\indexgame)$ of all games within any round induce a partition of~$\setteams$. 

\subsection{Bracket}

A \textit{bracket} $\indexbracket \in \{0,1\}^{\nteams \times \ngames}$ is a binary matrix where~$\indexbracket_{\indexteam,\indexgame} = 1$ indicates the assignment of team~$\indexteam$ to game $\indexgame$ (or, in other words, the selection of~$\indexteam$ as the winner of~$\indexgame$).
A bracket is \textit{feasible} if it satisfies the following conditions:

\begin{enumerate}
    \item Exactly one team is assigned to each game {$\indexgame~\in~\setgames$}, i.e.,
    \[
    \sum_{\indexteam \in \setteams}\indexbracket_{\indexteam,\indexgame} = 1;
    \]

    \item A team can only be assigned to games it may play, i.e., $\indexbracket_{\indexteam,\indexgame} = 1$ only if  $\indexteam \in \setteams(\indexgame)$;  

    \item After round 1, any team assigned to a game 
    %in an advanced round 
    must also be assigned to the game leading to it 
    %preceding game
    %of its game 
    in the previous round, i.e., if $\indexround(\indexgame) > 1$ and $\indexbracket_{\indexteam,\indexgame} = 1$, then we must have $\indexbracket_{\indexteam,\gamma^{-}(\indexgame,\indexteam)} = 1$.  
    
\end{enumerate}
We use~$\setbrackets$ to denote the set of all feasible brackets. Finally, a feasible bracket can also be represented by an ordered list of~$\ngames$ elements, where each position represents a game, and the value of each entry is associated with a team.

\subsection{Outcomes and Entries}
 
A feasible bracket $\indexbracket \in \setbrackets$ can represent  either the \emph{outcome} of the tournament or an \emph{entry} submitted by a participant in a betting pool. We use~$\indexoutcome$ when referring to a bracket as an outcome and~$\indexentry$ to represent entries that bettors create and submit.  

Let $\nentries$ denote the number of entries that any given participant may enter. 
As with teams, rounds, and games, we 
define:
\begin{itemize}
    \item $\setentries \coloneqq \{1, \ldots, \nentries\}$, with cardinality $\nentries$ and elements indexed by $\indexentryi$.   
\end{itemize}

Since entries are matrices, we must use additional notation to refer to the object itself, where we use $\newindexentry{\indexentryi}$ to refer to the entry matrix for entry $\indexentryi$. With a slight abuse of notation, we also refer to~$\setentries$ as the set of entry matrices, i.e., $\setentries = \{\newindexentry{1}, \ldots, \newindexentry{\nentries}\}$.

A bettor aims to select entries that are as similar as possible to the actual outcome. The outcome is uncertain, so we leverage historic data and use the pairwise probabilities of teams beating other teams.  We define a collection of probability matrices used for various arguments later on:

\begin{itemize}
    \item $\Pteam \in [0,1]^{\nteams \times \nteams}$: the \emph{team-by-team win matrix}, where $\Pteami_{\indexteam,\indexteam'}$ is the probability team $\indexteam$ would beat team $\indexteam'$ in a game;
    \item $\Pgame \in [0,1]^{\nteams \times \ngames}$: the \emph{team-by-game win matrix}, where $\Pgamei_{\indexteam,\indexgame}$ is the probability team $\indexteam$ wins game $\indexgame$; 
    \item $\Pround \in [0,1]^{\nteams \times \nrounds}$: the \emph{team-by-round win matrix}, where $\Proundi_{\indexteam,\indexround}$ is the probability team $\indexteam$ wins a game in round~$\indexround$.
\end{itemize}
Similarly to previous work in the area, we assume that~$\tourney$ is a Markov tournament, i.e., the probabilities~$\Pteami_{\indexteam,\indexteam'}$ are defined before the tournament starts and do not change based on the outcome of the games \citep{david1959tournaments,horen1985comparing,kaplan2001march}. We discuss the estimation of $\Pteam$ as well as how to calculate $\Pgame$ and $\Pround$ in Appendix~\ref{appendix:Win Probability Matrices}.

\subsection{Single Entry Score}

The score of an entry is the weighted sum of the entries in the Hadamard product of the outcome bracket and the entry bracket, i.e., a weighted sum of the number of games for which the entry bracket and the actual outcome bracket coincide on 1s. The number of points awarded is weighted by the round of the games, with later rounds having heavier weight. Here, we adopt the \textit{ESPN Tournament Challenge scoring system}, where a game in round $\indexround \in \setrounds$ is worth $2^{\indexround-1}$ points (this is the standard scoring system described in Section~\ref{sec:introduction}). Therefore, the \emph{score} $s(\indexentry,\indexoutcome)$ of an entry $\indexentry \in \setbrackets$ given an outcome $\indexoutcome \in \setbrackets$ is given by
\[
s(\indexentry,\indexoutcome) \coloneqq \sum_{\indexteam \in \setteams} \sum_{\indexgame \in \setgames} 2^{\indexround(\indexgame)-1} \cdot \indexentry_{\indexteam,\indexgame} \cdot \indexoutcome_{\indexteam,\indexgame}.
\]
Let $S(\indexentry)$ denote the random variable representing the score of an entry~$\indexentry$.   For $\indexteam \in \setteams$ and $\indexgame \in \setgames$,  if we let $X_{\indexteam,\indexgame} \sim \text{Bernoulli}(\Pgamei_{\indexteam,\indexgame})$ be the random variable indicating whether team $\indexteam$ wins game $\indexgame$, we can then express $S(\indexentry)$ by
\[
S(\indexentry) \coloneqq \sum_{\indexteam \in \setteams} \sum_{\indexgame \in \setgames} 2^{\indexround(\indexgame)-1} \cdot \indexentry_{\indexteam,\indexgame} \cdot X_{\indexteam,\indexgame}.
\]
Therefore, the expected score of a single entry can be expressed as 
\begin{align}
    \mathbb{E}\left[S\paren{E}\right]  
    \coloneqq  
    \sum_{ \indexgame \in \setgames} \sum_{ \indexteam \in \setteams} 
    2^{\indexround(\indexgame)-1} \cdot\indexentry_{\indexteam, \indexgame}  \cdot \Pgamei_{\indexteam,\indexgame}.
 \label{eq:e2}
\end{align} 
Using~\eqref{eq:e2}, 
\cite{kaplan2001march} present a \emph{dynamic programming} approach that identifies in polynomial time an entry~$\indexentry$ that maximizes~$\mathbb{E}[S(\indexentry)]$.

\subsection{Multiple Entry Score}

In this paper, we evaluate the quality of a collection of entries by the maximum score achieved by at least one of the entries. To this end, we extend the function~$S$ to a set function parameterized by a set of entries $\setentries \subseteq \setbrackets$. For consistency with the definitions above and succinctness, we abuse notation slightly and use $S(E)$ and $S(\{E\})$ interchangeably for singleton sets. For entry sets~$\mathcal{E}$ with two or more elements, the random variable~$S(\setentries)$ representing the score of~$\setentries$ is defined as follows:
\[
S(\setentries) := \max_{E \in \setentries} S(\{E\}).
\]

\subsection{Decision Problem}

In our framework, a participant of a betting pool wishes to select a subset $\mathcal{E}$ of~$\setbrackets$ of size $\nentries$ that maximizes the expected score of the maximum-scoring entry, which we herein refer to as the \emph{expected maximum score} (EMS). Therefore, the goal is to solve the following problem:
\begin{align}
\max_{\mathcal{E} \subseteq \mathcal{B} : |\mathcal{E}| = \nentries} \mathbb{E}\left[S(\mathcal{E})\right].\label{eq:MultiEntryProblem}
\end{align}

\section{Calculating and Estimating the EMS}
\label{sec:EMS_Calculation}

\begin{figure}[htb]
  \centering
  \begin{minipage}[c]{0.45\linewidth}
  \centering
  \includegraphics[width=0.75\linewidth]{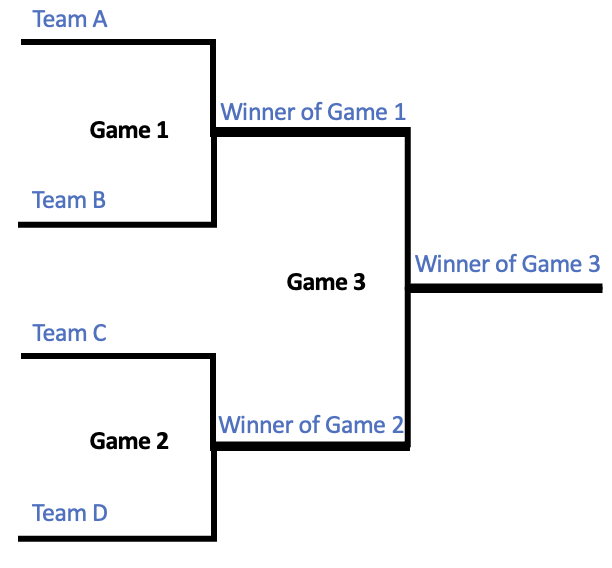}
  \end{minipage}
  \begin{minipage}[c]{0.45\linewidth}
   \begin{tabular}{c|llllllll}
   & \multicolumn{8}{c}{Feasible Brackets}\\
   \hline
 Game & $\indexbracket_1$ & $\indexbracket_2$ & $\indexbracket_3$ & $\indexbracket_4$ & $\indexbracket_5$ & $\indexbracket_6$ & $\indexbracket_7$ & $\indexbracket_8$ \\
\hline
1  & A & A & A & A & B & B & B & B \\
2 & C & D & C & D & C & D & C & D\\
3 & A & A & C & D & B & B & C & D
\end{tabular}
  \end{minipage}
  \caption{Example Four-Team Bracket Structure and Feasible Bracket Outcomes}\label{fig:example}
\end{figure}

%Given two entries, t
The calculation of the EMS is difficult for two or more entries because the objective function involves an order statistics expression where the random variables have intricate correlations. We demonstrate the challenge of explicitly computing the EMS by considering the structure of a tournament of two rounds, three games, and four teams. Figure~\ref{fig:example} shows the tournament structure (left) and the eight feasible brackets (right) given by the team choice in each game. The maximum score in this tournament is four points, with two points awarded in the first round (one per game) and the other two in the championship game.
The other parameters are the win matrices $\Pteam$, $\Pround$, and $\Pgame$, which are presented in Table~\ref{tab:ProbMatrices}.
\begin{table}[h!]
\begin{minipage}{.33\textwidth}
    \centering
    \begin{tabular}{|c|c|c|c|c|}
\hline
  & A              & B              & C              & D              \\ \hline
A & -              & $\Pteami_{A,B}$ & $\Pteami_{A,C}$ & $\Pteami_{A,D}$ \\ \hline
B & $\Pteami_{B,A}$ & -              & $\Pteami_{B,C}$ & $\Pteami_{B,D}$ \\ \hline
C & $\Pteami_{C,A}$ & $\Pteami_{C,B}$ & -              & $\Pteami_{C,D}$ \\ \hline
D & $\Pteami_{D,A}$ & $\Pteami_{D,B}$ & $\Pteami_{D,C}$ & -              \\ \hline
\end{tabular}
  \end{minipage}%
  \begin{minipage}{.33\textwidth}
    \centering
    \begin{tabular}{|c|c|c|}
    \hline
      & Round 1         & Round 2         \\ \hline
    A & $\Proundi_{A,1}$ & $\Proundi_{A,2}$ \\ \hline
    B & $\Proundi_{B,1}$ & $\Proundi_{A,2}$ \\ \hline
    C & $\Proundi_{C,1}$ & $\Proundi_{C,2}$ \\ \hline
    D & $\Proundi_{D,1}$ & $\Proundi_{D,2}$ \\ \hline
    \end{tabular}
\end{minipage}
\begin{minipage}{.33\textwidth}
    \centering
    
    \begin{tabular}{|c|c|c|c|}
    \hline
      & Game 1          & Game 2          & Game 3          \\ \hline
    A & $\Pgamei_{A,1}$ & 0               & $\Pgamei_{A,3}$ \\ \hline
    B & $\Pgamei_{B,1}$ & 0               & $\Pgamei_{B,3}$ \\ \hline
    C & 0               & $\Pgamei_{C,2}$ & $\Pgamei_{C,3}$ \\ \hline
    D & 0               & $\Pgamei_{D,2}$ & $\Pgamei_{D,3}$ \\ \hline
    \end{tabular}
\end{minipage}
    \caption{Probability Matrices $\Pteam, \Pround,$ and $\Pgame$}
\label{tab:ProbMatrices}
\end{table}

For any $\indexoutcome\in\setbrackets$, let $P_\indexoutcome$ be the probability that $\indexoutcome$ is the tournament outcome. An expression for calculating the EMS of $\setentries=\{\indexbracket_1, \indexbracket_6\}$ is
\begin{eqnarray}
\mathbb{E}\left[S(\{\indexbracket_1, \indexbracket_6\})\right] 
    &=& 
        \sum_{\indexoutcome \in \completebrackets}P_{\indexoutcome}\max(s(\indexbracket_1, \indexoutcome),s(\indexbracket_6,\indexoutcome)) \nonumber \\
    % \qquad\quad\;\qquad\;\, 
    &=& 
        P_{\indexbracket_1}\cdot \max(4,0) + P_{\indexbracket_2}\cdot \max(3,1) +P_{\indexbracket_3}\cdot \max(2,0) +P_{\indexbracket_4}\cdot \max(1,1) +
    \nonumber \\
    &&
        % \qquad \qquad \qquad\qquad  \quad 
        P_{\indexbracket_5}\cdot \max(1,3) +P_{\indexbracket_6}\cdot \max(0,4) + P_{\indexbracket_7}\cdot \max(1,1) + P_{\indexbracket_8}\cdot \max(0,2)
    \nonumber \\
    &=&
        % \qquad\qquad\qquad\qquad = 
        4 \cdot \Pteami_{A,B}\cdot \Pteami_{C,D}\cdot \Pteami_{A,C} + 
        3 \cdot \Pteami_{A,B}\cdot \Pteami_{D,C}\cdot \Pteami_{A,D}  + 
        2 \cdot \Pteami_{A,B}\cdot \Pteami_{C,D}\cdot \Pteami_{C,A}  +
    \nonumber \\
    &&
        % \qquad\qquad\qquad\qquad\qquad\; 
        \Pteami_{A,B}\cdot \Pteami_{D,C}\cdot \Pteami_{D,A} +
        3 \cdot \Pteami_{B,A}\cdot \Pteami_{C,D}\cdot \Pteami_{B,C} + 
        4 \cdot \Pteami_{B,A}\cdot \Pteami_{D,C}\cdot \Pteami_{B,D} +
    \nonumber\\
    &&
        % \qquad\qquad\qquad\qquad\qquad\;
        \Pteami_{B,A}\cdot \Pteami_{C,D}\cdot \Pteami_{C,B}  + 
        2 \cdot \Pteami_{B,A}\cdot \Pteami_{D,C}\cdot \Pteami_{D,B}.
\end{eqnarray}
\label{eq:expectedFunction}
In general, the EMS for the two-entry solution $\mathcal{E}=\{\indexentry^1, \indexentry^2\}$ is given by 
\begin{eqnarray*}
    \mathbb{E}\left[S(\mathcal{E})\right]
    &=& 
        \sum_{\indexoutcome \in \completebrackets} P_{\indexoutcome}\max \Biggl(s(\indexentry^1,\indexoutcome), s(\indexentry^2,\indexoutcome)\Biggr)\nonumber
    \\
    &=&
        \sum_{\indexoutcome \in \completebrackets} P_{\indexoutcome}\max \Biggl(\sum_{\indexteam \in \setteams} \sum_{\indexgame \in \setgames} 2^{\indexround(\indexgame)-1} \cdot \indexentry^{1}_{\indexteam,\indexgame} \cdot O_{\indexteam,\indexgame},\quad \sum_{\indexteam \in \setteams} \sum_{\indexgame \in \setgames} 2^{\indexround(\indexgame)-1} \cdot \indexentry^{2}_{\indexteam,\indexgame} \cdot O_{ \indexteam,\indexgame}\Biggr). \label{eq:MultiEntries_eq1}
\end{eqnarray*}

The rest of this section
shows how one can evaluate, exactly, the EMS for any collection of entries.  The time complexity of our procedure is 
%worst-case run time is 
exponential in the number of entries, and our numerical evaluation in Section~\ref{sec:DPTime} shows that the computational runtime is prohibitively long, even for two entries. Therefore, we conclude this section with a simulation-based estimation approach for EMS that is computationally efficient and sufficiently accurate in practice.

\subsection{Exact Algorithm}
\label{sec:DPexact}

We now introduce the notation and the algorithm to prove  the following theorem:
\begin{theorem}\label{thm: Exact EMS}
    The EMS can be computed in time~$O\left(\nteams \cdot (\nteams \cdot \log_2{(\nteams))^{2\nentries+1}} \right) 
% \nteams^{2(\nentries+1)} 
 % \log_2{(\nteams)^{2\nentries+1}} %\cdot \nteams^2 \cdot (\frac{\nteams}{2}\log_2(\nteams))^{2|\setentries|}
$.
\end{theorem}
For any game $\indexgame$ and tournament~$\tourney$, we define the sub-tournament $\tourney(\indexgame) \coloneqq 
\angles{
\setteams(\indexgame),\setrounds(\indexgame),\setgames(\indexgame)}$, where %, where:
\begin{itemize}
    \item $\setteams(\indexgame)$ is as above, i.e., it contains %defined as above to be 
    all teams that might play in game $\indexgame$; 
    \item $\setrounds(\indexgame) \coloneqq \{1, \ldots, \indexround(\indexgame)\}$, i.e., the number of rounds is restricted to~$\indexround(\indexgame)$; and
    \item $\setgames(\indexgame) = \{\indexgame':\indexround(\indexgame') \leq \indexround(\indexgame),\setteams(\indexgame') \subseteq \setteams(\indexgame)\}$, i.e., only games leading to~$\indexgame$ are considered.
\end{itemize}
From the definition, $\tourney(\indexgame)$ can be interpreted as the restriction of a tournament~$\tourney$ to the games leading to~$\indexgame$. Similarly, $\tourney(\indexgame)$ is the sub-tournament of~$\tourney$ where~$\indexgame$ is the championship game.
%and consisting of all games and teams that lead from the first round to game $\indexgame$.

Similarly to tournaments, our algorithm also works with the projection of entries induced by games. Namely, for any game~$\indexgame$ and entry~$\indexentry$, we define the projection~$\indexentry(\indexgame)$ by restricting the entries of bracket~$\indexentry$ to rows and columns indexed by teams $\setteams(\indexgame)$ and games $\setgames(\indexgame)$, respectively. We use~$\setentries(\indexgame)$ to denote the projections in~$\indexgame$ of all entries in~$\setentries$.

\newcommand{\maxscore}{\ensuremath{\overline{s}}}

\subsubsection{Dynamic Programming Formulation}
% \noindent \textbf{Dynamic Programming Formulation:}
We present a dynamic programming (DP) formulation to compute the expected score of~$\setentries$ for a given tournament~$\tourney$ and team-by-team win matrix~$\Pteam$. We only consider the case of two entries for simplicity of exposition, but the extension to any given~$\nentries$ is straightforward. 

The state space of our formulation is a set of tuples~$(g,t,x,y)$ in~$\setgames \times \setteams \times \mathbb{N} \times \mathbb{N}$. Each state is associated 

with the joint probability~$Z\left[g,t,x,y\right]$ that
\begin{itemize}
    \item team $\indexteam$ wins game $\indexgame$; and
    \item entries $E^1(\indexgame)$ and $E^2(\indexgame)$ score $x$ and $y$ points in~$\tourney(\indexgame)$, respectively.
\end{itemize}

For any game~$\indexgame^*$, including the final game, the value of $\mathbb{E}(S(\setentries(\indexgame^*)))$ can be computed based on the values of~$Z\left[\indexgame^*,\indexteam,x,y\right]$ for every~$\indexteam$  in~$\setteams(g^*)$ and every pair of scores $(x,y)$ in~$[0,\maxscore(\indexgame^*)]$ as follows: %r(g^*) \cdot 2^{r(g^*) - 1}]$ 

\[
\mathbb{E}\left[S(\setentries(g^*))\right] = \sum_{t \in \setteams(g^*)} \sum\limits_{x = 0}^{r(g) \cdot 2^{r(g) - 1}} \, \sum\limits_{y = 0}^{r(g) \cdot 2^{r(g) - 1}} \max (x,y) \cdot Z\left[g^*,t,x,y\right].
\]

We show how to calculate $Z\left[g^*,t^*,x^*,y^*\right]$ given the values of~$Z\left[g,t,x,y\right]$ for all games in~$\setgames(\indexgame^*) \setminus \{\indexgame^*\}$, teams in~$\indexteam \in \setteams(\indexgame^*)$, and possible scores $x$ and $y$ for the two entries. The construction directly yields an inductive procedure to calculate $\mathbb{E}(S(\setentries))$ for~$\tourney$.  

Let $\indexteam^1$ and $\indexteam^2$ be the teams selected by the two entries in game $\indexgame^*$. These teams \textit{may or may not} each coincide with $\indexteam^*$. For ease of notation, let $g_\gamma = \gamma^{-1}(\indexgame^*,\indexteam^*)$ be the game that~$\indexteam^*$ must win to play~$\indexgame^*$, and let $g_\delta = \delta^{-1}(\indexgame^*,\indexteam^*)$ be the game its adversary must win to play~$\indexgame^*$. Moreover, let
\begin{itemize}
    \item $x_\gamma$ be the score of $E^1\left(g_\gamma\right)$  in $\tourney\left(g_\gamma\right)$;
    \item $x_\delta$ be the score of $E^1\left(g_\delta\right)$  in $\tourney\left(g_\delta\right)$;
    \item $y_\gamma$ be the score of $E^2\left(g_\gamma\right)$  in $\tourney\left(g_\gamma\right)$; and
    \item $y_\delta$ be the score of $E^2\left(g_\delta\right)$  in $\tourney\left(g_\delta\right)$.
\end{itemize}

We break into four cases to enumerate the ways in which entries~$E^1(\indexgame)$ and~$E^2(\indexgame)$ score~$x^*$ points and $y^*$ points in $\tourney(g^*)$, respectively. The cases are defined based on whether or not the teams selected in both entries for game~$\indexgame^*$ coincide with $t^*$.

\noindent \textbf{CASE 1 - Both entries are different from $t^*$}: In this case, (a) $x_\gamma + x_\delta = x^*$, and (b) $y_\gamma + y_\delta = y^*$, therefore
\[
Z\left[g^*,t^*,x^*,y^* \right] = 
\sum_{x_\gamma = 0}^{x^*} 
\sum_{y_\gamma = 0}^{y^*}
Z[g_\gamma,t^*,x_\gamma,y_\gamma ]
\sum_{t \in \setteams(g_\delta)}
\Pteami_{t^*,t}\,
Z[g_\delta,t,x^* - x_\delta,y^*-y_\delta ]. 
\]

\noindent \textbf{CASE 2 - Both entries select $t^*$}: In this case, (a) $x_\gamma + x_\delta = x^*-2^{\indexround(\indexgame^*)-1}$, and (b) $y_\gamma + y_\delta = y^*-2^{\indexround(\indexgame^*)-1}$, because both entries receive $2^{r(g^*)-1}$ points for correctly selecting $t^*$ to win $g^*$. Therefore
\begin{equation} \label{eq1}
\begin{split}
Z\left[g^*,t^*,x^*,y^* \right]
& = \sum_{x_\gamma = 0}^{x^*-2^{\indexround(\indexgame^*)-1}} 
\sum_{y_\gamma = 0}^{y^* - 2^{\indexround(\indexgame^*)-1}}
Z[g_\gamma,t^*,x_\gamma,y_\gamma ] \cdot \\
 &  \quad \quad \quad \sum_{t \in \setteams(g_\delta)}\Pteami_{\indexteam^*,\indexteam}
Z[g_\delta,t,x^* - x_\gamma - 2^{\indexround(\indexgame^*)-1},y^*-y_\gamma - 2^{\indexround(\indexgame^*)-1}] .
\end{split} \nonumber
\end{equation} 

\noindent \textbf{CASE 3 - $\indexentry^1$ selects $\indexteam^*$ and $\indexentry^2$ selects a team different from $\indexteam^*$}: In this case, (a) $x_\gamma + x_\delta = x^*-2^{\indexround(\indexgame^*)-1}$, and (b) $y_\gamma + y_\delta = y^*$, because only entry $\indexentry^1$ receives $2^{\indexround(\indexgame^*)-1}$ points for correctly selecting $\indexteam^*$ to win $\indexgame^*$. Therefore

\[
Z\left[\indexgame^*,\indexteam^*,x^*,y^* \right] = 
\sum_{x_\gamma = 0}^{x^*-2^{\indexround(\indexgame^*)-1}} 
\sum_{y_\gamma = 0}^{y^*}
Z[g_\gamma,t^*,x_\gamma,y_\gamma ]
\sum_{\indexteam \in \setteams(\indexgame_\delta)}\Pteami_{\indexteam^*,\indexteam}
Z[\indexgame_\delta,\indexteam,x^* - x_\gamma - 2^{\indexround(\indexgame^*)-1},y^*-y_\gamma ]. 
\]

\noindent \textbf{CASE 4 - $\indexentry^1$ selects a team different from $\indexteam^*$ and $\indexentry^2$ selects $\indexteam^*$}: In this case, (a) $x_\gamma + x_\delta = x^*$, and (b) $y_\gamma + y_\delta = y^*-2^{\indexround(\indexgame^*)-1}$, because only entry $\indexentry^2$ receives $2^{\indexround(\indexgame^*)-1}$ points for correctly selecting $\indexteam^*$ to win $\indexgame^*$. Therefore
\[
Z\left[\indexgame^*,\indexteam^*,x^*,y^* \right] = 
\sum_{x_\gamma = 0}^{x^*} 
\sum_{y_\gamma = 0}^{y^* - 2^{\indexround(\indexgame^*)-1}}
Z[\indexgame_\gamma,\indexteam^*,x_\gamma,y_\gamma ]
\sum_{\indexteam \in \setteams(\indexgame_\delta)}\Pteami_{\indexteam^*,\indexteam}
Z[\indexgame_\delta,\indexteam,x^* - x_\gamma,y^*-y_\gamma - 2^{\indexround(\indexgame^*)-1}]. 
\]

For the first-round games, we compute the probabilities using the same four cases as above. Let $\indexteam'$ be the opposing team to $\indexteam^*$ for a given first-round game. 

\noindent \textbf{CASE 1 - Both entries are different from $t^*$:} 
\begin{align*}
    Z[\indexgame^*, \indexteam^*, 0,0]&=\Pteami_{\indexteam^*, \indexteam'}\\
Z[\indexgame^*, \indexteam^*, 1,1]=Z[\indexgame^*, \indexteam^*, 1,0]=Z[\indexgame^*, \indexteam^*, 0,1]&=0
\end{align*}

\noindent \textbf{CASE 2 - Both entries select $t^*$:} 
\begin{align*}
    Z[\indexgame^*, \indexteam^*, 1,1]&=\Pteami_{\indexteam^*, \indexteam'}\\
Z[\indexgame^*, \indexteam^*, 0,0]=Z[\indexgame^*, \indexteam^*, 1,0]=Z[\indexgame^*, \indexteam^*, 0,1]&=0
\end{align*}

\noindent \textbf{CASE 3 - $\indexentry^1$ selects $t^*$ and $\indexentry^2$ selects a team different from $t^*$: } 
\begin{align*}
    Z[\indexgame^*, \indexteam^*, 1,0]&=\Pteami_{\indexteam^*, \indexteam'}\\
Z[\indexgame^*, \indexteam^*, 1,1]=Z[\indexgame^*, \indexteam^*, 0,0]=Z[\indexgame^*, \indexteam^*, 0,1]&=0
\end{align*}

\noindent \textbf{CASE 4 - $\indexentry^1$ selects a team different from $t^*$ and $\indexentry^2$ selects $t^*$: } 
\begin{align*}
    Z[\indexgame^*, \indexteam^*, 0,1]&=\Pteami_{\indexteam^*, \indexteam'}\\
Z[\indexgame^*, \indexteam^*, 1,1]=Z[\indexgame^*, \indexteam^*, 0,0]=Z[\indexgame^*, \indexteam^*, 1,0]&=0.
\end{align*}

\subsubsection{Space and Time Complexity of the Algorithm} 

The state space contains one element for each combination of scores, teams, and games, so the space complexity of the algorithm is~$O\left(\nteams^2(\frac{\nteams}{2}\log_2(\nteams))^{\nentries}\right)$. For the evaluation of $\mathbb{E}\left[S(\mathcal{E})\right]$, we need to compute all entries of~$Z$, an operation that consumes time~$O\left(\log_2{(\nteams)} \cdot \nteams^2 \cdot (\frac{\nteams}{2}\log_2(\nteams))^{2\nentries}\right)
$ (we present the derivation in Appendix~\ref{appendix:spacetime}). Therefore, the complexity of the algorithm is polynomial in the number of teams but exponential in the number of entries. Our numerical experiments show that this algorithm is very time-consuming for two entries and essentially intractable for three or more entries.

\subsection{Estimation by Simulation}
\label{sec:obj_evaluation}
Due to the difficulty in calculating the EMS, we use a Monte Carlo procedure to estimate~$\mathbb{E}\left[S(\mathcal{E})\right]$. The algorithm relies on~\SIM, a randomized simulation algorithm that generates brackets based on the win matrix~$\Pteam$. \SIM is an iterative procedure that makes the assignment round by round,
%We sample a bracket by iteratively 
selecting the winners of the games based on~$\Pteam$. Namely, for a game involving teams~$\indexteam$ and~$\indexteam'$, \SIM chooses team~$\indexteam$ as the winner of this game with probability~$\Pteami_{\indexteam,\indexteam'}$. After choosing all winners in round~$r$, \SIM proceeds to round~$r+1$ and repeats the same procedure until winners for all games have been selected. By construction, \SIM naturally favors results that are more likely to happen. Namely, any arbitrary element~$\indexoutcome$ of~$\setbrackets$ is generated with the same probability with which~$\indexoutcome$ is the observed outcome of~$\tourney$. 

We obtain an estimation~$\widehat{\mathbb{E}\left[S(\mathcal{E})\right]}$ of~$\mathbb{E}\left[S(\mathcal{E})\right]$ by applying the Monte Carlo method to a set of simulations generated by~\SIM. More precisely, let~$\setsimulations \coloneqq \{1, \ldots, \nsimulations\}$ be a set of~$\nsimulations$ simulations; we also refer to~$\setsimulations$ as the respective set of matrices, i.e., $\setsimulations \coloneqq \{\newsimulation{1}, \ldots, \newsimulation{\nsimulations}\}$. We obtain~$\widehat{\mathbb{E}\left[S(\mathcal{E})\right]}$ by computing the average score of the best performing entry in~$\mathcal{E}$ for each outcome in~$\setsimulations$, i.e., \begin{align}
\widehat{\mathbb{E}\left[S(\mathcal{E})\right]}
    \coloneqq 
        \frac{1}{\nsimulations}\sum_{\indexsimulation \in \setsimulations} \left(  \max_{\indexentry \in \mathcal{E} } \,s(\indexentry,\indexsimulation)\right).
    \label{eq:montecarlo}
\end{align}

\section{Structural Results}
\label{sec:StructuralProperties}

Section~\ref{sec:DPexact} presents an exact polynomial-time algorithm to calculate the EMS for a \textit{fixed} set of entries. With respect to actually identifying an \textit{optimal} set of entries, we conjecture that this problem is NP-hard even for~$\nentries = 2$. Therefore, this section examines the structural properties of the problem, including the characterization of optimal solutions for special cases. We explore these insights in the following sections to design high-quality heuristic algorithms. The proofs are presented in Appendices~\ref{appendix:subEMS} and~\ref{appendix:MultiEntryLB}.

\subsection{Submodularity}

Intuitively, the incorporation of entries can only help bettors interested in maximizing the expected score of the best-performing entry. However, one can show that the impact of adding new entries diminishes as the entry set grows; this observation is formally characterized in Proposition~\ref{prop:submodular}.

\begin{proposition}\label{prop:submodular}
The function $\mathbb{E}\left[S(\mathcal{E})\right]$ is monotone submodular.
\end{proposition}

Monotonic submodularity is a desirable property of set functions with significant algorithmic implications. Namely, a simple greedy algorithm has approximation guarantees under these conditions \citep{nemhauser1978analysis}. We could not obtain similar performance guarantees (for technical reasons we discuss below), but this observation is explored in one of our algorithms.

\subsection{Level of Diversification}

This subsection explores how much diversity is required in optimal entry sets.  On one end of the spectrum, an optimal solution consists of several copies of the same entry; on the other, one might diversify as much as possible.  We show that this range of diversity in optimal entries is largely related to the relative strength of the teams in the tournament. 

We start with a few simple examples highlighting how some apparently intuitive properties of multi-entry bets actually \textit{do not} hold in practice. Remark~\ref{sp:1} shows that an optimal $\nentries$-entry set cannot always be derived from an optimal $(\nentries-1)$-entry set. In particular, a bettor may not want to pick the best single entry as part of a multi-entry strategy. 
\begin{Remark}
    \label{sp:1}
An entry with the highest expected score does not necessarily compose an optimal $\nentries$-entries solution for~$\nentries \geq 2$.
\end{Remark}

We illustrate Remark~\ref{sp:1} using the 4-team tournament illustrated in Figure~\ref{fig:example} parameterized by the win matrix~$\Pteam$ presented in Table~\ref{tab:VictoryMatrix1}. Entry~$\indexbracket_1$ has the highest expected score of  2.0515 points, and the best complementary entry to $\indexbracket_1$ is $\indexbracket_7$, yielding an EMS of 2.72275 points. However, the entry set $\{\indexbracket_2, \indexbracket_3\}$ attains an EMS of 2.83425 points.

\medskip
\begin{minipage}[c]{0.5\textwidth}
\centering
    \begin{tabular}{|c|c|c|c|c|}
\hline
       & Team A & Team B & Team C & Team D \\ \hline
Team A & -      & 70\%   & 55\%   & 60\%   \\ \hline
Team B & 30\%   & -      & 40\%   & 45\%   \\ \hline
Team C & 45\%   & 60\%   & -      & 55\%   \\ \hline
Team D & 40\%   & 55\%   & 45\%   & -      \\ \hline
\end{tabular}
\captionof{table}{$\Pteam$ Matrix for Remark~\ref{sp:1}}\label{tab:VictoryMatrix1}
\end{minipage}
\begin{minipage}[c]{0.5\textwidth}
\centering
\begin{tabular}{|c|c|c|c|c|}
\hline
       & Team A & Team B & Team C & Team D \\ \hline
Team A & -      & 80\%   & 80\%   & 80\%   \\ \hline
Team B & 20\%   & -      & 40\%   & 45\%   \\ \hline
Team C & 20\%   & 60\%   & -      & 55\%   \\ \hline
Team D & 20\%   & 55\%   & 45\%   & -      \\ \hline
\end{tabular}
\captionof{table}{$\Pteam$ Matrix for Remark~\ref{sp:3}}\label{tab:VictoryMatrix2}
\end{minipage}
\medskip

The next remark shows that entry diversification strategies are nuanced. In the scoring system adopted by most \textit{March Madness} pools, the entries that choose the correct champion score at least 63 points (almost 33\% of the maximum score). This observation advocates for diversification in the championship game, i.e., bettors have a good reason to use entries with different champions. However, Remark~\ref{sp:3} shows that this strategy is sub-optimal in general. One example involves a 4-team tournament parameterized by the win  matrix~$\Pteam$ presented in Table~\ref{tab:VictoryMatrix2}, where $\{\indexbracket_1,\indexbracket_2\}$ is the optimal two-entry solution, and both select Team A as the winner of the tournament.

\begin{Remark}
    \label{sp:3}
    Optimal multi-entry sets may select the same tournament winner across all entries. 
\end{Remark}

Examining the matrices in Table~\ref{tab:VictoryMatrix1} and Table~\ref{tab:VictoryMatrix2}, we see a difference in the relative strengths of the teams. In particular, in the matrix in Table~\ref{tab:VictoryMatrix1}, the win probabilities are much closer to 0.5 than they are in Table~\ref{tab:VictoryMatrix2}.  This is why Remark~\ref{sp:1} and Remark~\ref{sp:3} use them respectively; Remark~\ref{sp:1} relates to more diversity while Remark~\ref{sp:3} relates to less diversity. 

We now show how one can formally identify optimal entries for extreme scenarios of strength differential among teams. For the case where all elements of the team-by-team win matrix are binary, a bettor does not need multiple entries, as the problem is rendered trivial in these cases; there is just one possible outcome for the tournament, and the score for a participant is maximized by selecting that outcome as the entry.  This means that diversity is not important at all.

\begin{Remark} \label{Binary Case}
    If the win matrix is binary, there is no need for diversity in entries. Moreover, the optimal entry selects the winner in each game.
\end{Remark}

In practice, many elements in the win matrix may be closer to 0.5 than 1.0. Therefore, multi-entry sets for \textit{March Madness} pools benefit from diversification strategies, whereby different winners are selected for one or more games across entries. 

One extreme diversification strategy consists of adopting \textit{disjoint} entries; two entries are disjoint if they make different choices for all games, i.e., entries~$\indexentry^1$ and~$\indexentry^2$ are disjoint if $\indexentry^1_{\indexteam, \indexgame}+\indexentry^2_{\indexteam,\indexgame}\leq 1$ for every team~$\indexteam$ in $\setteams$ and game~$\indexgame$ in $\setgames$. This relatively simple idea is surprisingly powerful in the special case where all the entries of the team-by-team win matrix equal 0.5; Theorem~\ref{sp:2} shows that \textit{any} pair of disjoint entries compose an optimal two-entry solution in these cases.

\begin{theorem}
    \label{sp:2}
    If $\Pteam_{\indexteam,\indexteam'} =0.5$ for all teams~$\indexteam,\indexteam'$ in~$\setteams$, an entry set~$\setentries = \{\indexentry,\indexentry'\}$ is optimal if and only if $\indexentry$ and~$\indexentry'$ are  disjoint.
\end{theorem}

Based on Remark~\ref{sp:3} and Remark~\ref{Binary Case} versus Remark~\ref{sp:1} and Theorem~\ref{sp:2}, we see how diversity in optimal entries manifests; when the probabilities are closer to binary, less diversification is necessary, but as the probabilities near 0.5, diversity becomes critical. 

\subsection{Provable Lower Bounds}

In the worst-case scenario, the minimum score attained by any single entry is zero (which happens if all choices for the first-round games are incorrect). This outcome is unlikely, but one cannot derive better \textit{deterministic} guarantees for single entries. In contrast, stronger deterministic lower bounds can be achieved with multiple entries. For this, we explore the idea of \textit{complementary entries}; we say that a set of entries is complementary if it contains at least one entry for each possible outcome of a particular game. Proposition~\ref{prop:det_lb} shows that there exists a two-entry strategy focused on the round-one games that always score at least $\frac{\nteams}{4}$ points. 
\begin{proposition}\label{prop:det_lb}
There is a set $\mathcal{E}$ composed of two entries such that $S(\mathcal{E}) \geq \frac{\nteams}{4}$.
\end{proposition}

The same idea can be extended for even stronger bounds (e.g., by incorporating more entries and achieving guarantees for later rounds; see Appendix~\ref{appendix:MultiEntryLB}). However, the number of entries makes this strategy impractical even for very modest improvements. 

Proposition~\ref{prop:det_lb2} uses an alternative approach focused on games played in later tournament rounds. For a moderate number of entries, this strategy dominates the one in Proposition~\ref{prop:det_lb}.

\begin{proposition}\label{prop:det_lb2}
For every~$\indexround$ in $\setrounds$, there is a set $\mathcal{E}$ composed of $2^{r}$ entries such that
$S(\mathcal{E}) \geq 2^{r}-1.$
\end{proposition}

In practice, the strategies used in Propositions~\ref{prop:det_lb} and~\ref{prop:det_lb2} do not perform well, especially because upsets do not happen so frequently (recall Table~\ref{tab:upsetTable}). Nevertheless, we present multi-entry strategies that explore the idea of complementary entries. % more effectively (see~\S\ref{sec:SIP}).

\section{Optimization}
\label{sec:Optimization}
This section details our models and algorithms for solving~\eqref{eq:MultiEntryProblem}.  We also discuss the parameters used by each algorithm to streamline the discussion in Section~\ref{sec:ExperimentalResults} and Section~\ref{sec:DKContest},
where we detail our numerical results. First, we introduce~\ref{ipmm}, an exact MIP formulation for the single-entry problem. \citet{kaplan2001march} present an exact \emph{dynamic programming} algorithm to solve the same problem in polynomial time, but~\ref{ipmm} is more convenient for the extension to the multi-entry case.
\[
\begin{array}{lllllll}\label{ipmm}
\tag{IP} \max  & \sum\limits_{\indexgame\in\setgames} \sum\limits_{\indexteam\in \setteams(\indexgame)} 2^{\indexround(\indexgame)-1} \cdot
x_{\indexteam, \indexround(\indexgame)} \cdot
\Proundi_{\indexteam,\indexround(\indexgame)}  \\[2ex] 
    \text{s.t}. 
    & \sum\limits_{\indexteam \in \setteams(\indexgame)} x_{\indexteam, \indexround(\indexgame)}=1, && \forall 
    \indexgame\in\setgames &&& (\text{\ref{ipmm}}.1) \\[2ex]
    & x_{\indexteam, \indexround}\leq x_{\indexteam, \indexround-1}, && \forall t\in\setteams, \indexround\in\setrounds\setminus\{1\} &&&(\text{\ref{ipmm}}.2) \\[2ex]
    & x_{\indexteam, \indexround}\in \{0,1\}, &&\forall \indexteam\in\setteams, \indexround\in\setrounds.
%    & (\texttt{IP\textsuperscript{MM}}-3) 
\end{array}
\]

Formulation~\ref{ipmm} uses binary variables~$x_{\indexteam,\indexround}$, where~$x_{\indexteam,\indexround} = 1$ if team $\indexteam$ is selected in round $\indexround$ and~$x_{\indexteam,\indexround} = 0$ otherwise. Constraints~\ref{ipmm}.1 assert that exactly one team is selected for each game. Constraints~\ref{ipmm}.2 enforce coherence in the selection process by forcing the selection of team~$\indexteam$ in round~$\indexround-1$ whenever~$\indexteam$ is selected in round~$\indexround$. The first two conditions assert the feasibility of the chosen bracket, so every feasible solution to~\ref{ipmm} directly maps to a feasible bracket and vice-versa. Finally, the objective function calculates the expected number of points for the selected entry. %Note that one could also define an MIP with binary variables indexed by team and game, but using the variables above is more compact and solves quickly. 

\subsection{Sample Average Approximation}
\label{sec:SAA}

The multi-entry case has no known closed-formula expression, so a natural approach uses a sample average approximation algorithm (\texttt{SAA}). Our implementation of~\texttt{SAA} relies on the solution of~\ref{SAA}, the following mathematical programming formulation based on~\ref{ipmm}:
\[
\begin{array}{llllllll}\label{SAA}
% \tag{\texttt{SAA}($\nentries$, $\indexoutcome_{sim}$)} 
\tag{\texttt{SAA}} 
\max  & 
    \frac{1}{\nsimulations} \sum\limits_{\indexsimulationi=1}^{\nsimulations} s^{\max}_{\indexsimulationi} \\ [2ex]
    & \sum\limits_{\indexteam \in T(g)} x_{\indexteam, \indexround(\indexgame), \indexentryi} = 1 && \forall \indexgame \in \setgames, \forall \indexentryi \in [\nentries] &&& \text{(\ref{SAA}.1)}\\[2ex] %\label{eq:c2_2}\\
    &    x_{\indexteam, \indexround, \indexentryi} \leq x_{\indexteam, \indexround-1, \indexentryi} && \forall \indexteam \in \setteams, \forall \indexround \in \setrounds \backslash \{1\}, \forall \indexentryi \in [\nentries] &&& \text{(\ref{SAA}.2)} \\ [2ex]
     &  s_{\indexsimulationi,\indexentryi} = \sum\limits_{\indexgame \in \setgames}2^{\indexround(\indexgame)-1} \cdot \indexsimulation^{\indexsimulationi}_{t,g} \cdot x_{\indexteam, \indexround(\indexgame), \indexentryi} \quad && \forall \indexsimulationi \in [\nsimulations], \forall \indexentryi \in [\nentries] &&& \text{(\ref{SAA}.3)} \\[2ex]

    & \sum\limits_{\indexentryi \in [\nentries]} z_{\indexsimulationi,\indexentryi}=1 && \forall \;  \indexsimulationi \in [\nsimulations]&&&\text{(\ref{SAA}.4)} \\[2ex]

    & s^{\text{max}}_{\indexsimulationi} \leq s_{\indexsimulationi,\indexentryi}z_{\indexsimulationi,\indexentryi} + M(1-z_{\indexsimulationi,\indexentryi}) && \forall \;  \indexsimulationi \in [\nsimulations],\forall \;  \indexentryi \in [\nentries] &&&\text{(\ref{SAA}.5)} \\[2ex]
     
    & x_{\indexteam,\indexround,\indexentryi}  \in \{0,1\}   && \forall \indexteam \in \setteams, \forall \indexround \in \setrounds, \forall \indexentryi \in [\nentries] &&&  \\[2ex] %\label{eq:c3_6}\\
    & s_{\indexsimulationi,\indexentryi}, s^{\max}_{\indexsimulationi} \in \mathbb{R}_{+}    && \forall \indexsimulationi \in [\nsimulations], \forall \indexentryi\in[\nentries]. &&&%\label{eq:c3_7}
\end{array}
\]
Formulation~\ref{SAA} uses binary variable $x_{\indexteam,\indexround,\indexentryi}$ to indicate whether team~$\indexteam$ is chosen in round $\indexround$ for entry $\newindexentry{\indexentryi}$, 
variable $s_{\indexsimulationi,\indexentryi}$ to keep the score of entry $\newindexentry{\indexentryi}$ for simulation $\newsimulation{\indexsimulationi}$, 
variable $s^{\max}_\indexsimulationi$ to determine the highest-scoring entry for~$\newsimulation{\indexsimulationi}$, and binary variable~$z_{\indexsimulationi,\indexentryi}$ to indicate if entry $\newindexentry{\indexentryi}$ is a highest-scoring entry for~$\newsimulation{\indexsimulationi}$.
Constraints~\ref{SAA}.1 and~\ref{SAA}.2 are similar to their counterparts in formulation~\ref{ipmm}. Constraints~\ref{SAA}.3 compute the score of each  simulation~$\newsimulation{\indexsimulationi}$. Finally, we use a set of big-$M$ constraints in~\ref{SAA}.4 and~\ref{SAA}.5 to set~$s^{max}_{\indexsimulationi}$ as the highest score attained by an entry for each~$\newsimulation{\indexsimulationi}$; we set $M=\nrounds\,2^{\nrounds-1}$ (i.e., the maximum score for a $\nrounds$-round tournament). 

\ref{SAA} can be used to deliver solutions with statistical optimality guarantees. However, our experiments show that~\ref{SAA} does not scale in the number of entries, mainly because of the complexity of the underlying optimization problem and the number of simulations required for tight confidence intervals. Nonetheless, we report the results identified by~\ref{SAA} when $\nentries$ is relatively small. In these instances, our approach involves running ten independently generated~\ref{SAA} formulations, each with a time limit of $500 \cdot \indexteam$ seconds, and selecting the solution with the highest EMS evaluated out-of-sample. This motivates the investigation of faster and more scalable solution approaches that sacrifice global guarantees in favor of performance.

\subsection{Sequential Algorithms}
\label{sec:SequentialAlgo}
In this subsection, we present two algorithms that adopt a sequential approach to construct entries. Both algorithms are iterative greedy algorithms that explore the idea of complementary entries discussed in Section~\ref{sec:StructuralProperties}. The first algorithm adapts the procedure proposed by~\cite{hunter2016picking} for daily fantasy sports. The second approach, \GreedySAA, is an iterative algorithm that mimics \ref{SAA} but only requires selecting one entry at a time, thereby leading to scalability. We show in our experiments that \GreedySAA is the better of the two approaches.

\subsubsection*{Sequential Integer Programming Model ($\mathtt{SIP}$) }\label{sec:SIP}
$\mathtt{SIP}$ is an iterative algorithm that solves the single-entry problem in each step. $\mathtt{SIP}$ uses Formulation~\ref{ipseq}, an extension of~\ref{ipmm} that incorporates a set of constraints %into \ref{ipmm} 
after each step to ensure diversification across entries.  Namely,
\[
\begin{array}{llllllll}\label{ipseq}
\tag{$\mathtt{SIP}(\mathbf{\overline{x}})$} \max  & \sum\limits_{g\in\ngames} \sum\limits_{t\in T(g)} 2^{\indexround(g)-1} \Pgamei_{t,\indexgame} x_{t, \indexround(g)} \\[2ex] %\label{eq:obj1} \\
    %\text{s.t}. 
     & \sum\limits_{\indexteam \in \setteams(\indexgame)} x_{\indexteam, \indexround(\indexgame)}=1 && \forall \indexgame\in\setgames &&&\text{(\ref{ipseq}.1)}  \\[2ex]
     & x_{\indexteam, \indexround}\leq x_{\indexteam, \indexround-1} && \forall \indexteam \in \setteams, \forall \indexround \in \setrounds \backslash \{1\} &&&\text{(\ref{ipseq}.2)} 
    \\[2ex]
    & D(\mathbf{\overline{x}})  && &&&\text{(\ref{ipseq}.3)}\\[2ex]
    & x_{\indexteam, \indexround} \in \{0,1\} &&\forall \indexteam \in \setteams, \forall \indexround \in \setrounds. \nonumber
\end{array}
\]
\ref{ipseq} is parameterized by a three-dimensional binary matrix~$\mathbf{\overline{x}}$, where each entry~$\overline{x}_{\indexteam, \indexround, \indexentryi'}$ indicates if team $\indexteam$ is selected in round $\indexround$ by entry~$\indexentryi'$, $1 \leq \indexentryi' < \indexentryi$. The set of variables and constraints used in Formulation~\ref{ipseq} are identical to those in Formulation~\ref{ipmm}, with the sole distinction being the inclusion of the set of constraints~$D(\mathbf{\overline{x}})$. In its first iteration, $\overline{x} = \boldsymbol{0}$, so $D(\mathbf{\overline{x}})$ is empty and~\ref{ipseq} selects an optimal single-entry solution. After each iteration, we update~$\overline{x}$ and incorporate constraints into~$D(\mathbf{\overline{x}})$ as follows.

First, as discussed before, an entry scores 63 points (out of 192) in a typical \textit{March Madness} pool if it selects the champion correctly. One diversification idea is to limit the number of times a team~$\indexteam$ is selected as the champion in an entry based on the probability with which~$\indexteam$ wins the competition.
Therefore, we consider the following diversification conditions for the last two rounds (i.e., semifinals and championship games):
\begin{itemize}
    \item \textit{Champion Constraint}: Team $\indexteam$ cannot be selected by more than $\left\lceil \nentries\cdot \Proundi_{\indexteam,\indexround}\right \rceil$ if~$\indexround \geq \nrounds-1$. For example, for $\nentries=2$, \SIP can only select team $\indexteam$ twice in round $\nrounds \geq \nrounds-1$ if $\Pteami_{\indexteam, \indexround}>0.5$. 
\end{itemize}

\begin{itemize}
    \item \textit{Finalist Constraint}: Each entry has a unique pair of teams reaching the final game. 
\end{itemize}
Additionally, for the first four rounds, we consider the following  conditions:

\begin{itemize}
    \item \textit{Global Constraint}:  Any pair of entries can have at most~$\sigma$ identical selections throughout the entire tournament.\label{constraint:sigma}
    %when compared to any other previously chosen entry; and
    \item \textit{Round Constraints}: Any pair of entries can have at most $\sigma_{\indexround}$ identical selections in round $\indexround \in \{1,2,3,4\}$.\label{constraint:sigmaR}
\end{itemize}

Hyperparameter tuning experiments (presented along with the calibration of the algorithms in Appendix~\ref{appendix:Calibration}) show that the best parameterization of~$D(\overline{x})$ depends on the number of entries~$\nentries$. We use the following configurations of \SIP in our experiments:
\begin{itemize}
    \item $\nentries\in\{2,\ldots,4\}$: \textit{Champion Constraint}, \textit{Finalist Constraint}, and \textit{Round Constraints} with $\sigma_1=30$, $\sigma_2=11$, $\sigma_3=7$, and $\sigma_4=1$;
    \item $\nentries\in\{5,\ldots,9\}$: \textit{Champion Constraint}, \textit{Finalist Constraint}, and \textit{Round Constraints} with $\sigma_1=31$, $\sigma_2=13$, $\sigma_3=6$, and $\sigma_4=2$;
    \item $\nentries\in\{10,\ldots,25\}$: \textit{Champion Constraint}, \textit{Finalist Constraint}, and \textit{Round Constraints} with $\sigma_1=32, \sigma_2=15, \sigma_3=7,$ and $\sigma_4=4$; and 
    \item $\nentries\geq26$: \textit{Champion Constraint}, \textit{Finalist Constraint}, and \textit{Global Constraint} with $\sigma=54$.
\end{itemize}

\subsubsection*{Greedy Sample Average Approximation (\GreedySAA)}
\label{sec:G-SAA}

Algorithm~\ref{algo:GSAA} presents \GreedySAA, an iterative multi-entry greedy policy based on~\ref{SAA}. In each step, \GreedySAA incorporates an entry that maximizes the average score of the new entry set for a set of simulated outcomes. 
\begin{figure}[ht!]
    \centering
\begin{algorithm}[H]
\SetAlgoLined
$\mathcal{E}_0 \gets \emptyset$ \\[1ex]
\For{$\indexentryi$ in $\{1,2,\ldots,\nentries\}$ }{\vspace{.2cm}
Generate $\setsimulations$ using \SIM\\\vspace{.2cm}
% $\Bar{S}\gets \vec{0}$\\
\For{$\indexsimulationi=[\nsimulations]$}{
$\displaystyle \Bar{s}_\indexsimulationi  \gets \max\left\{\max_{\indexentry \in \setentries}\left\{ s(\indexentry, \newsimulation{\indexsimulationi})\right\},0\right\}$ \\
}

$\indexentry^* \gets \texttt{G-SAA}(\Bar{s})$ \\\vspace{.2cm}
$\mathcal{E}_{\indexentryi} \gets \mathcal{E}_{\indexentryi-1} \cup \{\indexentry^*\}$
}
\caption{Greedy Sample Average Approximation Algorithm ($\texttt{G-SAA}$)}
\label{algo:GSAA}
\end{algorithm}
\end{figure}
More precisely, in each step $\indexentryi \in [\nentries]$, \GreedySAA starts with an entry set~$\setentries_{\indexentryi-1}$ and samples a set~$\setsimulations$ with $\nsimulations$ outcomes. Given the vector~$\Bar{s}$ containing the maximum score attained by the entries in~$\setentries_{\indexentryi-1}$ for each outcome in~$\setsimulations$, \GreedySAA solves~\ref{SAAMAX} to identify an entry~$\indexentry^*$ that maximizes~$\widehat{\mathbb{E}_{\setsimulations}\left[S(\setentries \cup \{\indexentry^*\})\right]}$. Namely,
\[
\begin{array}{llllllll}\label{SAAMAX}
% \tag{\texttt{SAA}($\nentries$, $\indexoutcome_{sim}$)} 
\tag{\texttt{G-SAA}($\Bar{s}$)} 
\max  & 
    \frac{1}{\nsimulations} \sum\limits_{\indexsimulationi=1}^{\nsimulations} s^{\max}_{\indexsimulationi} \\ [2ex]
    & \sum\limits_{\indexteam \in T(g)} x_{\indexteam, \indexround(\indexgame)} = 1, && \forall \indexgame \in \setgames &&& \text{(\ref{SAAMAX}.1)}\\[2ex] %\label{eq:c2_2}\\
    &    x_{\indexteam, \indexround} \leq x_{\indexteam, \indexround-1}, && \forall \indexteam \in \setteams, \forall \indexround \in \setrounds \backslash \{1\} &&& \text{(\ref{SAAMAX}.2)} \\ [2ex]
     & s_{\indexsimulationi} =  \sum\limits_{\indexgame \in \setgames}2^{\indexround(\indexgame)-1}\indexsimulation^{\indexsimulationi}_{\indexteam,\indexgame}x_{\indexteam, \indexround(\indexgame)}, && \forall \indexsimulationi \in [\nsimulations] &&& \text{(\ref{SAAMAX}.3)} \\[2ex]
    % &  z_{\indexsimulationi}+\Bar{z}_{\indexsimulationi}=1 && \forall \;  \indexsimulationi \in [\nsimulations]&&&\text{(\ref{SAAMAX}.4)} \\[2ex]

    & s^{\text{max}}_{\indexsimulationi} \leq s_{\indexsimulationi}z_{\indexsimulationi} + M(1-z_{\indexsimulationi}), && \forall \;  \indexsimulationi \in [\nsimulations]&&&\text{(\ref{SAAMAX}.4)} \\[2ex]
    & s^{\text{max}}_{\indexsimulationi} \leq \Bar{s}_{\indexsimulationi}(1-z_{\indexsimulationi}) + M z_{\indexsimulationi}, && \forall \;  \indexsimulationi \in [\nsimulations]&&&\text{(\ref{SAAMAX}.5)} \\[2ex]
     
    & x_{\indexteam,\indexround}  \in \{0,1\},   && \forall \indexteam \in \setteams, \forall \indexround \in \setrounds&&&  \\[2ex] %\label{eq:c3_6}\\
    & s_{\indexsimulationi}, s^{\max}_{\indexsimulationi} \in \mathbb{R}_{+},z_{\indexsimulationi}  \in \{0,1\},    && \forall \indexsimulationi \in [\nsimulations]. &&&
\end{array}
\]
Formulation~\ref{SAAMAX} uses binary variable~$x_{\indexteam,\indexround}$ indicating whether team $\indexteam$ is chosen in round $\indexround$,  
variable~$s_{\indexsimulationi}$ to keep the score of~$\indexentry^*$ for simulation $\newsimulation{\indexsimulationi}$, 
variable $s^{\max}_\indexsimulationi$ to determine the highest score attained by an entry in~$\setentries_{\nentries} \cup \{\indexentry^*\}$ for~$\newsimulation{\indexsimulationi}$, and binary variable~$z_{\indexsimulationi}$ indicating if~$\indexentry^*$ is the highest-scoring entry for~$\newsimulation{\indexsimulationi}$. Constraints~\ref{SAAMAX}.1 and~\ref{SAAMAX}.2 are similar to their counterparts in formulation~\ref{ipmm}. Constraints~\ref{SAAMAX}.3 compute the score of~$\indexentry^*$  for each simulation in~$\setsimulations$. Finally, we use a set of big-$M$ constraints in~\ref{SAAMAX}.4 and~\ref{SAAMAX}.5 to set~$s^{max}_{\indexsimulationi}$ as the highest score attained for~$\newsimulation{\indexsimulationi}$, where $M=\nrounds2^{\nrounds-1}$. 

Proposition~\ref{prop:submodular} shows that~$\texttt{G-SAA}$ could be transformed into a~$(1 - \frac{1}{e})$-approximation algorithm to~\eqref{eq:MultiEntryProblem} \citep{nemhauser1978analysis} if the simulation-based step were replaced with an exact approach, i.e., if we could select in each step the entry that maximizes the expected score given the pre-selected set of entries. It is unclear whether such an algorithm exists, though, and it follows from the results in~\cite{hassidim2017submodular} that~$\texttt{G-SAA}$ can deliver arbitrarily bad solutions if a sub-optimal algorithm (such as our simulation-based algorithm) is used instead. However, our experiments show that Algorithm~\ref{algo:GSAA} performs well in practice.

Differently from \ref{SAA}, the performance of \GreedySAA is relatively robust, so we perform a single execution of the algorithm for each instance in our experiments. Additionally, we set a time limit of 500 seconds per entry and parameterize the algorithm with 250 random samples.

\subsection{Proportion Algorithm}
\label{sec:ProportionAlgorithm}

The shortcoming of \GreedySAA is that it iteratively finds entries, thereby losing global information in each step.  To adapt the quality of the solutions found by \GreedySAA into an algorithm that selects entries in a more integrated way, we present \Prop, a simple problem-specific algorithm that selects teams as the winner of games based on the probability with which they reach and win in that round. Insights derived from~\GreedySAA show that the selection probability slightly deviates from these estimates, especially in the early and late rounds of the tournament (when favorites and underdogs must be over-selected, respectively), so we incorporate these ideas to enhance~\Prop and derive~\Propplus, a heuristic that outperforms the other algorithms presented in this paper.

\subsubsection{Baseline procedure (\Prop)}

\begin{figure}[ht!]
    \centering
    \includegraphics[width=0.7\linewidth]{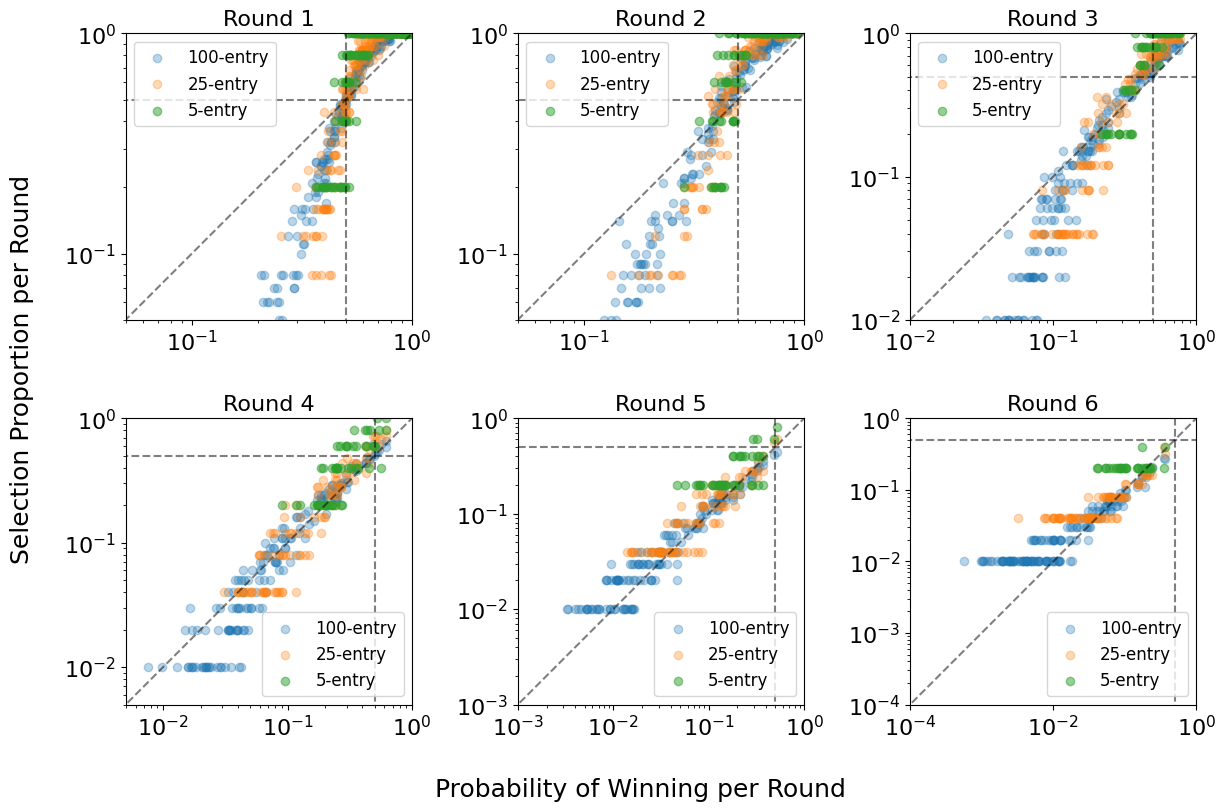}
    \caption{Probability of Winning vs Selection Distribution of \texttt{G-SAA} by Round}
    \label{fig:ProbWinningvsPickDist}
\end{figure}

The Proportion Algorithm relies on a key insight derived from Figure~\ref{fig:ProbWinningvsPickDist}, which compares the frequency \GreedySAA selects a team $\indexteam$ on 5-entry, 25-entry, and 100-entry solutions ($y$-axis) in each round versus~$\Proundi_{\indexteam,\indexround}$ ($x$-axis). A superficial analysis of these results shows that~$\Pround$ is a good predictor of how~\GreedySAA chooses its entries. Based on this observation, we design~\Prop, an iterative procedure that constructs all entries sequentially while maintaining a certain level of integration in the decision-making process. \Prop is presented in Algorithm~\ref{algo:ProundHeuristic}. 
\begin{figure}[ht!]
    \centering
    \footnotesize
\begin{algorithm}[H]
\label{algo:ProundHeuristic}
\SetAlgoLined
$q \gets \vec{0}^{\nentries}$ \\
$\mathbf{N} \gets \mathbf{0}^{\nteams\times \nrounds} $ \\
\For{$\indexentryi \in [\nentries]$}{
    $\newindexentry{\indexentryi} \gets \mathbf{0}^{\nteams \times \ngames}$ \\
    }
\For{$\indexround = \nrounds, \nrounds-1, \ldots, 1$}{
    $\mathcal{C} \gets [\nentries]$ \\
    \While{$\mathcal{C} \neq \emptyset$}{
        $\tilde{\indexentryi} = \arg\min \limits_{\indexentryi \in \mathcal{C}} q_{\indexentryi}$ \\
        $\tilde{\setgames} = \left\{
            \indexgame \in \setgames : 
                \indexround(\indexgame) = \indexround,
                \sum \limits_{\indexteam \in \setteams(\indexgame)} \newindexentry{\tilde{\indexentryi}}_{\indexteam,\indexgame} = 0
        \right\}$ \\ 
        $\tilde{\setteams} = \bigcup \limits_{\indexgame \in \tilde{\setgames}} \setteams(\indexgame) $ \\
        %\text{ and } \Proundi_{\indexteam,\indexround}>\widehat{P_{\indexround}}  \right\}$\\ 
        $\tilde{\indexteam} = 
            \arg\max \limits_{\indexteam \in \tilde{\setteams} } 
            \left[
            (N_{\indexteam,\indexround}-\Proundi_{\indexteam,\indexround}\nentries)^2 -
                (N_{\indexteam,\indexround}+1-\Proundi_{\indexteam,\indexround}\nentries)^2
                \right]
        $ \\
            \For{$\tilde{\indexgame} \in \setgames : r(\tilde{g}) \leq \indexround \text{ and } \tilde{\indexteam} \in \mathcal{T}(\tilde{g})$}{$\newindexentry{\tilde{\indexentryi}}_{\tilde{\indexteam},\tilde{\indexgame}} = 1$ \\
            $q_{\tilde{\indexentryi}} = q_{\tilde{\indexentryi}} + 2^{\indexround(\tilde{\indexgame})-1}\cdot \Pround_{\tilde{\indexteam}, \indexround(\tilde{\indexgame})}$\\
            $N_{\tilde{\indexteam},\indexround(\tilde{\indexgame})}=N_{\tilde{\indexteam},\indexround(\tilde{\indexgame})}+1$
            } 
        \If{$\sum_{\indexteam\in \setteams}\newindexentry{\tilde{\indexentryi}}_{\indexteam,\indexround}=2^{\nrounds-\indexround}$}{
  $\mathcal{C} \gets \mathcal{C} \backslash \{\tilde{\indexentryi}\}$
}
        }
} 
\caption{Proportion Algorithm (\Prop)}
\end{algorithm}
    % \caption{Caption}
\end{figure}

\Prop uses variables~$N_{\indexteam,\indexround}$ to store the number of times team~$\indexteam$ is chosen for a game in round~$\indexround$, $q_{\indexentryi}$ to indicate the current score of entry~$\indexentryi$, and~$\newindexentry{\indexentryi}$ to represent entry~$\indexentryi$; all these variables are initialized to zero. Each iteration is associated with a round~$\indexround$, starting from~$\nrounds$ (the championship round) and moving backward to the first round. In each iteration and for each entry, \Prop selects some winner~$\indexteam$ for each game in~$\indexround$ that is still unfilled and back-propagates this choice by selecting~$\indexteam$ for all the previous rounds, thus enforcing consistency. Therefore, except for the first iteration (last round), half of the games in round~$\indexround$ will have already been filled by the time the iteration associated with~$\indexround$ starts. \Prop sets~$\nentries\cdot\Proundi_{\indexteam,\indexround}$ as the target value for each~$N_{\indexteam,\indexround}$, and it proceeds with the selection for all entries based on their current scores. Namely, \Prop sorts the entries in ascending order based on their current expected scores (zero points are awarded to games without a selection). Then, for each entry, the algorithm assigns a team to its specific game in the current round such that it minimizes the distance between the target value~$\nentries\cdot\Proundi_{\indexteam,\indexround}$ and $N_{\indexteam,\indexround}$, using a squared error metric.  All variables are updated accordingly, and the algorithm proceeds until all entries are filled. 

\subsubsection{Enhanced procedure (\Propplus)}

Our numerical results show that~\Prop underperforms compared to the other algorithms. However, a more careful inspection of Figure~\ref{fig:ProbWinningvsPickDist} shows significant deviations between the selections of \GreedySAA and the entries of~$\Pround$. The diagonal line indicates the points where these two values coincide, and points above (below) show teams that are over-selected (under-selected) with respect to~$\Pround$.  The plots reveal that the proportion of entries for which~\GreedySAA selects each team in each round is \emph{nearly} identical to the probability that the team wins that round, with a slight caveat; it is risk-averse in the early rounds and relatively aggressive toward the end of the competitions. Namely, \GreedySAA over-selects the favorite teams in the first rounds, slowly changing its strategy and favoring underdog teams in later rounds. Moreover, the rate of change in the strategy depends on the number of entries;  Figure~\ref{fig:ProbWinningvsPickDist} shows that this transition is slower for solutions with fewer entries, i.e., \GreedySAA becomes more aggressive earlier when more entries are available. 

Therefore, to mimic the deviations uncovered in Figure~\ref{fig:ProbWinningvsPickDist}, we let $\widehat{P_{\indexround}}$ be the minimum probability~$\Proundi_{\indexteam,\indexround}$ for team~$\indexteam$ to be selected in round $\indexround$. Therefore, set~$\tilde{\setteams}$ is redefined as follows:
\[
\tilde{\setteams} = \bigcup\limits_{\indexgame \in \tilde{\setgames}}   \left\{ \indexteam\in \setteams(\indexgame) \text{ and } \Proundi_{\indexteam,\indexround}>\widehat{P_{\indexround}} \right\}. 
\]
We use hyper-parameter tuning to set~$\widehat{P_{\indexround}}$ for each $\nentries$ in $\{2,3,5,10,25,50,100\}$; Table~\ref{tab:LB_Prop} presents the best values.

\begin{table}[h!]
\begin{tabular}{|c|c|c|c|c|c|c|}
\hline
          & $\widehat{P_{1}}$ & $\widehat{P_{2}}$ & $\widehat{P_{3}}$ & $\widehat{P_{4}}$ & $\widehat{P_{5}}$ & $\widehat{P_{6}}$ \\ \hline
2-entry	&	0.479	&	0.390	&	0.332	&	0.226	&	0.094	&	0.030	\\	\hline
3-entry	&	0.476	&	0.392	&	0.334	&	0.213	&	0.055	&	0.032	\\	\hline
5-entry	&	0.453	&	0.380	&	0.312	&	0.215	&	0.069	&	0.018	\\	\hline
10-entry	&	0.474	&	0.382	&	0.340	&	0.195	&	0.053	&	0.036	\\	\hline
25-entry	&	0.438	&	0.383	&	0.338	&	0.195	&	0.031	&	0.012	\\	\hline
50-entry	&	0.450	&	0.373	&	0.326	&	0.202	&	0.018	&	0.002	\\	\hline
100-entry	&	0.450	&	0.372	&	0.305	&	0.186	&	0.000	&	0.000	\\	 \hline
\end{tabular}
\caption{Lower Bounds per Round on $\Pround$ as the Number of Entries Varies}
\label{tab:LB_Prop}
\end{table}
We call the resulting algorithm \Propplus.

\section{Experimental Results}
\label{sec:ExperimentalResults}
We report on a series of experiments conducted to provide insights about \textit{March Madness} competitions and evaluate the efficacy of the proposed models. The algorithms are implemented in~\texttt{Python 3.8.15}. We use \texttt{Gurobi 10.0.0} \citep{gurobi} to solve the mathematical programming formulations. All experiments ran on an AMD Ryzen 5 3600X CPU at 3.80GHz, limited to a single thread. The code and the instances used in our experiments will be available online upon the acceptance of the paper.
%\blue{The code and the instances used in our experiments are available at \texttt{https://github.com/jeffSylvestre-Decary/Multi-Entry-Sports-Tournament} Do we need this???} .  

\subsection{Instances}
Our experiments utilize \textit{March Madness} tournament data spanning from 2017-2023, except for 2020 (canceled because of the COVID-19 pandemic). The tournament structures are sourced from Kaggle's \textit{March Machine Learning Mania 2023} contest (\url{https://kaggle.com/competitions/march-machine-learning-mania-2023}).
%\footnote{Kaggle: \url{https://kaggle.com/competitions/march-machine-learning-mania-2023} (Accessed: February 3, 2024)}.
 %\citep{march-machine-learning-mania-2023}. 
 Our focal win matrices are derived from data released by the widely used analytics portal \texttt{538} (\url{https://abcnews.go.com/538}); we discuss the estimation of the win matrices and compare the accuracy of the \texttt{538} data with other estimation methods in Appendix~\ref{appendix:pteamMatrix}. 

\subsection{Computing Time for the Exact Evaluation of the EMS}
\label{sec:DPTime}
We analyze the running time of the DP algorithm presented in Section~\ref{sec:DPexact} to evaluate the EMS of two-entry solutions on tournaments of different sizes. The instances used in these experiments are the sub-tournaments with 4, 8, 16, 32, and 64 teams extracted from the upper part of the
\textit{March Madness} 2023 tournament structure (see Figure~\ref{fig:2019_MM}). 
For each sub-tournament, we randomly generate  ten two-entry solutions using \SIM.

\begin{table}[h!]
\footnotesize
\begin{tabular}{|c|ccccc|}
\hline
& \multicolumn{5}{c|}{Tournament size} \\
               & 4 teams & 8 teams & 16 teams  & 32 teams & 64 teams \\ \hline
Running time (seconds) & $9\cdot10^{-5}$ & $10^{-3}$ & 0.06 & 11.17      & 2,548.42     \\ \hline
\end{tabular}
\caption{Time to Calculate the EMS for Two-entry Solutions with Various Tournament Sizes}
\label{tab:ExactEMS}
\end{table}

Table~\ref{tab:ExactEMS} presents the average solution time, measured in seconds. The results show that the DP is time-consuming, and in particular, the running time for 64-team tournaments hinders the direct incorporation of this algorithm into an exact optimization procedure. 

\subsection{Quality of the Monte Carlo Estimation of the EMS}
\label{sec:ExpectedValue}

\begin{figure}[ht!]
    \centering
    \subfigure[2-entry solutions]{\includegraphics[width=0.4\textwidth]{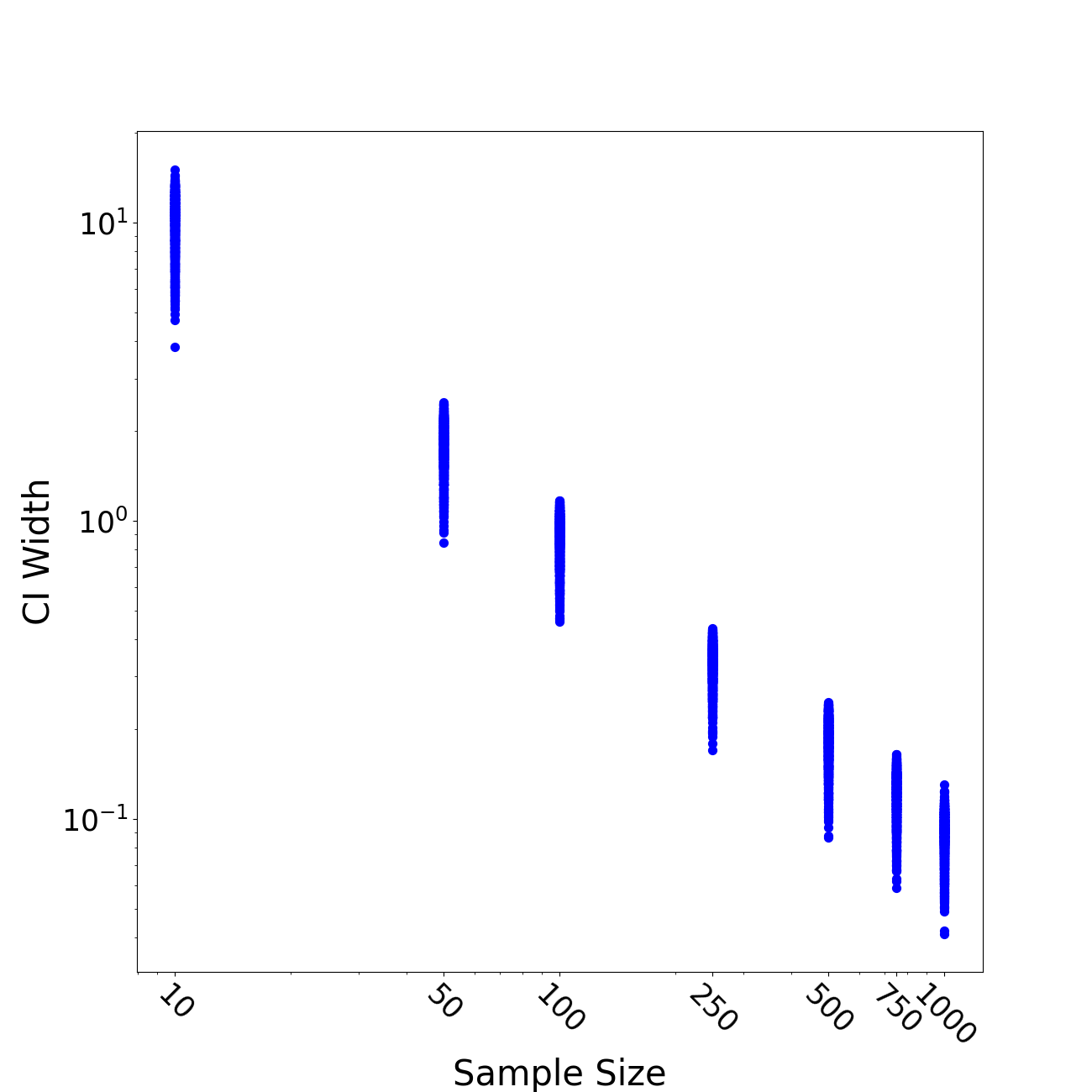}} 
    \subfigure[100-entry solutions]{\includegraphics[width=0.4\textwidth]{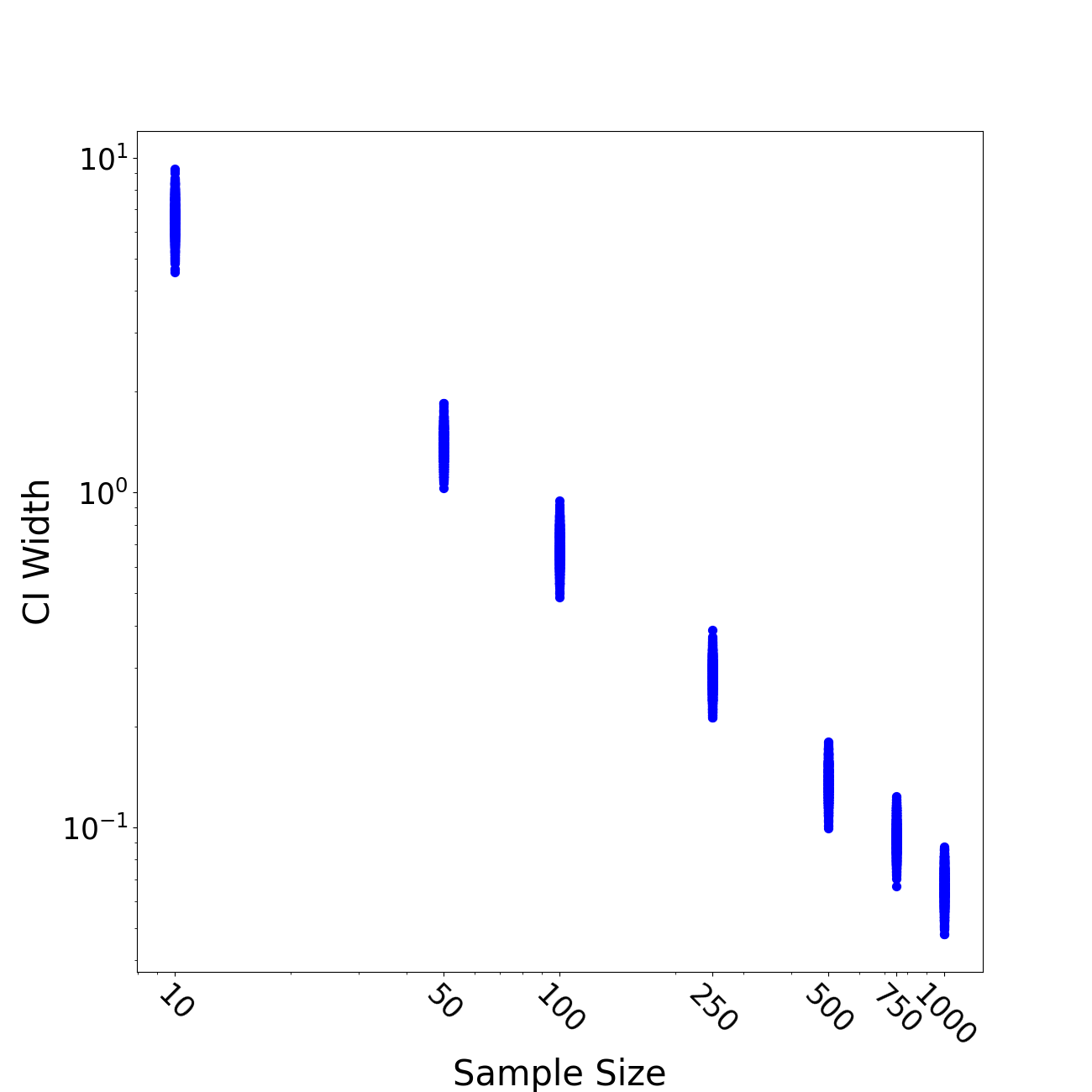}}
    \caption{Width of the 95\% CI for 2-entry and 100-entry Solutions for Different Sample Sizes}
    \label{fig:CI2entry}
\end{figure}

We investigate the accuracy of the Monte Carlo estimates of the EMS using \textit{March Madness} 2023 data; other \textit{March Madness} tournaments deliver the same results. We generate 1,000 random sets of 2-entry and 100-entry solutions using \SIM, and evaluate their EMS on batches of 10, 50, 100, 250, 500, 750, and 1,000 randomly generated outcomes brackets, also generated by~\SIM. We repeated this experiment 50 times for each solution to build 95\% confidence intervals (CI) for the EMS. The widths of the CIs are shown in Figure~\ref{fig:CI2entry}.

%Figure~\ref{fig:CI2entry} depicts the width of the $95^{\text{th}}$ percentile of the CI for the sample mean across all 2-entry and 100-entry solutions.
The results show that the width of the CI quickly decreases with the sample size. In particular, for 250 outcome brackets, the Monte Carlo approach achieves, in the worst-case, CI widths of 0.44 and 0.39 (out of 192 points) for the 2-entry and 100-entry solutions, respectively. The number of samples directly impacts the computational performance of our models, and we thus use 250 simulations as they provide satisfactory accuracy. 

Finally, we evaluate the quality of the solutions produced by all algorithms in the upcoming sections using the Monte Carlo estimation. More precisely, we obtain the \textit{empirical EMS} of each solution by evaluating out-of-sample on 10,000 randomly generated outcome brackets.

\subsection{Empirical Performance of the Optimization Algorithms}
\label{sec:EMSCalculation}

Table~\ref{tab:EVTable} reports for each optimization model the exact EMS computed using the dynamic programming approach for the 2-entry solutions, along with the empirical EMS for the 3-entry solutions obtained for the \textit{March Madness} competitions composing our dataset. We indicate in bold the best-performing algorithm per year and entry size. 

\begin{table}[h!]
\footnotesize
\begin{tabular}{|c|cccccccccc|}
\hline
        & \multicolumn{10}{c|}{Number of Entries}                                                                                                                                                                                                                                                                                \\ \hline
        & \multicolumn{5}{c||}{2 entries (Exact EMS)}                                                                                                                                         & \multicolumn{5}{c|}{3 entries (Empirical EMS)}                                                                                                                    \\ \hline
Year    & \multicolumn{1}{c|}{\SAA}            & \multicolumn{1}{c|}{\SIP}   & \multicolumn{1}{c|}{\GreedySAA} & \multicolumn{1}{c|}{\Prop}  & \multicolumn{1}{c||}{\Propplus}       & \multicolumn{1}{c|}{\SAA}            & \multicolumn{1}{c|}{\SIP}   & \multicolumn{1}{c|}{\GreedySAA} & \multicolumn{1}{c|}{\Prop}  & \Propplus       \\ \hline
2017    & \multicolumn{1}{c|}{101.7}          & \multicolumn{1}{c|}{101.1} & \multicolumn{1}{c|}{101.5}     & \multicolumn{1}{c|}{95.5}  & \multicolumn{1}{c||}{\textbf{101.8}} & \multicolumn{1}{c|}{108.0}          & \multicolumn{1}{c|}{104.4} & \multicolumn{1}{c|}{108.2}     & \multicolumn{1}{c|}{100.8} & \textbf{108.4} \\ \hline
2018    & \multicolumn{1}{c|}{103.4}          & \multicolumn{1}{c|}{100.8} & \multicolumn{1}{c|}{101.2}     & \multicolumn{1}{c|}{99.3}  & \multicolumn{1}{c||}{\textbf{104.2}} & \multicolumn{1}{c|}{\textbf{110.3}} & \multicolumn{1}{c|}{106.7} & \multicolumn{1}{c|}{108.7}     & \multicolumn{1}{c|}{102.7} & 110.2          \\ \hline
2019    & \multicolumn{1}{c|}{\textbf{114.1}} & \multicolumn{1}{c|}{109.3} & \multicolumn{1}{c|}{113.4}     & \multicolumn{1}{c|}{106.8} & \multicolumn{1}{c||}{\textbf{114.1}} & \multicolumn{1}{c|}{119.9}          & \multicolumn{1}{c|}{116.3} & \multicolumn{1}{c|}{119.8}     & \multicolumn{1}{c|}{111.8} & \textbf{120.5} \\ \hline
2021    & \multicolumn{1}{c|}{\textbf{111.9}} & \multicolumn{1}{c|}{106.5} & \multicolumn{1}{c|}{110.7}     & \multicolumn{1}{c|}{105.2} & \multicolumn{1}{c||}{111.4}          & \multicolumn{1}{c|}{\textbf{116.9}} & \multicolumn{1}{c|}{112.2} & \multicolumn{1}{c|}{116.9}     & \multicolumn{1}{c|}{107.4} & 116.5          \\ \hline
2022    & \multicolumn{1}{c|}{104.4}          & \multicolumn{1}{c|}{103.0} & \multicolumn{1}{c|}{103.5}     & \multicolumn{1}{c|}{100.5} & \multicolumn{1}{c||}{\textbf{105.3}} & \multicolumn{1}{c|}{\textbf{110.1}} & \multicolumn{1}{c|}{106.2} & \multicolumn{1}{c|}{107.7}     & \multicolumn{1}{c|}{102.6} & 109.6 \\ \hline
2023    & \multicolumn{1}{c|}{\textbf{98.6}}  & \multicolumn{1}{c|}{97.8}  & \multicolumn{1}{c|}{97.9}      & \multicolumn{1}{c|}{93.2}  & \multicolumn{1}{c||}{\textbf{98.6}}  & \multicolumn{1}{c|}{103.5}          & \multicolumn{1}{c|}{101.4} & \multicolumn{1}{c|}{102.9}     & \multicolumn{1}{c|}{96.0}  & \textbf{103.6} \\ \hline\hline
Average & \multicolumn{1}{c|}{105.7}          & \multicolumn{1}{c|}{103.1} & \multicolumn{1}{c|}{104.7}     & \multicolumn{1}{c|}{100.1} & \multicolumn{1}{c||}{\textbf{105.9}} & \multicolumn{1}{c|}{\textbf{111.5}} & \multicolumn{1}{c|}{107.9} & \multicolumn{1}{c|}{110.7}     & \multicolumn{1}{c|}{103.5} & \textbf{111.5} \\ \hline
\end{tabular}
\caption{Empirical EMS of the Optimization Algorithms}
\label{tab:EVTable}
\end{table}

Table~\ref{tab:EVTable} shows that~\SAA and~\Propplus consistently outperform the other algorithms in cases involving 2-entry and 3-entry solutions. A common feature of these methods is that they select all entries simultaneously, thus emphasizing the significance of avoiding a myopic selection process. Finally, the contrast in performance between~\Prop and~\Propplus is intriguing; the latter significantly outperforms the former, thereby showing the necessity of incorporating lower bounds.

\begin{figure}[ht!]
    \centering
    \subfigure[Empirical EMS]{\includegraphics[width=0.4\textwidth]{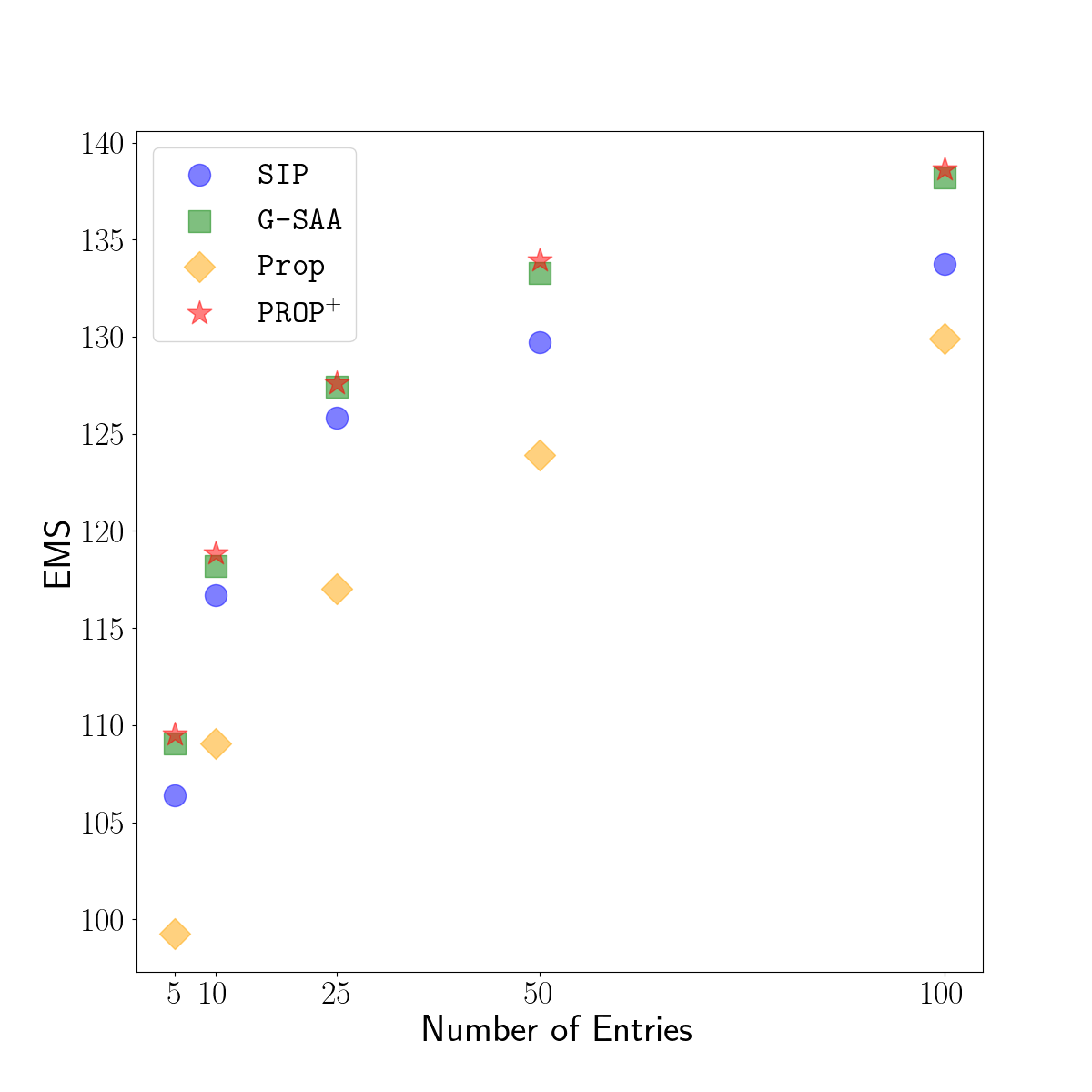}} 
    \subfigure[Standard deviation of the empirical EMS]{\includegraphics[width=0.4\textwidth]{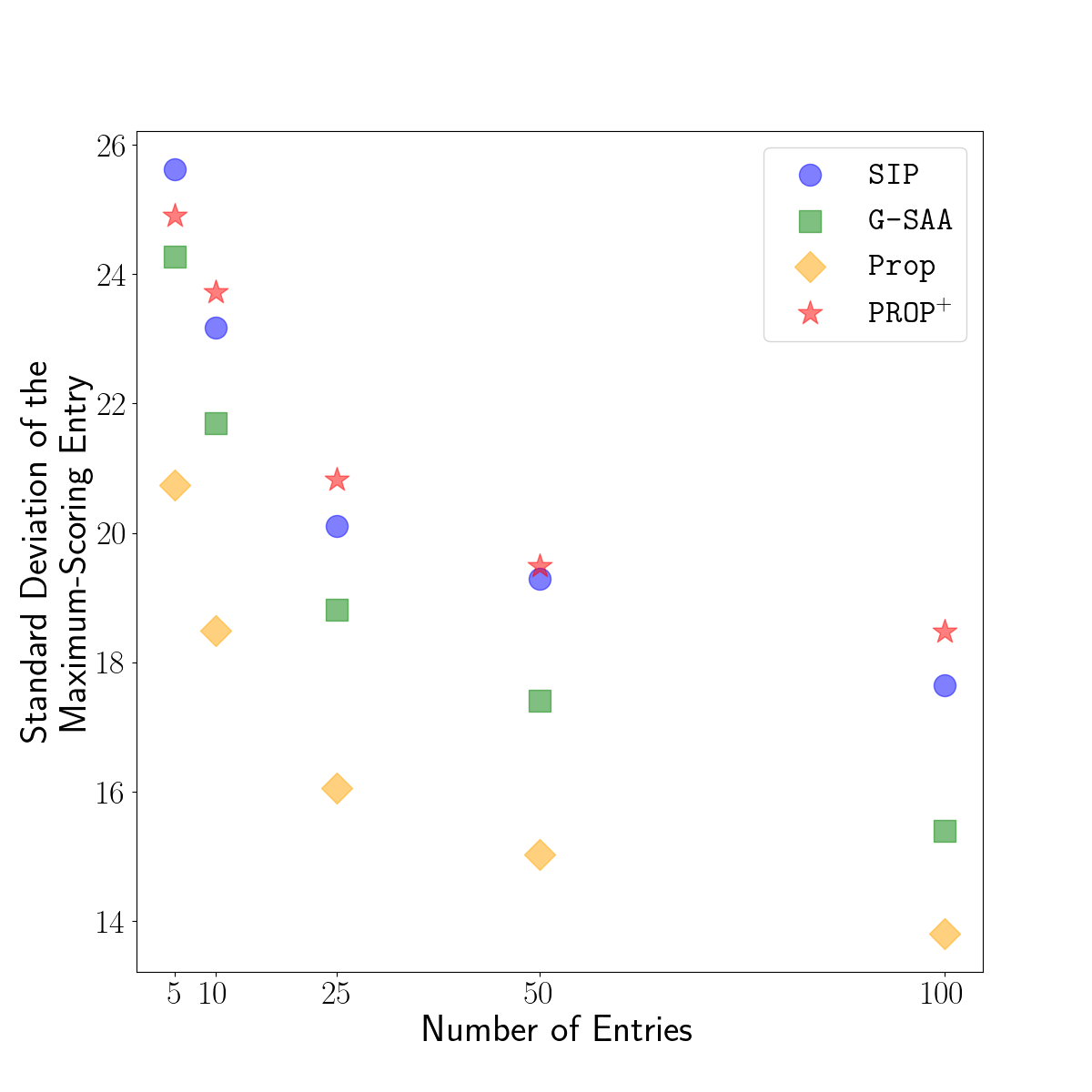}}
    \caption{Comparative Analysis for More Entries}
    \label{fig:MultiEntriesPlot}
\end{figure}

Figure~\ref{fig:MultiEntriesPlot} reports the results of similar experiments where the algorithms could generate up to 100 entries. We could not conduct these experiments with~\SAA, as the model is intractable when~$\nentries$ is large. Figure~\ref{fig:MultiEntriesPlot}(a)-(b) display
 the average and standard deviation (SD) of the empirical EMS for the \textit{March Madness 2023} tournament, respectively, for 5, 10, 25, 50, and 100 entries.

%For less than five entries, \SIP and \GreedySAA are comparable in terms of EMS. 
As the number of entries increases,~\GreedySAA and~\Propplus become more dominant in comparison with~\SIP and~\Prop. In particular, the 100-entry solution of~\Propplus achieves an empirical EMS of 138.62, considerably surpassing the solution of~\SIP in terms of EMS. Moreover, \Propplus achieves the highest SD for any number of entries compared to all other models; note that a higher SD might be preferred in contests with a top-heavy payout structure.

\section{Empirical Performance on a Real-World Betting Pool}
\label{sec:DKContest}
\emph{DraftKings} held a betting pool for the \textit{March Madness} 2023 competition with a \$1 million prize to the winner. Anyone could enter the pool at the cost of \$100 per entry, with a maximum of 100 entries per participant. We evaluate the performance of~\Propplus in this betting pool using solutions consisting of 2, 3, 5, 10, 25, 50, and 100 entries.\footnote{We note that the entries identified by~\Propplus were not registered in the official \emph{DraftKings} competition.}

%The entries generated by~\GreedySAA and~\Prop were not registered in the official betting pool, but we nevertheless consider them when evaluating the performance of the original 12,605 entries.

Tables~\ref{tab:DK_Contest} and~\ref{tab:PayoffTable} show the number of entries per participant and the payoff structure of the contest, respectively.
\begin{table}[ht!]
\footnotesize
\begin{tabular}{|c|c|c|c|c|c|c|c|c|c|c|c|}
\hline
Entries     & 1     & 2   & 3   & 4-5  & 6-10 & 11-25 & 26-50 & 51-99 & 100 & Total  \\ \hline
Contestants & 7,756 & 790 & 179 & 122 & 72   & 33    & 6     & 3     & 6   & 8,967 \\ \hline
\end{tabular}
\caption{Number of Contestants per Number of Entries}
\label{tab:DK_Contest}
\end{table}
There were 12,605 total entries in the contest submitted by 8,967 participants, with around 38.4\% registering at least two entries and 48 bettors entering more than ten.  \emph{DraftKings} retained a small portion as a management fee, and based on the assumption that each entry has an equal chance of winning, the expected profit for every entry stood at -\$0.83.
\begin{table}[ht!]
\footnotesize
\begin{tabular}{|c|c|c|c|c|c|c|c|c|c|}
\hline
Position    & 1\textsuperscript{st} & 2\textsuperscript{nd} & 3\textsuperscript{rd} & 4\textsuperscript{th} & 5\textsuperscript{th} & 6\textsuperscript{th} & 7\textsuperscript{th} & \ldots & 201\textsuperscript{st}-1000\textsuperscript{th} \\ \hline
Payoff (\$) & 1,000,000                              & 50,000                                 & 20,000                                 & 10,000                                 & 5,000                                  & 2,000                                  & 1,000                                  & \ldots & 150                                                                                \\ \hline
\end{tabular}
\caption{\emph{DraftKings} Contest Payoff Table}
\label{tab:PayoffTable}
\end{table}

\subsection{Expected scores}

\begin{figure}[ht!]
    \centering
    \subfigure[Empirical score of each entry]{\includegraphics[width=0.35\textwidth]{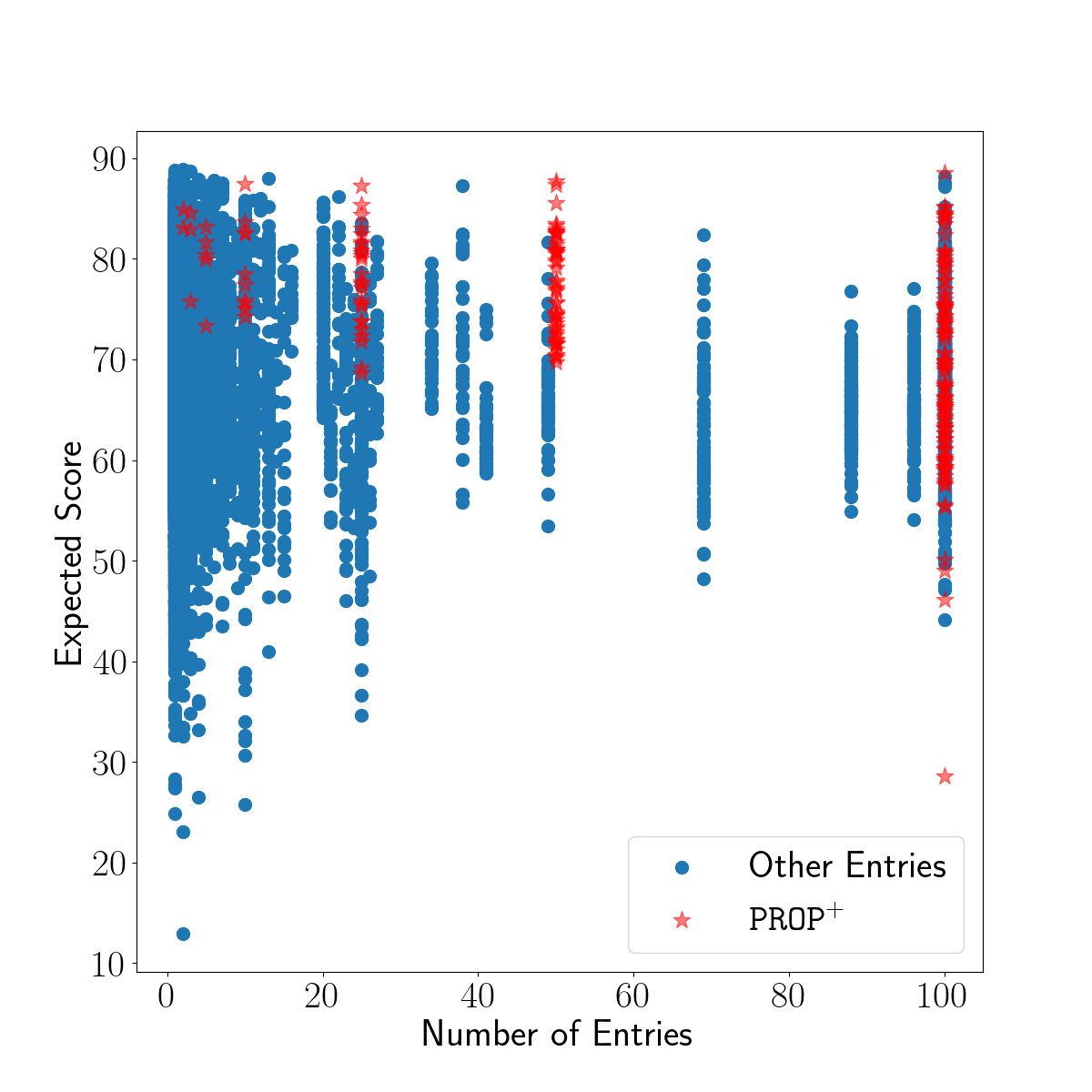}} 
    \subfigure[Empirical EMS for each participant]{\includegraphics[width=0.35\textwidth]{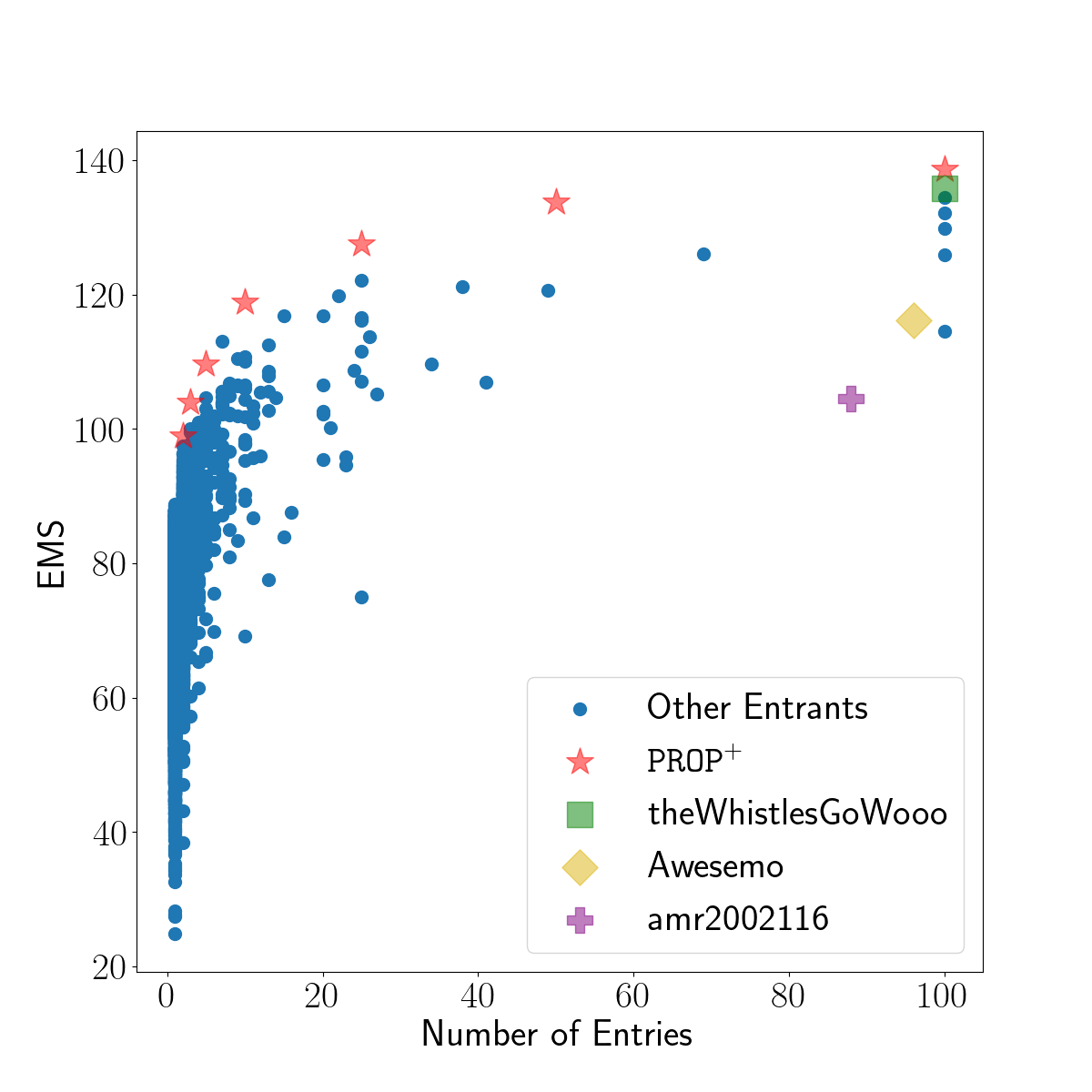}}
    \caption{Score Distribution on \emph{DraftKings}' \textit{March Madness} 2023 betting pool}
    \label{fig:DKPlot}
\end{figure}

Figure~\ref{fig:DKPlot} shows the expected score of each entry (Figure~\ref{fig:DKPlot}(a)) and participant (Figure~\ref{fig:DKPlot}(b)) based on our win matrix~$\Pteam$. The plots include the entries identified by~\Propplus and %for this competition and entries registered by 
\textit{Awesemo} and \textit{theWhistlesGoWooo}, who are among the top sports bettors in the world.\footnote{RotoGrinders: 
\url{https://rotogrinders.com/rankings}
(Accessed: February 3, 2024)} We also display the results of the entries of \textit{amr2002116}, a participant in the \emph{DraftKings} contest who achieves great results in our experiments. 
% Utilizing the win probability matrix~$\Pteam$, we conduct a more detailed analysis of each participant's expected score and expected profit. The results are presented in Figure~\ref{fig:DKPlot}, and include solutions generated by our algorithm (which were not entered in the competition) 
% and entries registered by \textit{Awesemo}, who is among the top sports bettors in the world\footnote{RotoGrinders: 
% \url{https://rotogrinders.com/rankings}
% (Accessed: February 3, 2024).
% }

Figure~\ref{fig:DKPlot}(a) illustrates the expected score for all single entries registered in the contest, organized based on the number of entries submitted by the associated participant. Figure~\ref{fig:DKPlot}(b) presents the empirical EMS of each participant. Notably, the best-performing entries identified by our algorithms\textemdash illustrated with red stars\textemdash attain higher empirical EMS than every other participant's entries with the same entry count. These results suggest that, while participants may have different objectives (e.g., maximize the number of entries finishing \textit{in the money}) and $\Pteam$, our solutions demonstrate great performance in addressing our initial problem.

\subsection{Victory Probability}

\begin{figure}[ht!]
\includegraphics[width=0.45\textwidth]{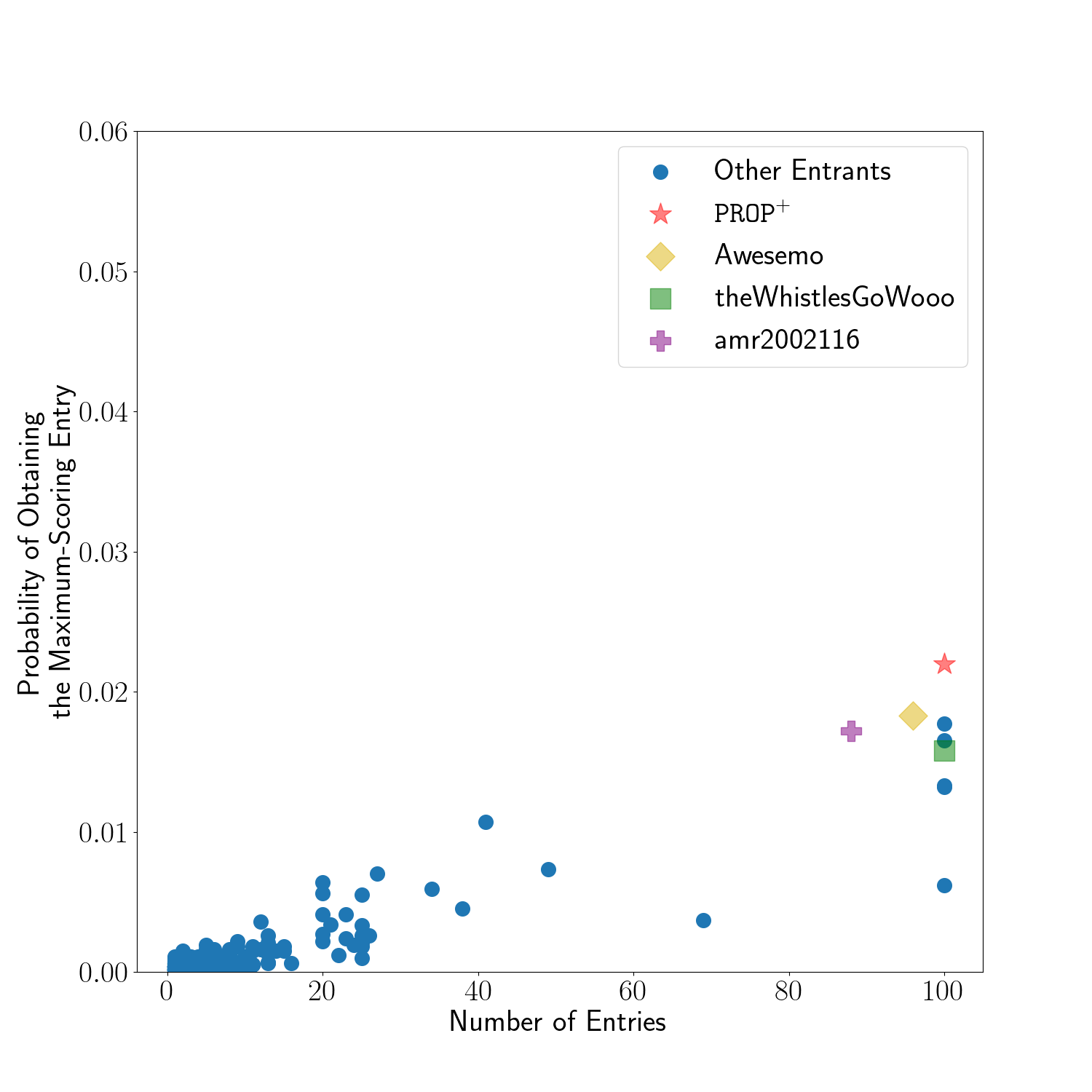}
    \caption{Victory Probability of Each Participant}
    \label{fig:PositionPlot}
\end{figure}

Figure~\ref{fig:PositionPlot} illustrates the empirical probability that each participant obtains the maximum-scoring EMS; in cases of draws among several entrants, we assume that all entries attain the maximum score, so the sum of the empirical probabilities may be larger than one. The 100-entry solution of \Propplus stands out with a 2.2\% chance of winning the pool, the highest probability among all participants. Following closely behind are \textit{Awesemo} and \textit{theWhistlesGoWooo}, with victory probabilities of 1.83\% and 1.58\%, respectively. Accounting solely for achieving the first position, the 2.2\% probability of winning the pool yielded by \Propplus holds an expected payoff of \$12,000.

Figure~\ref{fig:PositionPlot} shows that multi-entry strategies provide better chances of winning. Therefore, we evaluate how the victory probability of \Propplus changes as the number of entries varies. The results are presented in Table~\ref{tab:TopScoringTable}, showing that the probability of obtaining a top-scoring entry increases almost linearly with the number of entries.

\begin{table}[h!]
\footnotesize
\begin{tabular}{|c|c|c|c|c|c|c|c|}
\hline
Number of Entries                                                                                    & 2    & 3   & 5   & 10   & 25   & 50  & 100 \\ \hline
\begin{tabular}[c]{@{}c@{}}Chances of Obtaining the Maximum-Scoring Entry (\%)\end{tabular} & 0.03 & 0.1 & 0.1 & 0.24 & 0.59 & 1.2 & 2.2 \\ \hline
\end{tabular}
\caption{Probability of Obtaining Maximum-Scoring Entry given an $\nentries$-entry Solution}
\label{tab:TopScoringTable}
\end{table}

\subsection{Robustness Analyses}
\label{sec:RobustnessAnalysis}
Our analysis relies on the win matrices derived from \texttt{538}. However, these matrices are estimates, and each participant may work with a different estimation of $\Pteam$. Therefore, we examine the robustness of our findings by evaluating the quality of the optimal entries generated by \Propplus based on the \texttt{538} win matrix using two alternative estimation techniques for~$\Pteam$. The first utilizes the R package \texttt{cbbdata}, which is affiliated with the popular collegiate basketball website \url{https://barttorvik.com/} and includes a prediction function that forecasts the win probabilities of any match-up at a given date. The other is a \texttt{seed-based} logistic regression model, which uses the seeds of two teams~$\indexteam$ and~$\indexteam'$ to predict~$\Pteami_{\indexteam,\indexteam'}$; we use data from all \textit{March Madness} games played since 1985 for training the regression model. 
\begin{figure}[ht!]
    \centering
    \subfigure[Empirical EMS (\texttt{cbbdata})]{\includegraphics[width=0.35\textwidth]{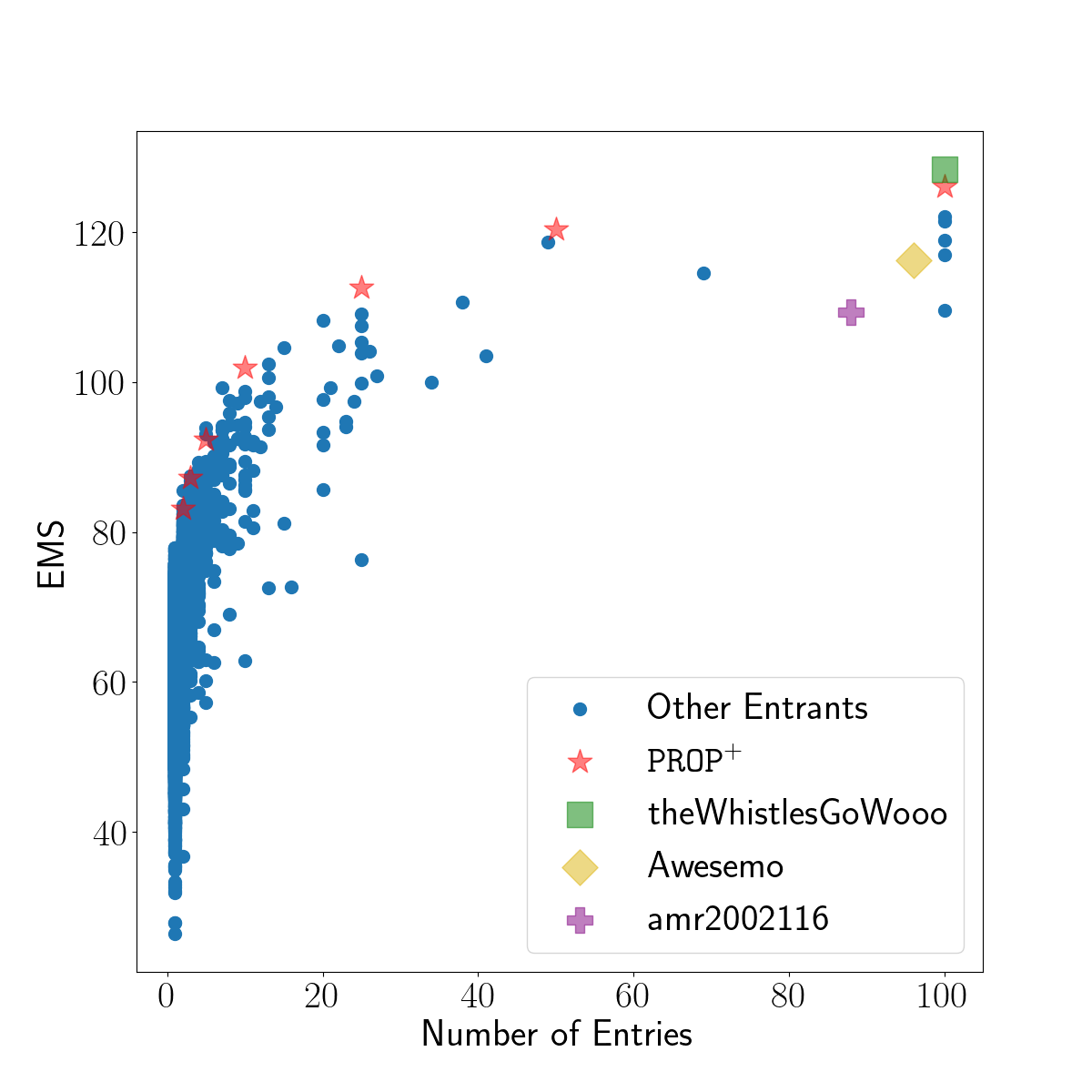}} 
    \subfigure[Empirical EMS (\texttt{seed-based})]{\includegraphics[width=0.35\textwidth]{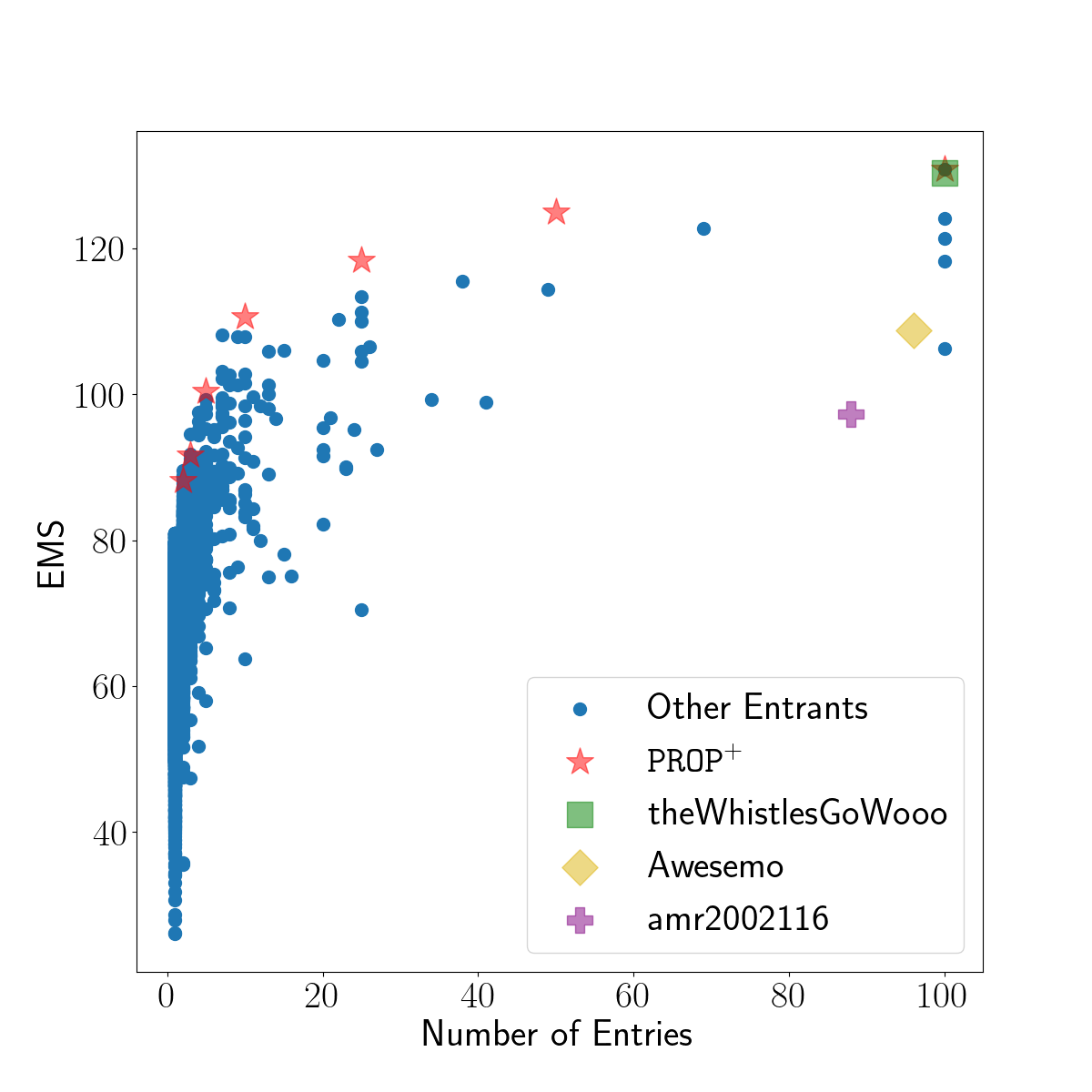}}\\
    \subfigure[Victory Probability (\texttt{cbbdata})]{\includegraphics[width=0.35\textwidth]{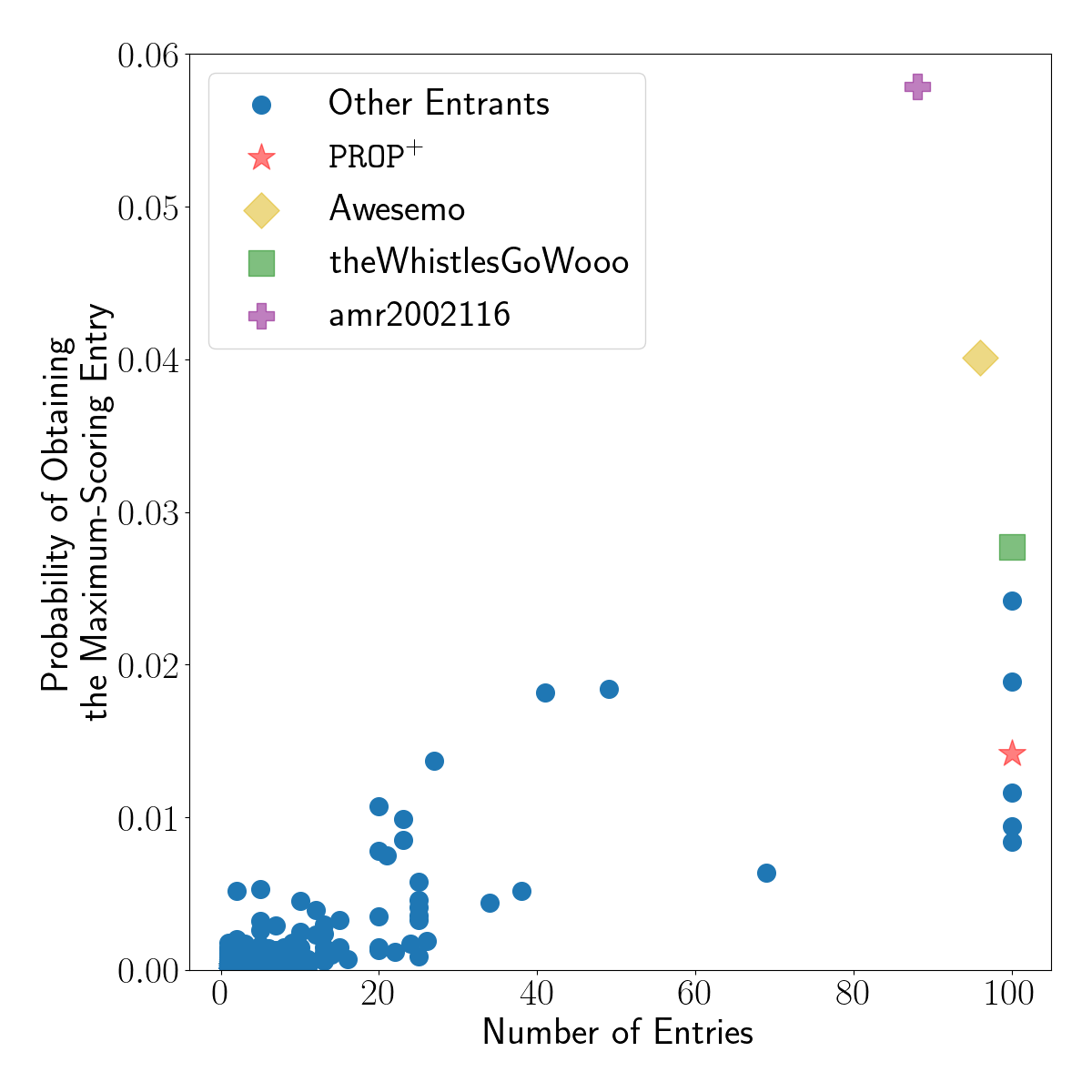}} 
    \subfigure[Victory Probability (\texttt{seed-based})]{\includegraphics[width=0.35\textwidth]{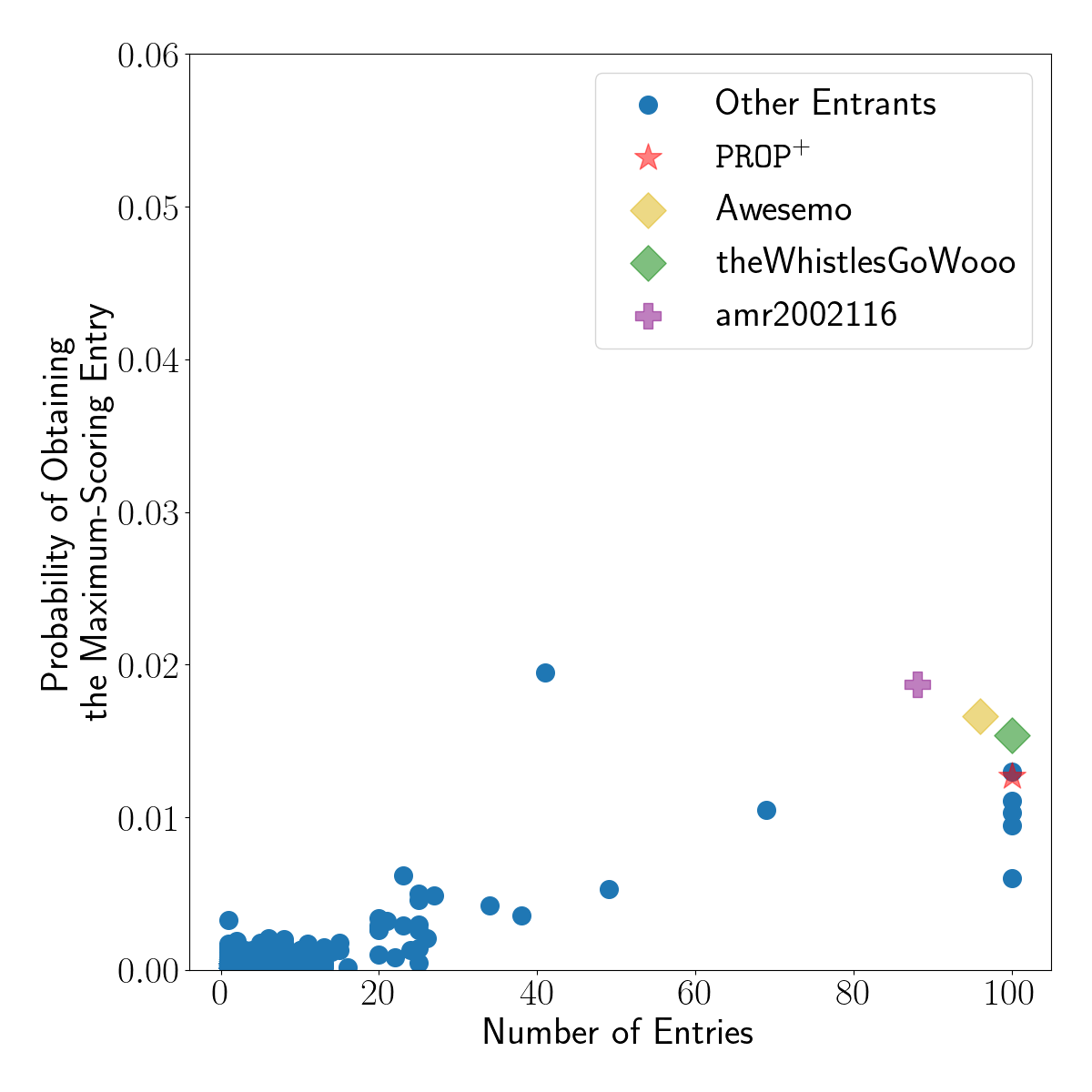}}
    \caption{Empirical EMS Evaluated on Different Win Matrices}
    \label{fig:RobustPlots}
\end{figure}

Figures~\ref{fig:RobustPlots}(a)-(b) show the empirical EMS of all entrants according to the win matrices derived from \texttt{cbbdata} and \texttt{seed-based}. We observe that the EMS of the best-performing entries produced by~$\Propplus$ based on the \texttt{538} matrix is still high, although it is no longer the best compared to other 100-entry solutions. The win probabilities reported in the plots depicted in Figure~\ref{fig:RobustPlots}(c)-(d) are interesting, as the results are clearly different from the ones observed in Figure~\ref{fig:PositionPlot}. A few participants, such as \textit{Awesemo} and \textit{theWhistleGoWooo}, achieve notably higher probabilities of obtaining the maximum-scoring entry. Nevertheless, the 100-entry solution of~\Propplus still has 1.42\% and 1.27\% victory probabilities according to the~\texttt{cbbdata} and \texttt{seed-based} data, respectively. Finally, \textit{amr2002116} had a 5.79\% chance of securing the maximum-scoring entry according to the \texttt{cbbdata} matrix. This result is interesting, especially because the participants' entries have a lower empirical EMS than \Propplus.  These results show significant sensitivity to the win probability estimates.  Some bettors appear to find solutions that are robust to several estimates, while others seem to focus on deriving the best solution to a particular model. 

\section{Conclusions}
\label{sec:Conclusion}
This paper investigates multi-entry betting strategies for single-elimination tournaments focusing on \textit{March Madness}. Given the top-heavy payoff structure of betting pools related to these contests, we focus on maximizing the expected score of the maximum-scoring entry. To achieve this, we introduce a dynamic programming approach, precisely evaluating the expected score of the maximum-scoring entry for any given multi-entry solution. However, due to the inherent complexity of the exact approach, we transition to employing simulation-based algorithms, which are more adequate and computationally tractable for real-world scenarios of the problem.

We propose several algorithms to solve the problem. Experiments on data from recent \textit{March Madness} competitions demonstrate that \Propplus, a heuristic tailored for the problem, consistently outperforms all other algorithms. We evaluate \Propplus's performance in a betting pool organized by \emph{DraftKings}. Remarkably, it exhibits a 2.2\% chance of winning the \$1,000,000 prize.

A clear direction for future work involves studying the problem's computational complexity of the multi-entry problem (we conjecture it to be at least NP-hard) and designing an exact algorithm to solve it. In particular, even though the dynamic programming algorithm is not scalable, we believe it may be effectively embedded into an efficient decomposition method.

\bibliographystyle{plainnat}
\bibliography{ArxivSubmission.bib}

\begin{thebibliography}{28}
\providecommand{\natexlab}[1]{#1}
\providecommand{\url}[1]{\texttt{#1}}
\expandafter\ifx\csname urlstyle\endcsname\relax
  \providecommand{\doi}[1]{doi: #1}\else
  \providecommand{\doi}{doi: \begingroup \urlstyle{rm}\Url}\fi

\bibitem[{American Gaming Association}(2019)]{MMAgamingA}
{American Gaming Association}.
\newblock \emph{Americans will wager \$8.5 billion on {M}arch {M}adness}, 2019.
\newblock URL \url{https://www.americangaming.org/new/americans-will-wager-8-5-billion-on-march-madness/}.

\bibitem[Bergman and Imbrogno(2017)]{bergman2017surviving}
David Bergman and Jason Imbrogno.
\newblock Surviving a {N}ational {F}ootball {L}eague survivor pool.
\newblock \emph{Operations Research}, 65\penalty0 (5):\penalty0 \:1343--1354, 2017.
\newblock \doi{10.1287/opre.2017.1633}.
\newblock URL \url{https://doi.org/10.1287/opre.2017.1633}.

\bibitem[Bergman et~al.(2023)Bergman, Cardonha, Imbrogno, and Lozano]{bergman2022optimizing}
David Bergman, Carlos Cardonha, Jason Imbrogno, and Leonardo Lozano.
\newblock Optimizing the expected maximum of two linear functions defined on a multivariate {G}aussian distribution.
\newblock \emph{INFORMS Journal on Computing}, 35\penalty0 (2):\penalty0 \:304--317, 2023.
\newblock \doi{10.1287/ijoc.2022.1259}.

\bibitem[Caudill(2003)]{caudill2003predicting}
Steven Caudill.
\newblock Predicting discrete outcomes with the maximum score estimator: The case of the {NCAA} men’s basketball tournament.
\newblock \emph{International Journal of Forecasting}, 19\penalty0 (2):\penalty0 \:313--317, 2003.
\newblock \doi{10.1016/S0169-2070(02)00008-0}.

\bibitem[Chung(2017)]{Chung2017}
Doug Chung.
\newblock How much is a win worth? {A}n application to intercollegiate athletics.
\newblock \emph{Management Science}, 63\penalty0 (2):\penalty0 \:548--565, 2017.
\newblock \doi{10.1287/mnsc.2015.2337}.
\newblock URL \url{https://doi.org/10.1287/mnsc.2015.2337}.

\bibitem[Clair and Letscher(2007)]{clair2007optimal}
Bryan Clair and David Letscher.
\newblock Optimal strategies for sports betting pools.
\newblock \emph{Operations Research}, 55\penalty0 (6):\penalty0 \:1163--1177, 2007.
\newblock \doi{10.1287/opre.1070.0448}.
\newblock URL \url{https://doi.org/10.1287/opre.1070.0448}.

\bibitem[David(1959)]{david1959tournaments}
Herbert David.
\newblock Tournaments and paired comparisons.
\newblock \emph{Biometrika}, 46\penalty0 (1/2):\penalty0 \:139--149, 1959.
\newblock \doi{10.2307/2332816}.

\bibitem[Gumm et~al.(2015)Gumm, Barrett, and Hu]{gumm2015machine}
Jordan Gumm, Andrew Barrett, and Gongzhu Hu.
\newblock A machine learning strategy for predicting {M}arch {M}adness winners.
\newblock In \emph{2015 IEEE/ACIS 16th International Conference on Software Engineering, Artificial Intelligence, Networking and Parallel/Distributed Computing (SNPD)}, pages 1--6. IEEE, 2015.

\bibitem[Gurobi~Optimization(2020)]{gurobi}
L~L~C Gurobi~Optimization.
\newblock Gurobi optimizer reference manual, 2020.
\newblock URL \url{http://www.gurobi.com}.

\bibitem[Hassidim and Singer(2017)]{hassidim2017submodular}
Avinatan Hassidim and Yaron Singer.
\newblock Submodular optimization under noise.
\newblock In \emph{Conference on Learning Theory}, pages 1069--1122. PMLR, 2017.

\bibitem[Haugh and Singal(2021)]{haugh2018play}
Martin Haugh and Raghav Singal.
\newblock How to play fantasy sports strategically (and win).
\newblock \emph{Management Science}, 67\penalty0 (1):\penalty0 \:72--92, 2021.
\newblock \doi{10.1287/mnsc.2019.3528}.
\newblock URL \url{https://doi.org/10.1287/mnsc.2019.3528}.

\bibitem[Hoegh et~al.(2015)Hoegh, Carzolio, Crandell, Hu, Roberts, Song, and Leman]{hoegh2015nearest}
Andrew Hoegh, Marcos Carzolio, Ian Crandell, Xinran Hu, Lucas Roberts, Yuhyun Song, and Scotland Leman.
\newblock Nearest-neighbor matchup effects: {A}ccounting for team matchups for predicting {M}arch {M}adness.
\newblock \emph{Journal of Quantitative Analysis in Sports}, 11\penalty0 (1):\penalty0 \:29--37, 2015.
\newblock \doi{10.1515/jqas-2014-0054}.

\bibitem[Horen and Riezman(1985)]{horen1985comparing}
Jeff Horen and Raymond Riezman.
\newblock Comparing draws for single elimination tournaments.
\newblock \emph{Operations Research}, 33\penalty0 (2):\penalty0 \:249--262, 1985.
\newblock URL \url{https://www.jstor.org/stable/170742}.

\bibitem[Hunter et~al.(2016)Hunter, Vielma, and Zaman]{hunter2016picking}
David Hunter, Juan Vielma, and Tauhid Zaman.
\newblock Picking winners in daily fantasy sports using integer programming.
\newblock \emph{arXiv preprint}, 2016.
\newblock \doi{10.48550/arXiv.1604.01455}.

\bibitem[Kaplan and Garstka(2001)]{kaplan2001march}
Edward Kaplan and Stanley Garstka.
\newblock March {M}adness and the office pool.
\newblock \emph{Management Science}, 47\penalty0 (3):\penalty0 \:369--382, 2001.
\newblock URL \url{https://www.jstor.org/stable/2661505}.

\bibitem[Knuth and Lossers(1987)]{knuth1987random}
DE~Knuth and OP~Lossers.
\newblock A random knockout tournament.
\newblock \emph{SIAM Review}, 29\penalty0 (1):\penalty0 \:127--129, 1987.
\newblock \doi{10.1137/1029011}.

\bibitem[Knysh and Korkolis(2016)]{Knysh2016}
Paul Knysh and Yannis Korkolis.
\newblock Blackbox: A procedure for parallel optimization of expensive black-box functions.
\newblock \emph{arXiv preprint}, 2016.
\newblock \doi{10.48550/arXiv.1605.00998}.

\bibitem[Kvam and Sokol(2006)]{kvam2006logistic}
Paul Kvam and Joel Sokol.
\newblock A logistic regression/{M}arkov chain model for {NCAA} basketball.
\newblock \emph{Naval Research Logistics}, 53\penalty0 (8):\penalty0 \:788--803, 2006.
\newblock \doi{10.1002/nav.20170}.

\bibitem[Liu et~al.(2023)Liu, Liu, and Teo]{liu2023picking}
Ju~Liu, Changchun Liu, and Chung Teo.
\newblock Picking winners: Diversification through portfolio optimization.
\newblock \emph{Production and Operations Management}, 32\penalty0 (9):\penalty0 \:2864--2884, 2023.
\newblock \doi{10.1111/poms.14013}.

\bibitem[Metrick(1996)]{metrick1996march}
Andrew Metrick.
\newblock March {M}adness? {S}trategic behavior in {NCAA} basketball tournament betting pools.
\newblock \emph{Journal of Economic Behavior \& Organization}, 30\penalty0 (2):\penalty0 \:159--172, 1996.
\newblock \doi{10.1016/S0167-2681(96)00855-4}.

\bibitem[Morewedge et~al.(2018)Morewedge, Tang, and Larrick]{Morewedge2018}
Carey Morewedge, Simone Tang, and Richard Larrick.
\newblock Betting your favorite to win: Costly reluctance to hedge desired outcomes.
\newblock \emph{Management Science}, 64\penalty0 (3):\penalty0 \:997--1014, 2018.
\newblock \doi{10.1287/mnsc.2016.2656}.
\newblock URL \url{https://doi.org/10.1287/mnsc.2016.2656}.

\bibitem[Nemhauser and Trick(1998)]{NemTri98}
George Nemhauser and Michael Trick.
\newblock Scheduling a major college basketball conference.
\newblock \emph{Operations Research}, 46\penalty0 (1):\penalty0 \:1--8, 1998.
\newblock \doi{10.1287/opre.46.1.1}.

\bibitem[Nemhauser et~al.(1978)Nemhauser, Wolsey, and Fisher]{nemhauser1978analysis}
George Nemhauser, Laurence Wolsey, and Marshall Fisher.
\newblock An analysis of approximations for maximizing submodular set functions—{I}.
\newblock \emph{Mathematical Programming}, 14\penalty0 (1):\penalty0 \:265--294, 1978.
\newblock \doi{10.1007/BF01588971}.

\bibitem[Palley and Soll(2019)]{Palley2019}
Asa Palley and Jack Soll.
\newblock Extracting the wisdom of crowds when information is shared.
\newblock \emph{Management Science}, 65\penalty0 (5):\penalty0 \:2291--2309, 2019.
\newblock \doi{10.1287/mnsc.2018.3047}.
\newblock URL \url{https://doi.org/10.1287/mnsc.2018.3047}.

\bibitem[Ribeiro(2012)]{ribeiro2012sports}
Celso Ribeiro.
\newblock Sports scheduling: Problems and applications.
\newblock \emph{International Transactions in Operational Research}, 19\penalty0 (1-2):\penalty0 \:201--226, 2012.
\newblock \doi{10.1111/j.1475-3995.2011.00819.x}.

\bibitem[Richter(2021)]{StatistaCollegeSports}
Felix Richter.
\newblock \emph{U.S. college sports are a billion-dollar game}, 2021.
\newblock URL \url{https://www.statista.com/chart/25236/ncaa-athletic-department-revenue/}.

\bibitem[{Supreme Court of the United States}(2018)]{Supreme18}
{Supreme Court of the United States}.
\newblock \emph{Murphy, Governor of New Jersey, et al. v. National Collegiate Athletic Association, et al.}, 2018.
\newblock URL \url{{https://www.supremecourt.gov/opinions/17pdf/16-476_dbfi.pdf}}.

\bibitem[Wright and Wiens(2016)]{wright2016method}
Mason Wright and Jenna Wiens.
\newblock Method to their {M}arch {M}adness: {I}nsight from mining a novel large-scale dataset of pool brackets.
\newblock In \emph{KDD Workshop on Large-Scale Sports Analytics}, 2016.

\end{thebibliography}

%\begin{APPENDICES}

\appendix % Switch to appendix mode

\section{Win Probability Matrices}
\label{appendix:Win Probability Matrices}

A participant in a betting pool can use historical data to construct a $\nteams \times \nteams$ \textit{team-by-team win matrix~$\Pteam$}, with each element $\Pteami_{\indexteam,\indexteam'}$ estimating the probability that team $\indexteam$ wins a game against team $\indexteam'$. All games must have a winner, so~$\Pteami_{\indexteam,\indexteam'} + \Pteami_{\indexteam',\indexteam} = 1.0$. 

Given $\Pteam$, one can generate $\Pgame$ by recursively computing the probability round by round. First, if team~$\indexteam$ cannot play game~$\indexgame$ (i.e., if~$\indexteam \notin \setteams(\indexgame)$), we have $\Pgamei_{\indexteam,\indexgame} = 0$. Otherwise, if~$\indexgame$ is a first-round game and~$\setteams(\indexgame) = \{\indexteam,\indexteam'\}$, we have
$\Pgamei_{\indexteam,\indexgame} = \Pteami_{\indexteam,\indexteam^{'}}$
and 
$\Pgamei_{\indexteam',\indexgame} = \Pteami_{\indexteam^{'},\indexteam}$. For~$\indexround(\indexgame) \geq 2$, we have
\[
\Pgamei_{\indexteam,\indexgame} \coloneqq
    \underbrace{ \Pgamei_{\indexteam,\gamma^-(\indexgame,\indexteam)}  }_{\indexteam \text{ wins the previous game}} 
    \cdot 
    \underbrace{\left( \sum_{\indexteam' \in \setteams(\delta^-(\indexgame,\indexteam))}
\Pgamei_{\indexteam',\delta^-(\indexgame,\indexteam')}  \cdot \Pteami_{\indexteam,\indexteam'} \right) }_{\indexteam \text{ wins game $\indexgame$}}. 
\]

One can also generate $\Pround$ from $\Pgame$ by only considering the games every team may play in for a given round. Given $\indexteam\in\setteams$ and $\displaystyle \indexround\in\setrounds$, we calculate \[\Proundi_{\indexteam,\indexround}= \sum_{\substack{\indexgame\in\setgames:\\
\indexround(\indexgame)=\indexround}}\Pgamei_{\indexteam,\indexgame}.\] Note that the sum always has exactly one non-zero element because a team may play in only one game per round by construction of the bracket.

\section{Space and Time Complexity of the Dynamic Programming Algorithm}
\label{appendix:spacetime}

Recall that a single-elimination tournament involving $\nteams$ teams consists of $\nrounds=\log_2(\nteams)$ rounds and $\ngames=\nteams-1$ games. Each entry can attain a score ranging from 0 to $\nrounds2^{\nrounds-1}=\frac{\nteams}{2}\log_2(\nteams)$.

\textit{Space Complexity}: The state space of our formulation contains one element for each combination of pairs of scores, teams, and games, so the space complexity of the algorithm is $O\left(\nteams^2(\frac{\nteams}{2}\log_2(\nteams))^{\nentries}\right)$.

\textit{Time Complexity}: Let us first separate the time complexity per round.
\begin{enumerate}
    \item First round: The first round consists of $\frac{\nteams}{2}$ games, each providing two outcomes per entry. The evaluation  of~$Z[\indexgame^*, \indexteam^*, x,y]$ consumes time~$O(1)$ for these games,
    so the time complexity of the first round amounts to $O\left(2^{\nentries-1}\nteams\right)$.
    \item Subsequent Rounds $(\indexround > 1)$: Each of the $2^{\nrounds-\indexround}$ games in round $\indexround$ can be played and won by $2^{\indexround}$ teams. The evaluation  of each~$Z[\indexgame^*, \indexteam^*, x,y]$ consumes time~$O( 2^{\indexround} (\frac{\nteams}{2}\log_2(\nteams))^{\nentries})$, and as there are %$O\left(2^{\nrounds}(\frac{\nteams}{2}\log_2(\nteams))^{\nentries}\right)$
    $O\left(\nteams(\frac{\nteams}{2}\log_2(\nteams))^{\nentries}\right)$
    entries per round, the time complexity of each round~$\indexround > 1$ is 
%    $O\left(2^{2\nrounds}(\frac{\nteams}{2}\log_2(\nteams))^{2\nentries}\right)$.
    $O\left(\nteams^{2}(\frac{\nteams}{2}\log_2(\nteams))^{2\nentries}\right)$.
\end{enumerate}
Finally, given~$Z$, the evaluation of $\mathbb{E}(S(\setentries))$ consumes time~$O(\nteams (\frac{\nteams}{2}\log_2(\nteams))^{2\nentries})$. 
As there are~$\nrounds = \log_2{(\nteams)}$ rounds in the tournament, the time complexity of computing~$Z$ is
$O\left(\log_2{(\nteams)} \cdot \nteams^2 \cdot (\frac{\nteams}{2}\log_2(\nteams))^{2\nentries}\right)
$.

\section{Submodularity of the EMS}
\label{appendix:subEMS}
\begin{customproof}[Proof of Proposition~\ref{prop:submodular}:] Incorporating an entry cannot reduce the EMS, so monotonicity follows directly. Submodularity follows from standard arguments over the discrete space~$\setbrackets$ of possible outcomes. Let~$\setentries'$ and~$\setentries''$ be sets of entries such that~$\setentries' \subset \setentries'' \subseteq \setbrackets \setminus \{\indexentry\}$, i.e., $\indexentry$ is a bracket (entry) that does not belong to the entry sets~$\setentries'$ and~$\setentries''$. For each possible outcome~$\indexoutcome^*$ of the tournament, if $\indexentry$ is the best-performing entry in~$\mathcal{E}'' \cup \{\indexentry\}$, then it must also be the best-performing entry in~$\mathcal{E}' \cup \{\indexentry\}$, whereas the converse does not necessarily hold; namely, there may exist some $\indexentry^*  \in \mathcal{E}'' \setminus \mathcal{E}'$ such that~$s(\indexentry^*,\indexoutcome) > s(\indexentry,\indexoutcome)$. 
$\hfill \Halmos$
\end{customproof}

\section{Multi-entry Strategies with Deterministic Guarantees}
\label{appendix:MultiEntryLB}

\begin{customproof}[Proof of Proposition~\ref{prop:det_lb}:] Let~$\indexentry$ be an arbitrary entry in $\setbrackets$. We construct a second entry~$\indexentry^{'}$ that flips all the first-round choices made by~$\indexentry$, i.e., for every game~$\indexgame$ such that~$\indexround(\indexgame) = 1$, if~$\setteams(\indexgame) = \{\indexteam,\indexteam'\}$ and~$\indexentry$ selects~$\indexteam$ as the winner of~$\indexgame$, then~$\indexentry'$ selects~$\indexteam'$ as the winner of~$\indexgame$ instead. The selections made by~$\indexentry'$ for all the other games can be arbitrarily chosen. By construction, $\mathcal{E}$ contains exactly one entry with the correct outcome for each first-round game. As there are two entries in total, it follows from the pigeonhole principle that there is at least one entry between and~$\indexentry'$ that makes the correct selection for at least half of the games in the first round, i.e., 
$s(\mathcal{E},\indexoutcome) = \max(\{s(\indexentry,\indexoutcome), s(\indexentry^{'},\indexoutcome)\}) \geq \frac{\nteams}{4}$ for every outcome $\indexoutcome\in\setbrackets$. Therefore, we must have 
$\mathbb{E}\left[S( \{\indexentry,\indexentry'\})\right] \geq \frac{\nteams}{4}$. $\hfill\Halmos$
\medskip
\end{customproof}

For a \textit{March Madness} pool with the traditional scoring system, Proposition~\ref{prop:det_lb} delivers a deterministic lower bound of 16 points. Extending the idea to derive multi-entry strategies with stronger deterministic guarantees requires a considerable increase in the number of entries. The next example shows that an analogous strategy requires 16 entries to increase the lower bound from 16 to 18 points.
%r \emph{March Madness}, as shown below.

 \begin{example}\label{ex:lbexample}
For any sub-tournament~$\tourney(\indexgame_3)$, where $\indexgame_3$ is played in round 2,
%is la of games in round 1 whose winner plays a common game $\indexgame_3$ in round 2, 
there are $2^3=8$ possible outcomes.  Let $\indexentry^{\indexentryi}\in\{\indexentry^1, \ldots, \indexentry^8\}$ be the set of arbitrary entries that pick unique options for these three games; for all the other games, the assigned teams may be arbitrarily chosen.  For $\indexentryi = 1, \ldots, 8$, let $\indexentry^{\indexentryi+8}$ be the entry that selects the same teams for games $g_1, g_2$ and $g_3$ as~$\indexentry^{\indexentryi}$ while making  complementary choices for all other first-round games.   This results in a collection of sixteen entries that will guarantee at least 18 points, as one pair of entries $\{\indexentry^{\indexentryi},\indexentry^{\indexentryi+8}\}$ will make the correct choices for all games in $g_1, g_2,$ and $g_3$ (thereby scoring 4 points), and at least one of these entries will score at least $\frac{\nteams}{4}-2$ points in the other first-round games.  Therefore, for any outcome, this set of entries scores at least $4 + \frac{\nteams}{4}-2 = \frac{\nteams}{4} + 2$ points.
 \medskip
\end{example}

\begin{customproof}[Proof of Proposition~\ref{prop:det_lb2}:] Every game $\indexgame \in \setgames$ has $2^{\indexround(\indexgame)}$ possible winners. Therefore, if $\mathcal{E}$ contains entries that cover all the $2^{\indexround(\indexgame)}$ possible winners of~$\indexgame$, then at least one bracket in~$\mathcal{E}$ correctly selects the winner of game~$\indexgame$ and all previous games for which the winner of game $\indexgame$ played in.  Therefore, for any $\indexgame$ in $\setgames$, we have $\displaystyle \sum_{\indexround\in [1,\ldots ,\indexround(\indexgame)]}2^{\indexround-1}=2^{\indexround(\indexgame)}-1$.
$\hfill \Halmos$
\medskip
\end{customproof}

\section{Proof  of Theorem~\ref{sp:2}}

Our proof of Theorem~\ref{sp:2} relies on the following lemma, which shows that we can always replace a solution~$\setentries =  \{E_1,E_2\}$ such that~$E_1$ and~$E_2$ have at least one common pick for another solution~$\setentries' = \{E_1,E_3\}$ such that~$E_1$ and~$E_3$ have fewer identical selections and the expected score of~$\setentries'$ is not smaller than the expected score of~$\setentries$.

\begin{lemma}
If $\Pteam$ is the $\nteams\times\nteams$ matrix with 0s on the diagonal and 0.5 elsewhere, for any non-disjoint 2-entry solution, there exists a game for which the two entries have an identical pick, and changing the selected team in one of the entries increases the expected score of the two-entry solution. \label{lemma:lemma2}
\end{lemma}
\begin{customproof}[Proof:]
Consider \{$E^a$,$E^b$\} as an arbitrary non-disjoint 2-entry solution. Identify the last round, denoted as $\indexround^{*}$, in which both $E^a$ and $E^b$ choose the same team for at least one of the games, and denote this selected team as $\indexteam'$ in the specific game $\indexgame^{*}$ within that round. By assumption, at least one of the entries does not select $\indexteam'$ in all subsequent rounds $\indexround>\indexround^{*}$; we assume w.l.o.g. that~$E^b$ is such an entry. Let $E^c$ be the entry identical to $E^b$ except that the team chosen for game $\indexgame^{*}$ is $\indexteam''\neq \indexteam'$ (which is uniquely defined based on the structure of the tournament). We show that $\expectation{S(\{\indexentry^a, \indexentry^b\})}\leq\expectation{S(\{\indexentry^a, \indexentry^c\})}$ to conclude our argument.

Let $\{E^a, E^b\}$ and $\{E^a, E^c\}$ be two 2-entry solutions. %Note that both 2-entry solutions have $E^a$. 
Because there exists exactly one pick that differs between the two 2-entry solutions, it follows that, for every outcome~$\indexoutcome$ in~$\setbrackets$,
\begin{align*}
%\forall~\indexoutcome~\in~\setbrackets:~
|s(\{E^a,E^c\}, \indexoutcome)-s(\{E^a,E^b\}, \indexoutcome)|\leq 2^{\indexround(\indexgame^{*})-1}.
\end{align*}

The following additional notation will be convenient in the rest of the proof. Let
\[
\displaystyle \alpha^\kappa(\indexoutcome) \coloneqq \displaystyle \sum_{\indexteam \in \setteams} \sum_{\indexgame \in \setgames\setminus{\{\indexgame^{*}\}}}2^{\indexround(\indexgame)-1} \cdot \indexentry^{k}_{\indexteam,\indexgame} \cdot O_{\indexteam,\indexgame}
\]
denote the score of entry $\indexentry^\kappa$ over all games except game $\indexgame^{*}$ in outcome $\indexoutcome$, and let
\[
\gamma^\kappa(\indexoutcome) \coloneqq \sum_{\indexteam \in \setteams}2^{\indexround(\indexgame^{*})-1}\cdot\indexentry^\kappa_{\indexteam, \indexgame^{*}}\indexoutcome_{\indexteam,\indexgame^{*}}
\]
be the score of entry $\indexentry^\kappa$ for game $\indexgame^{*}$ in outcome $\indexoutcome$. We omit  $\indexgame^{*}$ in the definitions of $\alpha^\kappa(\indexoutcome)$ and $\gamma^\kappa(\indexoutcome)$ to simplify the notation. 

The score of $\indexentry^\kappa$ for outcome $\indexoutcome$ is $S(\indexentry^\kappa, \indexoutcome)=\gamma^\kappa(\indexoutcome)+\alpha^\kappa(\indexoutcome)$. By construction, we have $\gamma^a(\indexoutcome)=\gamma^b(\indexoutcome)$, as entries $\indexentry^a$ and $\indexentry^b$ have an identical selection for game $\indexgame^{*}$. Moreover,  $\alpha^b(\indexoutcome)=\alpha^c(\indexoutcome)$, as entries $\indexentry^b$ and $\indexentry^c$ have identical selections for all games except for game $\indexgame^{*}$. Direct substitution shows that
\begin{align*}
    \mathbb{E}\left[S(\{\indexentry^a, \indexentry^b\})\right]&=0.5^\ngames\sum_{\indexoutcome \in \completebrackets}  \max \Biggl(\gamma^a(\indexoutcome)+\alpha^a(\indexoutcome), \gamma^b(\indexoutcome)+\alpha^b(\indexoutcome)\Biggr)\\
    &=0.5^\ngames\sum_{\indexoutcome \in \completebrackets}  \max \Biggl(\gamma^a(\indexoutcome)+\alpha^a(\indexoutcome), \gamma^a(\indexoutcome)+\alpha^b(\indexoutcome)\Biggr)
\end{align*}
and
\begin{align*}
    \mathbb{E}\left[S(\{\indexentry^a, \indexentry^c\})\right]&=0.5^\ngames\sum_{\indexoutcome \in \completebrackets}  \max \Biggl(\gamma^a(\indexoutcome)+\alpha^a(\indexoutcome), \gamma^c(\indexoutcome)+\alpha^c(\indexoutcome)\Biggr)\\
    &=0.5^\ngames\sum_{\indexoutcome \in \completebrackets}  \max \Biggl(\gamma^a(\indexoutcome)+\alpha^a(\indexoutcome), \gamma^c(\indexoutcome)+\alpha^b(\indexoutcome)\Biggr).
\end{align*}

We partition the set of outcomes $\setbrackets$ based on the outcome of game $\indexgame^{*}$ into sets~$\mathcal{\indexoutcome}^{'}$, $\mathcal{\indexoutcome}^{''}$, and $\mathcal{\indexoutcome}^{0}$, where:

\begin{itemize}
    \item $\indexoutcome'\in\mathcal{\indexoutcome}^{'}$: $\gamma^a(\indexoutcome')=2^{\indexround(\indexgame^{*})-1}$ (i.e., $\indexteam'$ wins $\indexgame^{*}$, so both $\indexentry^a$ and $\indexentry^b$ score points in $\indexgame^{*}$);
    \item $\indexoutcome''\in\mathcal{\indexoutcome}^{''}$: $\gamma^c(\indexoutcome'')=2^{\indexround(\indexgame^{*})-1}$ (i.e., $\indexteam''$ wins game $\indexgame^{*}$, so $\indexentry^c$ score points in $\indexgame^{*}$)
    \item $\indexoutcome\in\mathcal{\indexoutcome}^{0}$: $\gamma^a(\indexoutcome)=\gamma^c(\indexoutcome)=0$ (i.e., neither $\indexteam'$ nor $\indexteam''$ wins game $\indexgame^{*}$, so     neither $\indexentry^a$, $\indexentry^b$ nor $\indexentry^c$ score)
\end{itemize}
Given this partition of~$\setbrackets$, we have
\begin{align}
    \mathbb{E}\left[S(\{\indexentry^a, \indexentry^b\})\right]&=0.5^\ngames \Biggl( \sum_{\indexoutcome \in \mathcal{\indexoutcome}^{'}}  \max \Biggl(2^{\indexround(\indexgame^{*})-1}+\alpha^a(\indexoutcome), 2^{\indexround(\indexgame^{*})-1}+\alpha^b(\indexoutcome)\Biggr)+\nonumber\\&\qquad\qquad \sum_{\indexoutcome \in \mathcal{\indexoutcome}^{''}}  \max \Biggl(\alpha^a(\indexoutcome), \alpha^b(\indexoutcome)\Biggr)
+\nonumber\\&\qquad\qquad  \sum_{\indexoutcome \in \mathcal{\indexoutcome}^{0}}  \max \Biggl(\alpha^a(\indexoutcome), \alpha^b(\indexoutcome)\Biggr) \Biggr) \label{eq:prop2eq1}
\end{align}
and
\begin{align}
    \mathbb{E}\left[S(\{\indexentry^a, \indexentry^c\})\right]&=0.5^\ngames \Biggl( \sum_{\indexoutcome \in \mathcal{\indexoutcome}^{'}}  \max \Biggl(2^{\indexround(\indexgame^{*})-1}+\alpha^a(\indexoutcome), \alpha^b(\indexoutcome)\Biggr)+\nonumber \\&\qquad\qquad \sum_{\indexoutcome \in \mathcal{\indexoutcome}^{''}}  \max \Biggl(\alpha^a(\indexoutcome), 2^{\indexround(\indexgame^{*})-1}+\alpha^b(\indexoutcome)\Biggr)
+\nonumber \\&\qquad\qquad  \sum_{\indexoutcome \in \mathcal{\indexoutcome}^{0}}  \max \Biggl(\alpha^a(\indexoutcome), \alpha^b(\indexoutcome)\Biggr) \Biggr). \label{eq:prop2eq2}
\end{align}

It follows directly from \eqref{eq:prop2eq1} and \eqref{eq:prop2eq2} that for all outcomes $\indexoutcome\in \mathcal{\indexoutcome}^{0}$, $s(\{\indexentry^a, \indexentry^b\},\indexoutcome)=s(\{\indexentry^a, \indexentry^c\}, \indexoutcome)$. A simple symmetry argument shows that~$|\mathcal{\indexoutcome}^{'}| = |\mathcal{\indexoutcome}^{''}|$, so we will show that the EMS increases if we change from~$\{E^a, E^b\}$ to~$\{E^a, E^c\}$ by exploring a bijection between~$\mathcal{\indexoutcome}^{'}$ and~$\mathcal{\indexoutcome}^{''}$. Using the notation depicted in the left bracket of Figure~\ref{fig:transformation2},

we introduce four disjoint sets representing all games in $\setgames$ except $\indexgame^{*}$:
\begin{itemize}
    \item $\setgames^{'}:=\{\indexgame\in\setgames~:~\indexround(\indexgame)<\indexround^{*},~ \indexteam'\in\nteams(\indexgame)\}$ ~\text{(i.e., all games $\indexteam'$ may play before round $\indexround^{*}$)},
    
    \item $\setgames^{''}:=\{\indexgame\in\setgames~:~\indexround(\indexgame)<\indexround^{*},~ \indexteam''\in\nteams(\indexgame)\}$ \text{(i.e., all games $\indexteam''$ may play up to round $\indexround^{*}$)},
    
    \item $\setgames^{1}:=\{\indexgame\in\setgames: ~\indexround(\indexgame)>\indexround^{*}, ~\{\indexteam',~\indexteam''\}\subseteq \nteams(\indexgame)\}$~\text{(i.e., all games $\indexteam'$ and $\indexteam''$ may play after round $\indexround^{*}$)},
    
    \item $\setgames^{0}:=\{\indexgame\in\setgames:~\nteams(\indexgame)\subseteq \nteams(\indexgame)\setminus\{\indexteam', \indexteam''\}\}$~\text{(i.e., all games $\indexteam'$ and $\indexteam''$ can never play)}.
\end{itemize}

Furthermore, we use $\indexgame^{'}_\indexround$ and $\indexgame^{''}_\indexround$ to represent games in $\setgames^{'}$ and $\setgames^{''}$, respectively, 
with $\indexround$ denoting the round of the game $\indexteam'$ and $\indexteam''$ may play. Additionally, let $\indexteam^{w}_{\indexgame}$ be the winners of game $\indexgame$ in outcome $\indexoutcome^{'}$. 

Given an outcome $\indexoutcome'$ in $\mathcal{\indexoutcome}^{'}$, we construct a symmetric outcome $\indexoutcome''$ in $\mathcal{\indexoutcome}^{''}$ 
by assigning only the winner to each game in both $\indexoutcome'$ and $\indexoutcome''$ (e.g., $\indexoutcome_{\indexteam^1, \indexgame}=1 \iff \indexoutcome_{\indexteam^2, \indexgame}=1$). The construction is done in four steps:

\begin{enumerate}
    \item[1.] Set 
    $\indexoutcome^{''}_{\indexteam'', \indexgame^{*}}=1$, and for every~$\indexgame^{''}_{\indexround}$ in~$\setgames^{''}$, set    
    $\indexoutcome^{''}_{\indexteam'', \indexgame^{''}_{\indexround}}=1$ (see Figure~\ref{fig:transformation1}).
\end{enumerate}

\begin{figure}[h!]
            \centering
    \includegraphics[width=0.75\textwidth]{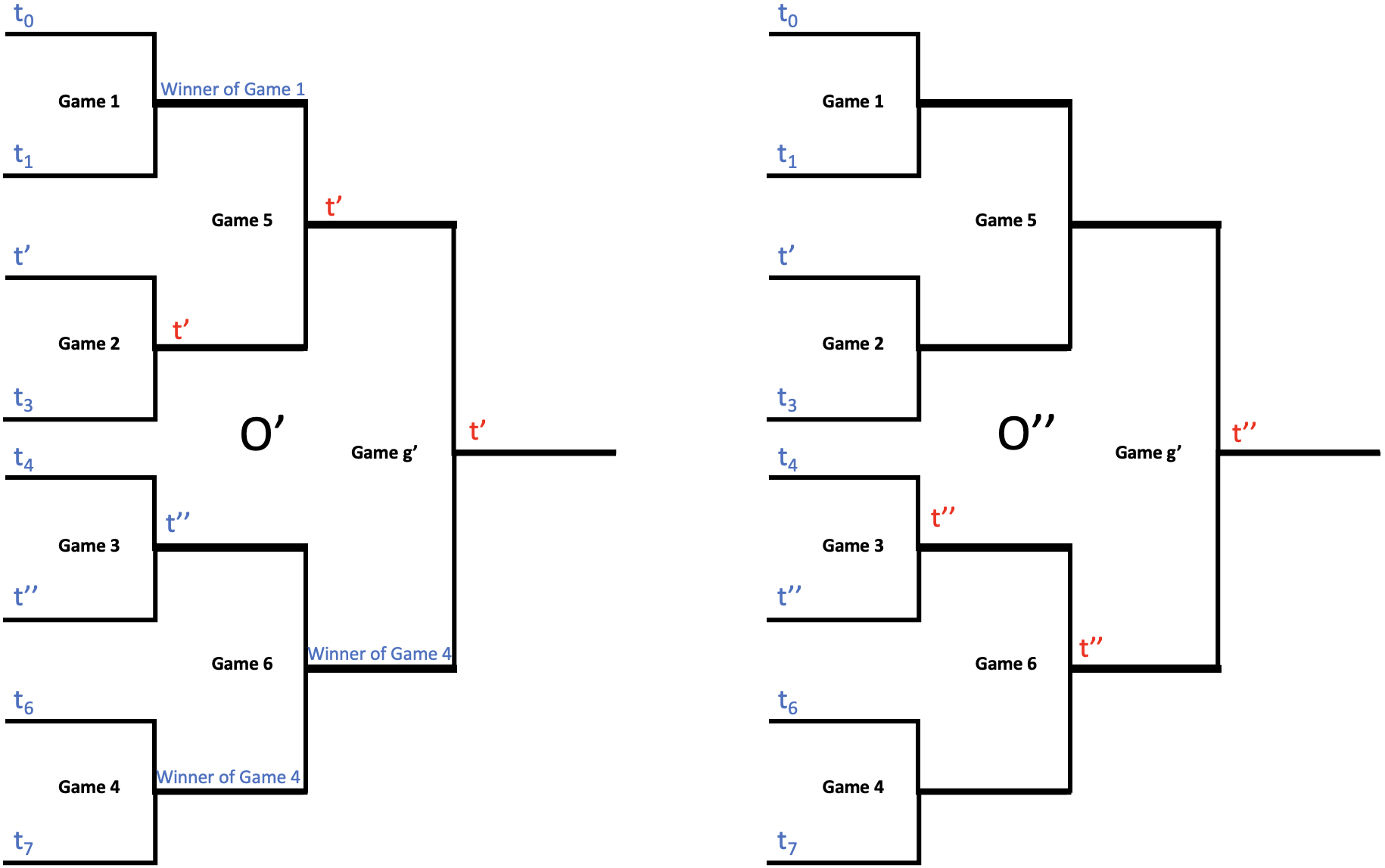}
            \caption{First Step (team $\indexteam''$ is assigned to $\indexgame^{*}$ and all games in $\setgames^{''}$) }
            \label{fig:transformation1}
\end{figure}
% \newpage
\begin{enumerate}
        \item[2.] For every $\indexteam$ in $\setteams$ and $\indexgame$ in $\setgames^0$, set $\indexoutcome^{''}_{\indexteam, \indexgame}=\indexoutcome^{'}_{\indexteam, \indexgame}$
        (see Figure~\ref{fig:transformation2}).
\end{enumerate}
\begin{figure}[h!]
            \centering
    \includegraphics[width=0.75\textwidth]{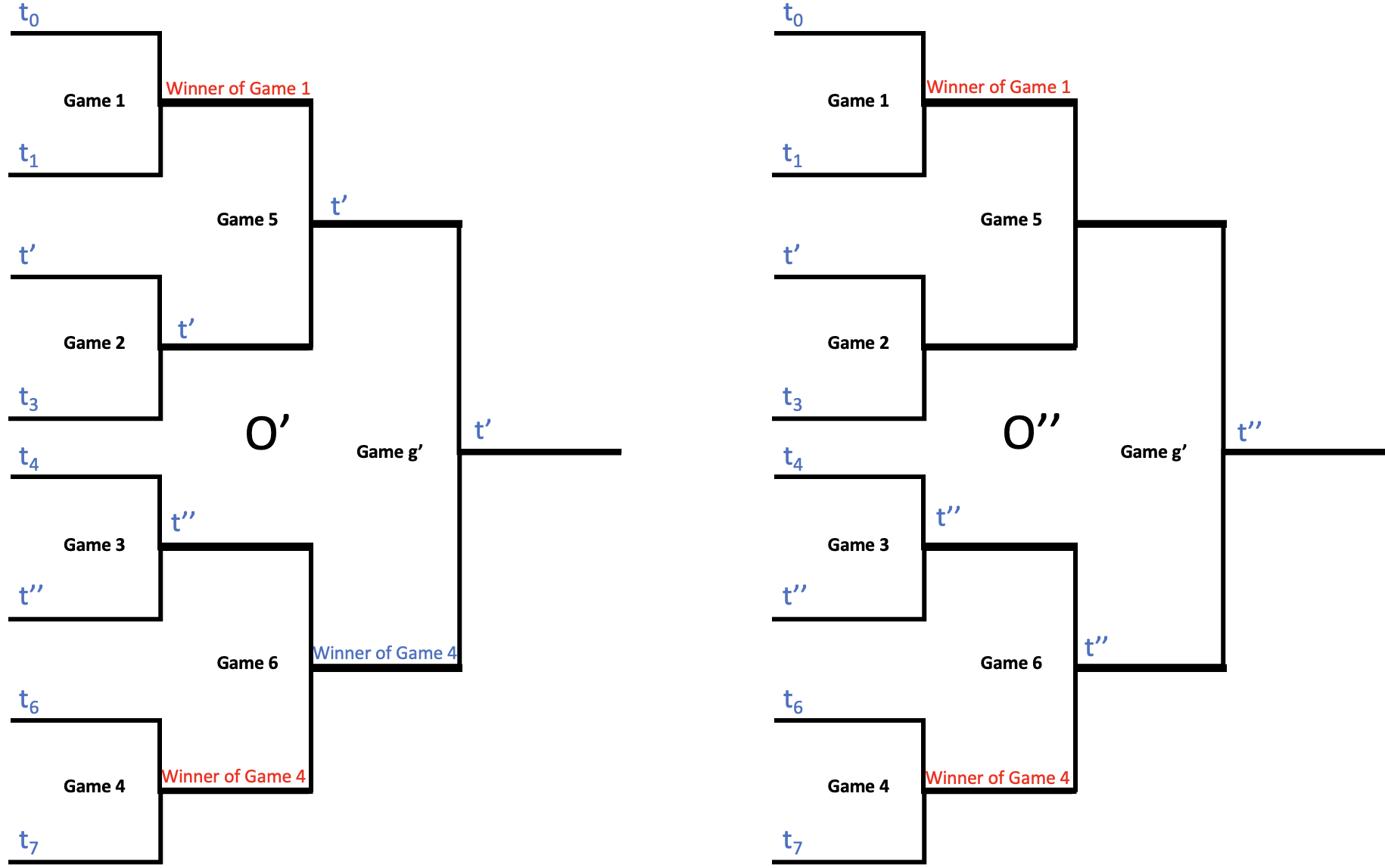}
            \caption{Second Step (copy the choice structure of $\indexoutcome^{'}$ for all games in $\setgames^0$)}
            \label{fig:transformation2}
\end{figure}
% \newpage

\begin{enumerate}
        \item[3.] For every $\indexgame$ in $\setgames^{1}$ 
        (see Figure~\ref{fig:transformation4}):
        \begin{enumerate}
            \item[3.1] If $\indexteam^{w}_{\indexgame}=\indexteam'$, then set $\indexoutcome^{''}_{\indexteam'',\indexgame}=1$.
            \item[3.2] Otherwise,  $\forall~\indexteam\in\setteams ~\indexoutcome^{''}_{\indexteam,\indexgame}=\indexoutcome^{'}_{\indexteam, \indexgame}$.
        \end{enumerate}
        \begin{figure}[h!]
            \centering
    \includegraphics[width=0.75\textwidth]{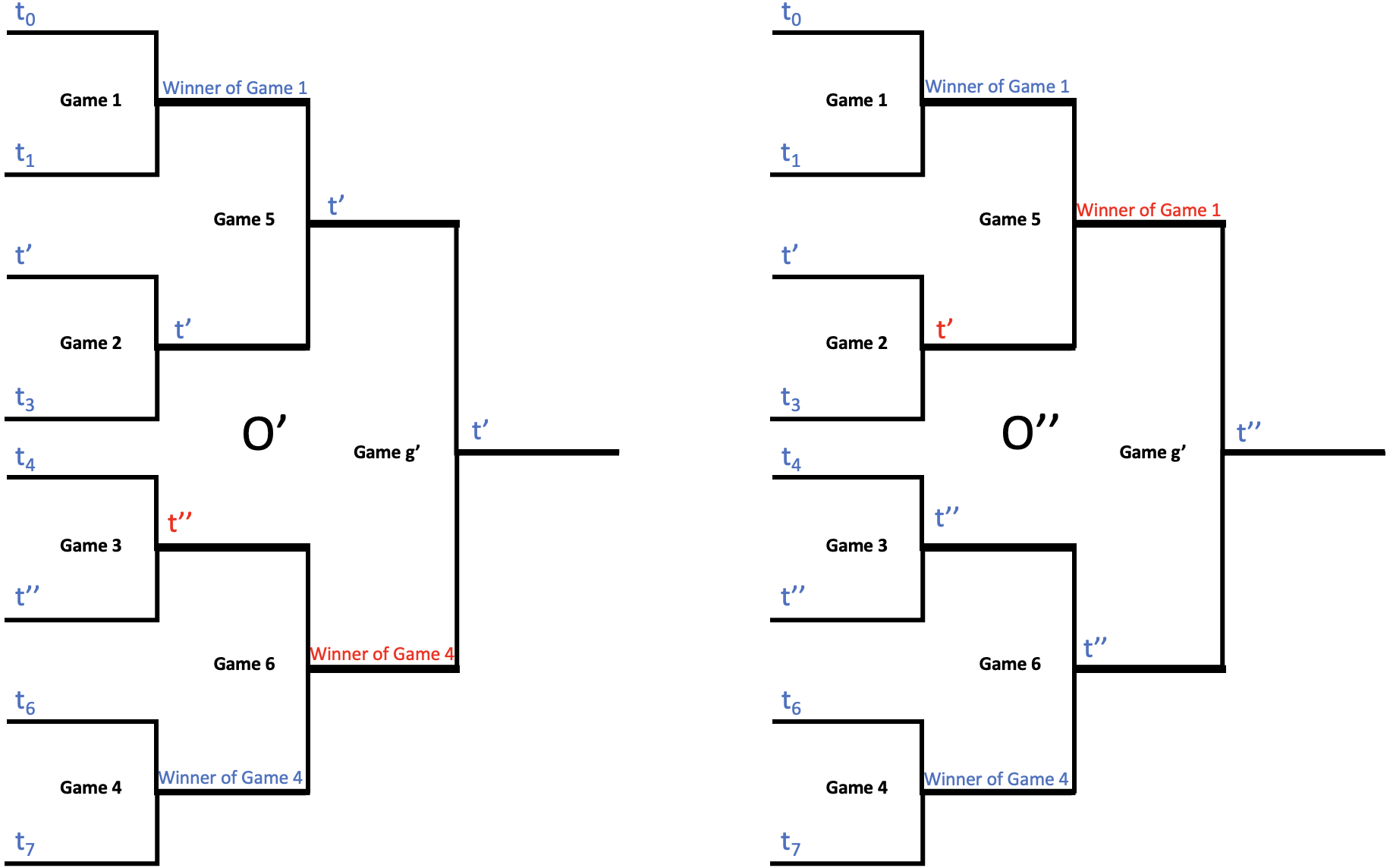}
            \caption{Third Step (if $\indexteam'$ was assigned to game $\indexgame$ in $\indexoutcome'$, then assign $\indexteam''$ to $\indexgame$ in $\indexoutcome''$; otherwise, copy the assignment of $\indexoutcome^{'}$ for game $\indexgame$) }
            \label{fig:transformation4}
\end{figure}

        \item[4.] For every
        $\indexround$ in $\setrounds$, $\indexgame^{'}_\indexround$ in $\setgames^{'},$ and $\indexgame^{''}_\indexround$ in $\setgames^{''}$:
        \begin{enumerate}
            \item[4.1] If $\indexround=1$:
            \begin{enumerate}
                \item[4.1.1] If $\indexteam^w_{\indexgame^{''}_1}=\indexteam''$,
                 set $\indexoutcome^{''}_{\indexteam', \indexgame^{'}_1}=1$  (i.e., if $\indexteam''$ is assigned to $\indexgame^{''}_{1}$ in $\indexoutcome'$, then assign $\indexteam'$ to $\indexgame^{''}_1$)
                \item[4.1.2] Otherwise, set $\indexoutcome^{''}_{\indexteam^*, \indexgame^{*}_1}=1$ for $\indexteam^*=\nteams(\indexgame^{'}_1)\setminus\{\indexteam'\}$  (i.e., assign the team playing $\indexteam'$ in the first round)
            \end{enumerate}
        \item[4.2] Otherwise:
        \begin{enumerate}
            \item[4.2.1] If $\indexteam^{w}_{\indexgame^{''}_r}=\indexteam^{w}_{\indexgame^{''}_{r-1}}$, set $\indexoutcome^{''}_{\indexteam,\indexgame^{'}_{\indexround}}=\indexoutcome^{''}_{\indexteam,\indexgame^{'}_{\indexround-1}}$ for every
            $\indexteam$ in $\setteams$
            %$\forall\indexteam\in\setteams$ $\indexoutcome^{''}_{\indexteam,\indexgame^{'}_{\indexround}}=\indexoutcome^{''}_{\indexteam,\indexgame^{'}_{\indexround-1}}$ 
            (i.e., if the same team is assigned in $\indexgame^{''}_\indexround$ and $\indexgame^{''}_{\indexround-1}$, then select the same team of $\indexgame^{'}_{\indexround-1}$ in $\indexgame^{'}_{\indexround})$
            \item[4.2.2] Otherwise, set $\indexoutcome^{''}_{\indexteam,\indexgame^{*}_{\indexround}}=\indexoutcome^{''}_{\indexteam,\indexgame^{*}}$ for            
            $\indexgame^* = \Gamma^{-}(\indexgame^{*}_\indexround)\setminus\{\indexgame^{*}_{\indexround-1}\}$ and every $\indexteam$ in $\setteams$ %$\indexoutcome^{''}_{\indexteam,\indexgame^{*}_{\indexround}}=\indexoutcome^{''}_{\indexteam,\indexgame^{*}}$ 
            (i.e., copy the assignment of the previous game to $\indexgame^{'}_{\indexround}$ which is not $\indexgame^{'}_{\indexround-1}$)
        \end{enumerate}
        \end{enumerate}
\end{enumerate}

\begin{figure}
    \centering
    \includegraphics[width=0.75\linewidth]{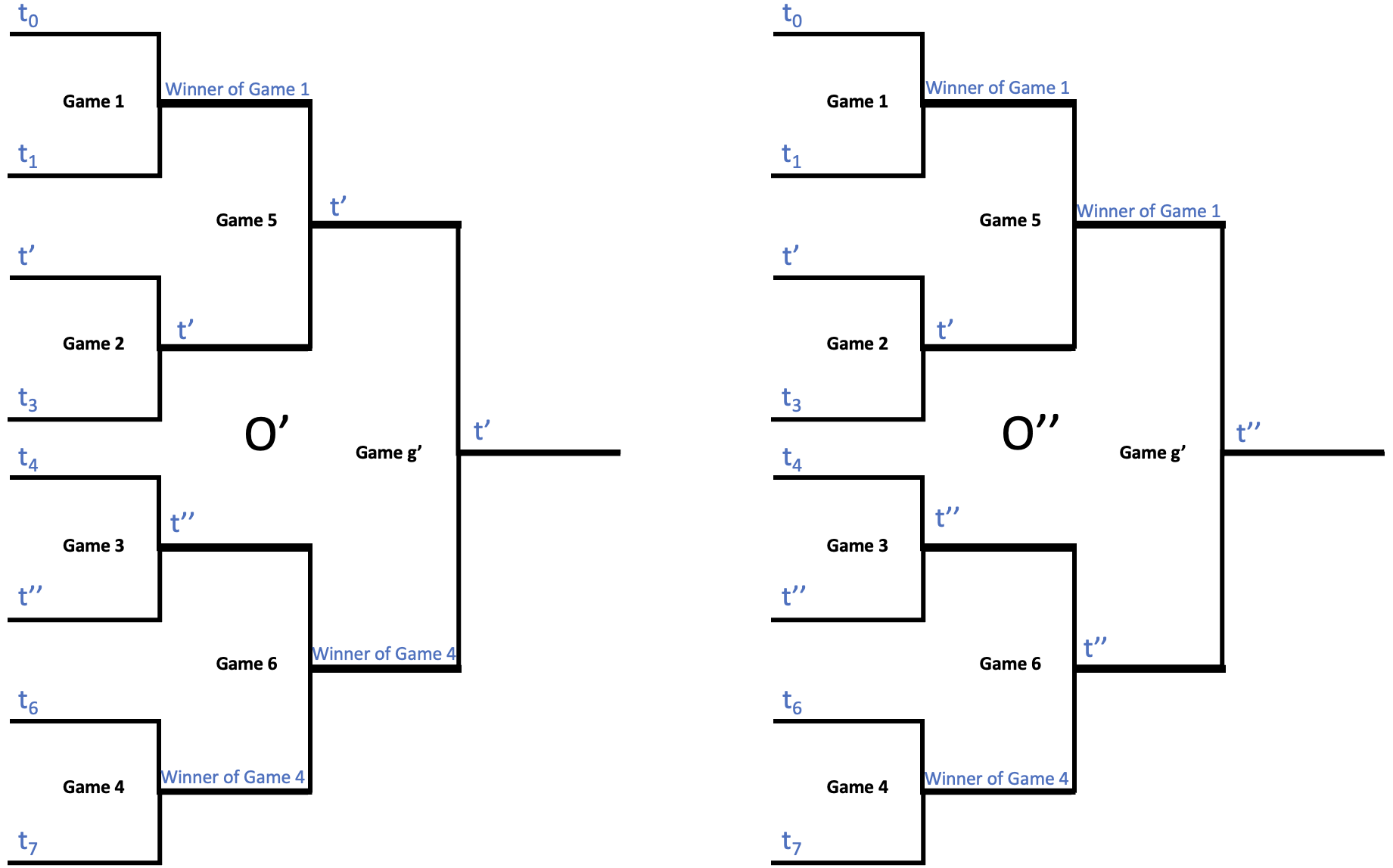}
    \caption{Symmetric Pair of Outcomes $(\indexoutcome', \indexoutcome'')$}
    \label{fig:SymmetricOutcomes}
\end{figure}

The results of our construction applied to our running example are depicted in Figure~\ref{fig:SymmetricOutcomes}. Next, we study the score of~$\indexentry^b$ for these symmetric pairs.

\begin{claim}
\label{claim:Lemma2Claim3}
Given a symmetrical pair of outcomes, denoted as $(\indexoutcome', \indexoutcome'')$, the score for entry $\indexentry^b$ in outcome $\indexoutcome'$ is identical to that in outcome $\indexoutcome''$ for all games except game $\indexgame^{*}$ (i.e., $\alpha^b(\indexoutcome')=\alpha^b(\indexoutcome'')$).
\end{claim}
\begin{customproof}[Proof:] We analyse the score of~$\indexentry^b$ for a symmetric pair $(\indexoutcome', \indexoutcome'')$ for the four (disjoint) categories of games 
%In this proof, we rely on the construction of a pair of symmetric outcomes which separates all games into four disjoint sets of games defined by 
$\setgames^{0}$, $\setgames^{1}$, $\setgames^{'}$, and $\setgames^{''}$ defined above.
% \medskip

\begin{enumerate}
    \item \textbf{Games in $\setgames^{0}$:} %Let's begin by analyzing all games in $\setgames^{0}$. 
By construction, 
%Given a pair of symmetric outcomes $(\indexoutcome', \indexoutcome'')$, 
the teams assignments for games in $\setgames^{0}$ are identical for both $\indexoutcome'$ and $\indexoutcome''$. Therefore, the score for $\indexentry^b$ is identical on the subset of games within $\setgames^{0}$.

    \item \textbf{Games in $\setgames^{1}$:}
    There are two cases to consider. In the first, there are games~$\indexgame$ assigned to $\indexteam'$ in $\indexoutcome'$ and to $\indexteam''$ in $\indexoutcome''$. By construction, $\indexentry^b$ does not select any of those teams after round $\indexround^{*}$, so the score of~$\indexentry^b$ on~$\indexgame$ must be the same for~$\indexoutcome'$ and~$\indexoutcome''$. In the second case, $\indexoutcome'$ and~$\indexoutcome''$ must select the same team, and it is clear that~$\indexentry^b$ scores the same number of points on both outcomes for such cases.

    \item \textbf{Games in $\setgames^{'}$:} By construction, teams $\indexteam'$ and $\indexteam''$ win game 
    %$\indexround^{*}$
    $\indexgame^{*}$
    in $\indexoutcome'$ and $\indexoutcome''$, respectively. As  $\indexentry^b$ assigns $\indexteam'$ and $\indexteam''$ in their respective games in round $\indexround^{*}-1$, the number of points $\indexentry^b$ obtains from  $\indexteam'$ for games in $\setgames'$ in $\indexoutcome'$ is identical to the number of points from $\indexteam''$ for games in $\setgames''$ in $\indexoutcome''$.

    \item \textbf{Games in $\setgames^{''}$:} %Finally, we need to demonstrate that the number of points in games $\setgames''$ in $\indexoutcome'$ is also identical to those in $\setgames'$ in $\indexoutcome''$. 
    By construction, 
%    The construction of the pair $(\indexoutcome', \indexoutcome'')$ ensures that 
if $\indexteam''$ is assigned in $\indexoutcome'$ to a given game 
in round $\indexround''<\indexround^{*}$, 
then
%the symmetric outcome 
$\indexoutcome''$ 
%guarantees that 
assigns $\indexteam'$ 
%is assigned 
to its respective game in round $\indexround''$. As a result, the score obtained by team $\indexteam''$ in outcome $\indexoutcome'$ is identical to the score obtained by team $\indexteam'$ in outcome $\indexoutcome'$.
    
\end{enumerate}

Therefore, it follows that $\alpha^b(\indexoutcome')=\alpha^b(\indexoutcome'')$, thereby concluding the proof of Claim~\ref{claim:Lemma2Claim3}.
$\hfill \Halmos$
\medskip
\end{customproof}

The next result shows that, differently from~$\indexentry^b$, $\indexentry^a$ obtains more points from~$\indexoutcome'$ than it does from~$\indexoutcome''$.

\begin{claim}
\label{claim:Lemma2Claim4}
    Given a symmetrical pair of outcomes, denoted as $(\indexoutcome', \indexoutcome'')$, the score for entry $\indexentry^a$ in outcome $\indexoutcome'$ is greater or equal to that in outcome $\indexoutcome''$ for all games except game $\indexgame^{*}$ (i.e., $\alpha^a(\indexoutcome')\geq\alpha^a(\indexoutcome'')$).
\end{claim}
\begin{customproof}[Proof:] This proof also analyzes the categories of games defined above. 

\begin{enumerate}
    \item \textbf{Games in $\setgames^{0}$:} The team assignments for games in $\setgames^{0}$ are identical for $\indexoutcome'$ and $\indexoutcome''$. Consequently, the score for $\indexentry^a$ remains identical on the subset of games within $\setgames^{0}$.

    \item \textbf{Games in $\setgames^{1}$:} There are two cases to consider for games~$\indexgame$ in~$\setgames^{1}$. If $\indexoutcome'$ and~$\indexoutcome''$ select the same team, $\indexentry^a$ scores the same number of points on both outcomes. Otherwise, $\indexoutcome'$ selects $\indexteam'$ and $\indexoutcome''$ selects $\indexteam''$. By construction, $\indexentry^a$ may select $\indexteam'$ in round $\indexround>\indexround^{*}$, but not  $\indexteam''$. Therefore, $\indexentry^a$ scores at least as many points  in~$\indexgame$ for~$\indexoutcome'$ as it does in $\indexoutcome''$. Moreover, if  $\indexentry^a$ assigns $\indexteam'$ in round $\indexround>\indexround^{*}$, then the score of $\indexentry^a$ decreases when $\indexteam''$ is assigned to the same round in $\indexoutcome''$.  

    \item \textbf{Games in $\setgames^{'}$ and $\setgames^{''}$:} By definition, $\indexteam'$ and $\indexteam''$ win game $\indexgame^{*}$ in $\indexoutcome'$ and $\indexoutcome''$, respectively. Additionally, we have no control over the teams
%    it's important to note that we don't know which team is 
assigned to games in $\setgames''$ in $\indexentry^a$,
and in particular, we do not know whether the selections of  $\indexentry^a$  for these games coincide with the ones in $\indexoutcome'$.

By construction, the number of games that $\indexteam''$ is assigned to in $\indexoutcome'$ is equal to the number of games to which $\indexteam'$ is assigned in $\indexoutcome''$. Assuming that $\indexentry^a$ has $\indexteam''$ correctly assigned in its respective games before round $\indexround'' \leq \indexround^{*} - 1$ in $\indexoutcome'$, then $\indexentry^a$ also correctly assigned $\indexteam'$ for $\indexround''$ games in $\indexoutcome''$. If $\indexround''=\indexround^{*}-1$, then the score of $\indexentry^a$ on games in $\setgames'$ and $\setgames''$ is the same for $\indexoutcome'$ and $\indexoutcome''$. Conversely, if $\indexround'' < \indexround^{*} - 1$,
then $\indexentry^a$ is incorrect on games $\{\indexgame^{''}_{\indexround''+1},\ldots,\indexgame^{''}_{\indexround^{*}-1}\}$ and $\{\indexgame^{'}_{\indexround''+1},\ldots,\indexgame^{'}_{\indexround^{*}-1}\}$ because $\indexteam'$ and $\indexteam''$ are respectively assigned up to $\indexround''$ and $\indexround^*$ in $\indexoutcome''$. Consequently, the score of $\indexentry^a$ for games in $\setgames'$ and $\setgames''$ in $\indexoutcome'$ either equals or surpasses the score in $\indexoutcome''$.

\end{enumerate}

Therefore, it follows that $\alpha^a(\indexoutcome')\geq \alpha^a(\indexoutcome'')$, thereby concluding the proof of Claim~\ref{claim:Lemma2Claim4}.
$\hfill \Halmos$
\medskip
\end{customproof}

Using the previous results, we conclude the proof by showing that
\begin{align}
    s(\{\indexentry^a, \indexentry^b\}, \indexoutcome')+s(\{\indexentry^a, \indexentry^b\}, \indexoutcome'')\leq s(\{\indexentry^a, \indexentry^c\}, \indexoutcome')+s(\{\indexentry^a, \indexentry^c\}, \indexoutcome'')\label{eq:prop2eq3}.
\end{align}

First, we partition the set of outcomes based on the respective values of $\alpha^a(\indexoutcome)$ and $\alpha^b(\indexoutcome)$ as follows:
\begin{itemize}
    \item \textbf{Case $\text{A}_1:$} $|\alpha^b(\indexoutcome)- \alpha^a(\indexoutcome)|< 2^{\indexround(\indexgame^{*})-1}$, i.e., the outcome of game $\indexgame^{*}$ matters:
    \begin{itemize}
        \item \textbf{$\text{A}_{1a}:$} $\alpha^a(\indexoutcome)\geq \alpha^b(\indexoutcome)$, i.e., $\indexentry^a$ scores at least as many points as $\indexentry^b$ on all games excluding $\indexgame^{*}$;
        \item \textbf{$\text{A}_{1b}:$} $\alpha^a(\indexoutcome)< \alpha^b(\indexoutcome)$, i.e., $\indexentry^b$ scores more points than $\indexentry^a$ on all games excluding $\indexgame^{*}$.
    \end{itemize}
    \item \textbf{Case $\text{A}_2:$} $|\alpha^b(\indexoutcome)- \alpha^a(\indexoutcome)|\geq 2^{\indexround(\indexgame^{*})-1}$, i.e., the outcome of game $\indexgame^{*}$ does not matter:
    \begin{itemize}
        \item \textbf{$\text{A}_{2a}:$} $\alpha^b(\indexoutcome)+ 2^{\indexround(\indexgame^{*})-1}\leq \alpha^a(\indexoutcome)$, i.e., $\indexentry^a$ scores at least as many points as $\indexentry^b$;
        \item \textbf{$\text{A}_{2b}:$} $\alpha^a(\indexoutcome)+ 2^{\indexround(\indexgame^{*})-1} \leq \alpha^b(\indexoutcome)$, i.e., $\indexentry^b$ scores at least as many points as $\indexentry^a$.
    \end{itemize}
\end{itemize}
Table~\ref{tab:Prop2ScoringTable} displays the score of both $S(\{\indexentry^a, \indexentry^b\}, \indexoutcome')$ and $S(\{\indexentry^a, \indexentry^c\}, \indexoutcome'')$ on all pairs of outcomes $(\indexoutcome', \indexoutcome'')$ given a specific partition. We now evaluate the results of Table~\ref{tab:Prop2ScoringTable} row by row to prove Equation~(\ref{eq:prop2eq3}).

\begin{table}[h!]
\footnotesize
\resizebox{\textwidth}{!}{
\begin{tabular}{|c|cc|cc|}
\hline
    & \multicolumn{2}{c|}{$s(\{\indexentry^a, \indexentry^b\}, \indexoutcome)$} & \multicolumn{2}{c|}{$s(\{\indexentry^a, \indexentry^c\}, \indexoutcome)$}  \\ \hline
\begin{tabular}[c]{@{}c@{}}$\forall (\indexoutcome^{'}, \indexoutcome^{''})$: \\ $\indexoutcome^{'}\in\mathcal{\indexoutcome^{'}}, \indexoutcome^{''} \in\mathcal{\indexoutcome}^{''}$\end{tabular} & \multicolumn{1}{c|}{$\indexoutcome^{'} \in~\mathcal{\indexoutcome}^{'}$}                      & $\indexoutcome^{''}~\in~\mathcal{\indexoutcome}^{''}$                       & \multicolumn{1}{c|}{$\indexoutcome^{'}~\in~\mathcal{\indexoutcome}^{'}$}                      & $\indexoutcome^{''}~\in~ \mathcal{\indexoutcome}^{''}$                                                      \\ \hline
$\text{A}_{1a}$                                                                                                                                        & \multicolumn{1}{c|}{$\alpha^a(\indexoutcome')+2^{\indexround(\indexgame^{*})-1}$} & $\alpha^b(\indexoutcome'')$ & \multicolumn{1}{c|}{$\alpha^a(\indexoutcome')+2^{\indexround(\indexgame^{*})-1}$} & $\alpha^b(\indexoutcome'')+2^{\indexround(\indexgame^{*})-1}$                                  \\ \hline
$\text{A}_{1b}$                                                                                                                                        & \multicolumn{1}{c|}{$\alpha^b(\indexoutcome')+2^{\indexround(\indexgame^{*})-1}$} & $\alpha^b(\indexoutcome'')$                                 & \multicolumn{1}{c|}{$\alpha^a(\indexoutcome')+2^{\indexround(\indexgame^{*})-1}$} & $\alpha^b(\indexoutcome'')+2^{\indexround(\indexgame^{*})-1}$                                 \\ \hline
$\text{A}_{2a}$                                                                                                                                        & \multicolumn{1}{c|}{$\alpha^a(\indexoutcome')+2^{\indexround(\indexgame^{*})-1}$} & $\max(\alpha^a(\indexoutcome''), \alpha^b(\indexoutcome''))$ & \multicolumn{1}{c|}{$\alpha^a(\indexoutcome')+2^{\indexround(\indexgame^{*})-1}$} & $\max(\alpha^a(\indexoutcome''), \alpha^b(\indexoutcome'')+2^{\indexround(\indexgame^{*})-1})$ \\ \hline
$\text{A}_{2b}$                                                                                                                                        & \multicolumn{1}{c|}{$\alpha^b(\indexoutcome')+2^{\indexround(\indexgame^{*})-1}$} & $\alpha^b(\indexoutcome'')$                                 & \multicolumn{1}{c|}{$\alpha^b(\indexoutcome')$}                                & $\alpha^b(\indexoutcome'')+2^{\indexround(\indexgame^{*})-1}$                                 \\ \hline
\end{tabular}}
\caption{$S(\{\indexentry^a, \indexentry^b\}, \indexoutcome)$ versus $S(\{\indexentry^a, \indexentry^c\}, \indexoutcome)$ on any Symmetric Pair  $(\indexoutcome', \indexoutcome'')$.}
\label{tab:Prop2ScoringTable}
\end{table}

\begin{enumerate}
    \item \textbf{Case $A_1$: } In both cases (\textbf{$A_{1a}$} and \textbf{$A_{1b}$}), %Let us first evaluate the partition defined by $\indexoutcome~\in~\text{A}_{1a}$, and $\indexoutcome~\in~\text{A}_{1b}$. First, 
    %$\alpha^a(\indexoutcome)$ and $\alpha^b(\indexoutcome)$
   % we observe that they have exactly the same function defining the score 
   %are identical for all $\indexoutcome'\in\mathcal{\indexoutcome}^{'}$ and $\indexoutcome''\in\mathcal{\indexoutcome}^{''}$, so
   % . For both partition, 
   Equation~\eqref{eq:prop2eq3} gives 
\begin{align*}
\alpha^a(\indexoutcome')+2^{\indexround(\indexgame^{*})-1}+\alpha^b(\indexoutcome'')&\leq \alpha^a(\indexoutcome')+2^{\indexround(\indexgame^{*})-1}+\alpha^b(\indexoutcome'')+2^{\indexround(\indexgame^{*})-1}\\
s(\{\indexentry^a,\indexentry^b\},\indexoutcome)&\leq s(\{\indexentry^a,\indexentry^c\},\indexoutcome).
\end{align*}

\item \textbf{Case $A_{2a}$: } For this set of outcomes, Equation~\eqref{eq:prop2eq3} gives  
\begin{align*}
\alpha^a(\indexoutcome')+2^{\indexround(\indexgame^{*})-1}+\max(\alpha^a(\indexoutcome''),\alpha^b(\indexoutcome''))&\leq \alpha^a(\indexoutcome)+2^{\indexround(\indexgame^{*})-1}+\max(\alpha^a(\indexoutcome''),\alpha^b(\indexoutcome'')+2^{\indexround(\indexgame^{*})-1})\\
s(\{\indexentry^a,\indexentry^b\},\indexoutcome)&\leq s(\{\indexentry^a,\indexentry^c\},\indexoutcome).
\end{align*}

\item \textbf{Case $A_{2b}$: } Finally, for $\indexoutcome~\in~\text{A}_{2b}$,  Equation~\eqref{eq:prop2eq3} gives
\begin{align*}
\alpha^b(\indexoutcome')+2^{\indexround(\indexgame^{*})-1}+\alpha^b(\indexoutcome'')&= \alpha^b(\indexoutcome')+\alpha^b(\indexoutcome'')+2^{\indexround(\indexgame^{*})-1}\\
s(\{\indexentry^a,\indexentry^b\},\indexoutcome)&= s(\{\indexentry^a,\indexentry^c\},\indexoutcome).
\end{align*}

\end{enumerate}

Therefore, Equation~\eqref{eq:prop2eq3} holds  for all $\indexoutcome$ in $\setbrackets$, thus proving that $\expectation{S(\{\indexentry^a, \indexentry^b\})}\leq \expectation{S(\{\indexentry^a, \indexentry^c\})}$. This completes the proof of Lemma~\ref{lemma:lemma2}. $\hfill \Halmos$
\medskip
\end{customproof}

Our construction of~$\indexentry''$ based on~$\{\indexentry,\indexentry'\}$ applies to any arbitrary pair of disjoint entries, and it follows from Lemma~\ref{lemma:lemma2} that one can always construct a pair of entries~$\{\indexentry,\indexentry''\}$ out of a pair of non-disjoint entries~$\{\indexentry,\indexentry'\}$ such that $\expectation{S(\{\indexentry, \indexentry'\})}\leq \expectation{S(\{\indexentry, \indexentry''\})}$. Moreover, by applying the same construction iteratively, we can obtain a solution~$\{\indexentry,\indexentry^* \}$ with disjoint entries
such that $\expectation{S(\{\indexentry, \indexentry'\})}\leq \expectation{S(\{\indexentry, \indexentry^*\})}$. Therefore, we conclude that for every solution with non-disjoint entries, a solution with disjoint entries and a larger expected score exists.

The final observation is that if $\Pteami_{\indexteam,\indexteam'} = 0.5$ for all~$\indexteam,\indexteam'$ in~$\setteams$, then all disjoint 2-entry solutions have the same expected score. This claim holds because, for any two solutions $\{\indexentry^1,\indexentry^2\}$ and $\{\indexentry^3,\indexentry^4\}$ with disjoint entries, the number of outcomes for which $E^1$ scores $x$ \emph{and} $E^2$ scores $y$ is identical to the number of outcomes for which $E^3$ scores $x$ \emph{and}  $E^4$ scores $y$. Therefore, it follows that if $\Pteam_{\indexteam,\indexteam'} =0.5$ for all teams~$\indexteam,\indexteam'$ in~$\setteams$, an entry set is optimal if and only if the entries are disjoint. This concludes the proof of Theorem~\ref{sp:2}. $\hfill \Halmos$

\section{Calibration of the Algorithms}
\label{appendix:Calibration}
\subsection{Calibration of \SIP}
Formulation~\ref{ipseq} includes the following set of constraints $D(\Bar{x})$:

\begin{itemize}
    \item \textit{Champion Constraint}: Team $\indexteam$ cannot be selected by more than $\left\lceil \nentries\cdot \Proundi_{\indexteam,\indexround}\right \rceil$ entries in the last two rounds;
    \item \textit{Finalist Constraint}: Each entry has a unique pair of teams reaching the final game;
    \item \textit{Global Constraint}: Any pair of entries have at most~$\sigma$ identical selections throughout the entire tournament;\label{constraint:Global}
    %when compared to any other previously chosen entry; and
    \item \textit{Round Constraints}: Any pair of entries can have at most $\sigma_{\indexround}$ identical selections in round $\indexround \in \{1,2,3,4\}$.
\end{itemize}

We always include the first two constraints, so we proceed to fine-tune the diversification parameters $\sigma$ and $\sigma_\indexround$, $\indexround \in \{1,2,3,4\}$, for different values of~$\nentries$ utilizing data from \textit{March Madness 2023}. 

\paragraph{Identification of~$\sigma$:}

\begin{figure}[h!]
    \centering
    \includegraphics[width=0.4\linewidth]{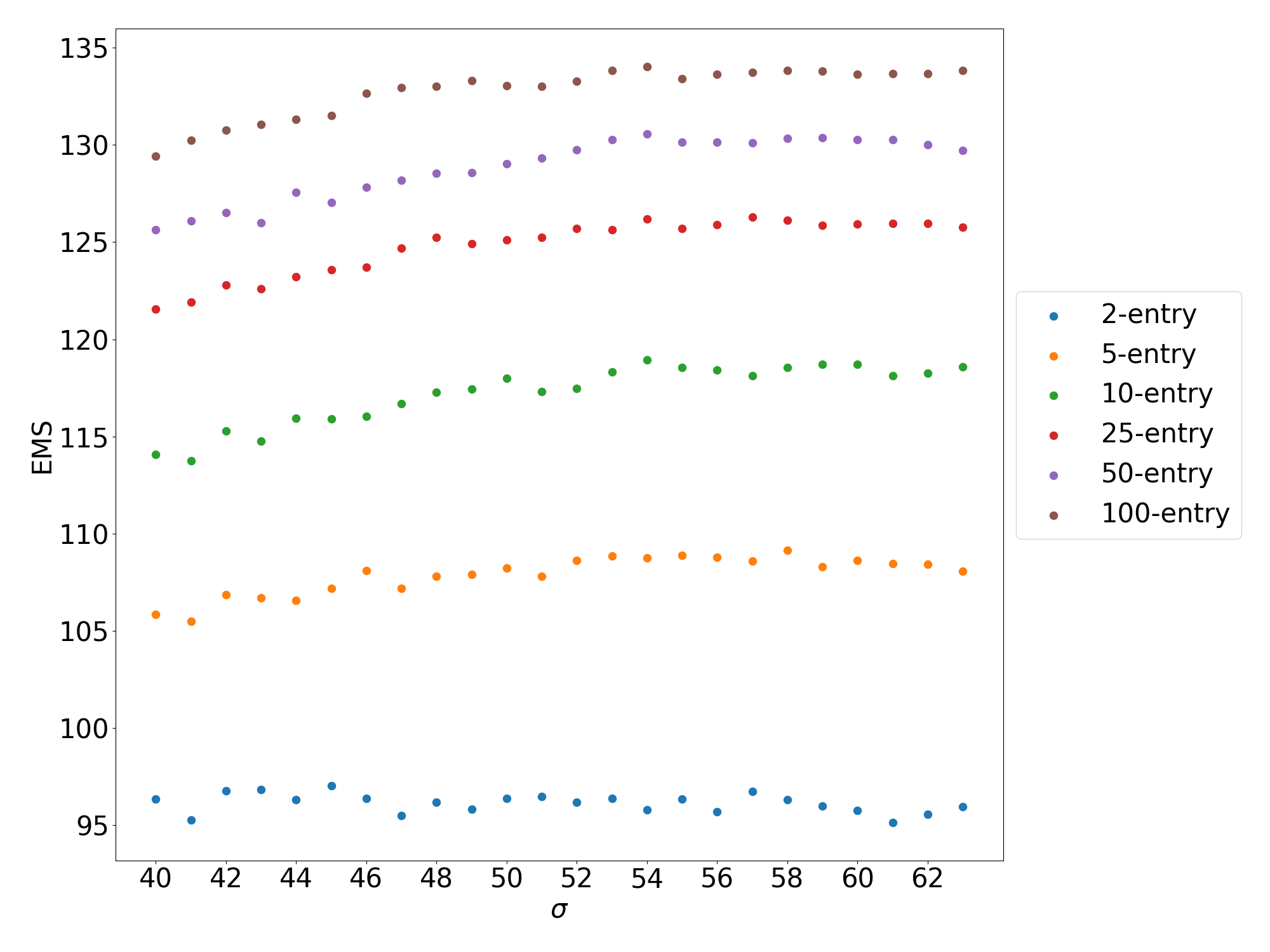}
    \caption{Empirical EMS for Different Values of~$\sigma$}
    \label{fig:sigma_tuning}
\end{figure}

In these experiments, we incorporate in $D(\Bar{x})$ the \textit{Champion Constraint}, the \textit{Finalist Constraint}, and the \textit{Global Constraint}, and test configuration of~\ref{ipseq} with all values of~$\sigma$ in~$\{40, \ldots, 62\}$. Figure~\ref{fig:sigma_tuning} reports the empirical EMS for solutions with 2, 10, 25, 50 and 100 entries. The results 
%Figure~\ref{fig:sigma_tuning} 
show that the best~$\sigma$ depends on the number of entries. For $\nentries=2$, the \textit{Global Constraint} does not significantly impact the empirical EMS (although $\sigma=45$ is marginally better). In contrast, the impact of the \textit{Global constraint} becomes more noticeable as $\nentries$ increases. For 5-entry solutions, setting $\sigma=57$ resulted in the best solutions, whereas $\sigma=54$ was the best fit for all collections of entries with more than 10 entries.

\paragraph{Identification of~$\sigma_\indexround$:}

\begin{figure}
    \centering
    \includegraphics[width=0.8\linewidth]{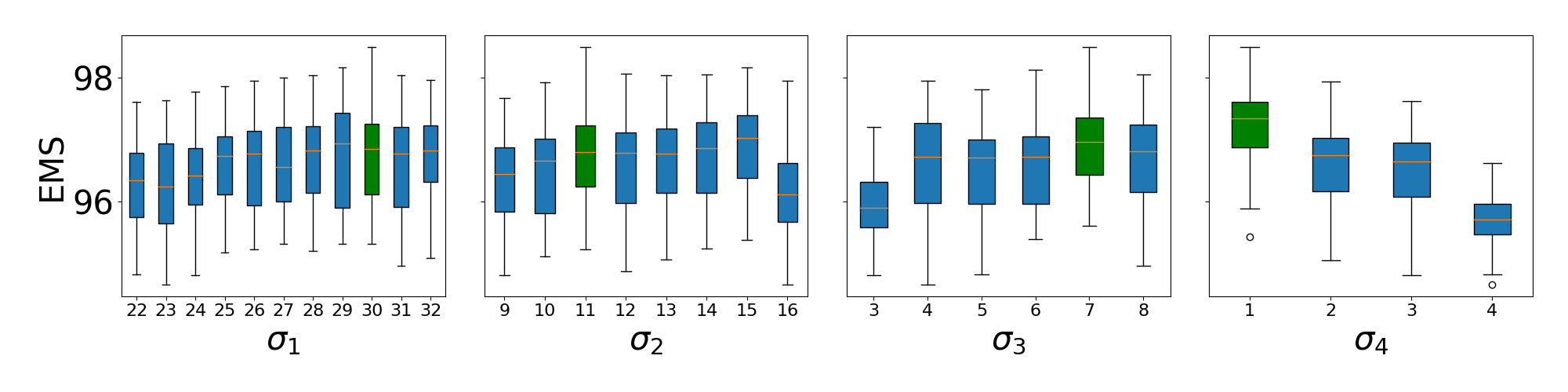}
    \caption{Hyperparameter Tuning for 2-entry Solutions}
    \label{fig:sigmaR_tuning_2entries}
\end{figure}

\begin{figure}
    \centering
    \includegraphics[width=0.825\linewidth]{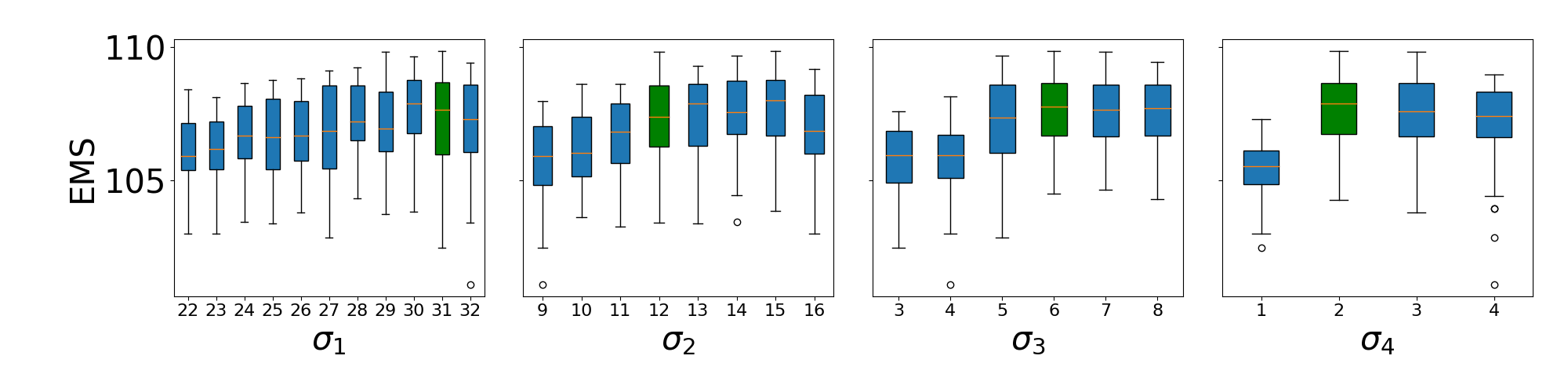}
    \caption{Hyperparameter Tuning for 5-entry Solutions}
    \label{fig:sigmaR_tuning_5entries}
\end{figure}

\begin{figure}
    \centering
    \includegraphics[width=0.8\linewidth]{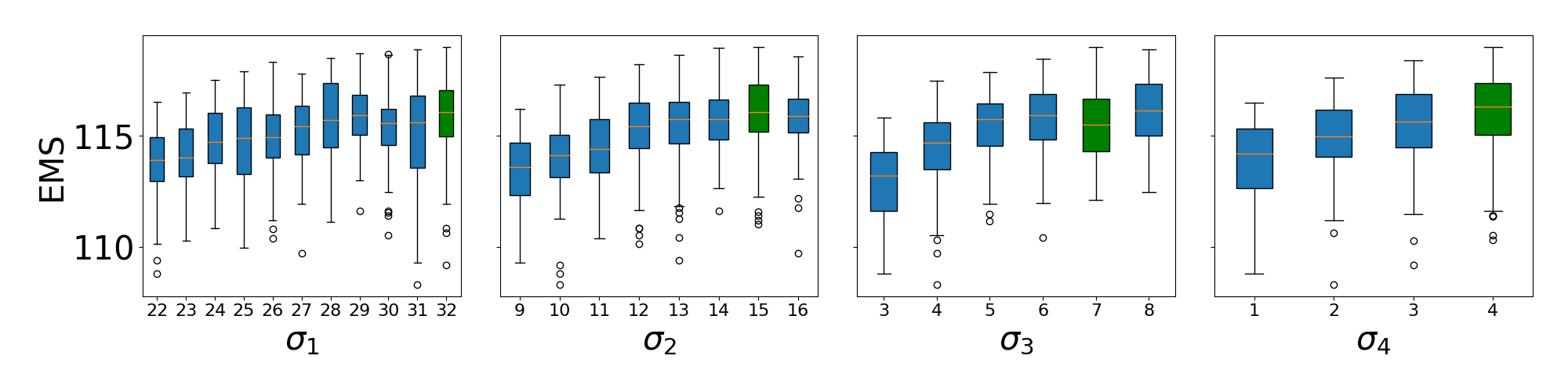}
    \caption{Hyperparameter Tuning for 10-entry Solutions}
    \label{fig:sigmaR_tuning_10entries}
\end{figure}

\begin{figure}
    \centering
    \includegraphics[width=0.8\linewidth]{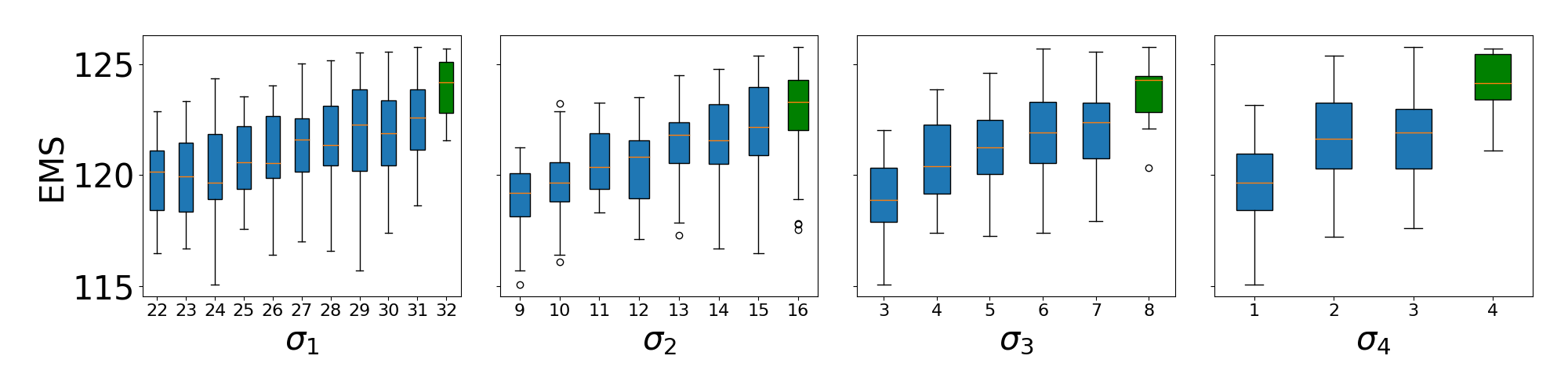}
    \caption{Hyperparameter Tuning for 25-entry Solutions}
    \label{fig:sigmaR_tuning_25entries}
\end{figure}

\begin{figure}
    \centering
    \includegraphics[width=0.8\linewidth]{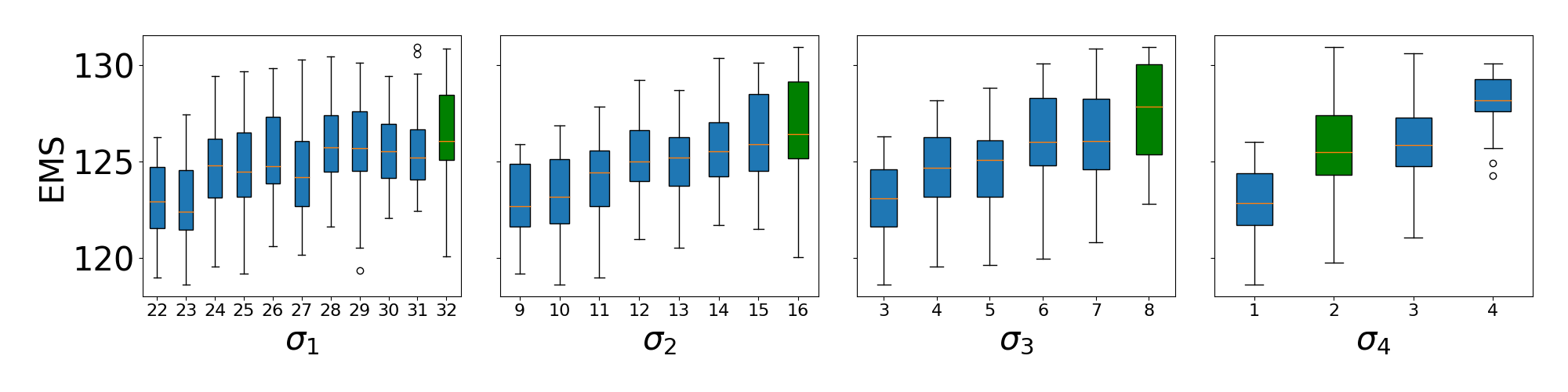}
    \caption{Hyperparameter Tuning for 50-entry Solutions}
    \label{fig:sigmaR_tuning_50entries}
\end{figure}

\begin{figure}
    \centering
    \includegraphics[width=0.8\linewidth]{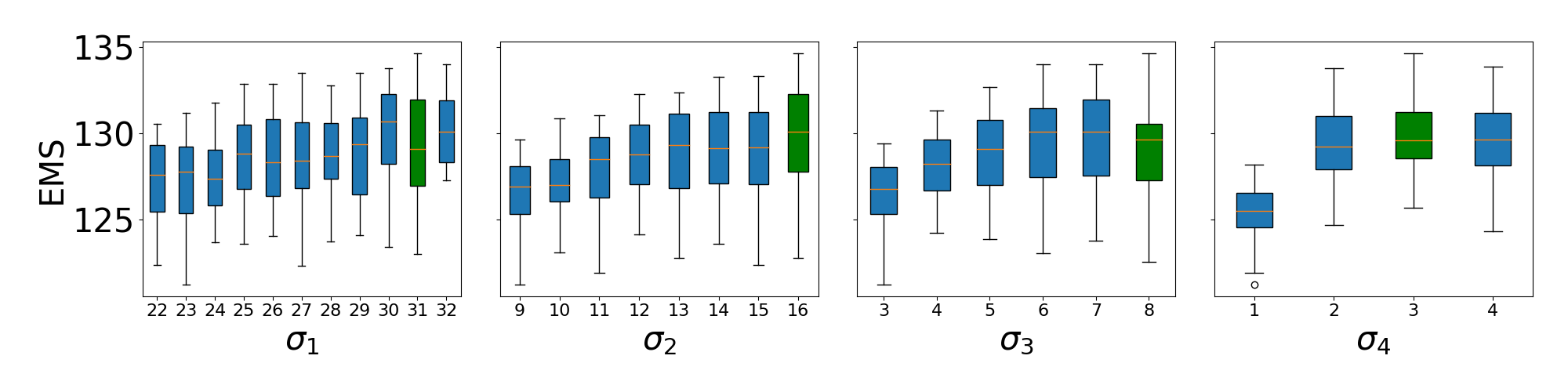}
    \caption{Hyperparameter Tuning for 100-entry Solutions}
    \label{fig:sigmaR_tuning_100entries}
\end{figure}

In these experiments, we incorporate in $D(\Bar{x})$ the \textit{Champion Constraint}, the \textit{Finalist Constraint}, and the \textit{Round Constraints}, and we test 400 randomly generated configurations of~\ref{ipseq} using combinations of values of $\sigma_1$ in $\{22,\ldots, 32\}$, $\sigma_2$ in $\{9,\ldots, 16\}$, $\sigma_3$ in $\{3,\ldots, 8\}$, and $\sigma_4$ in $\{1,\ldots, 4\}$. Figures~\ref{fig:sigmaR_tuning_2entries}, \ref{fig:sigmaR_tuning_5entries}, \ref{fig:sigmaR_tuning_10entries}, \ref{fig:sigmaR_tuning_25entries}, \ref{fig:sigmaR_tuning_50entries},~and \ref{fig:sigmaR_tuning_100entries} display boxplots of the empirical EMS for each $\sigma_\indexround$ for solutions with 2, 5, 10, 25, 50 and 100 entries. The green boxplots indicate the best-performing  $\sigma_\indexround$ for each round $\indexround$.

Figure~\ref{fig:sigmaR_tuning_2entries}-\ref{fig:sigmaR_tuning_100entries} shows that~$\sigma_\indexround$ increases for each round as the number of entries becomes larger, leaving more flexibility to \SIP. For 2 entries, we found that $\sigma_1=30$, $\sigma_2=11$, $\sigma_3=7$, and $\sigma_4=1$, which enforce significant diversification in round 4.
In particular, Figure~\ref{fig:sigmaR_tuning_25entries} shows that $\sigma_r$ equals the number of games in round~$\nrounds$ when $\nentries=25$, thus trivializing the constraint.

\begin{figure}[ht!]
    \centering
    \includegraphics[width=0.45\textwidth]{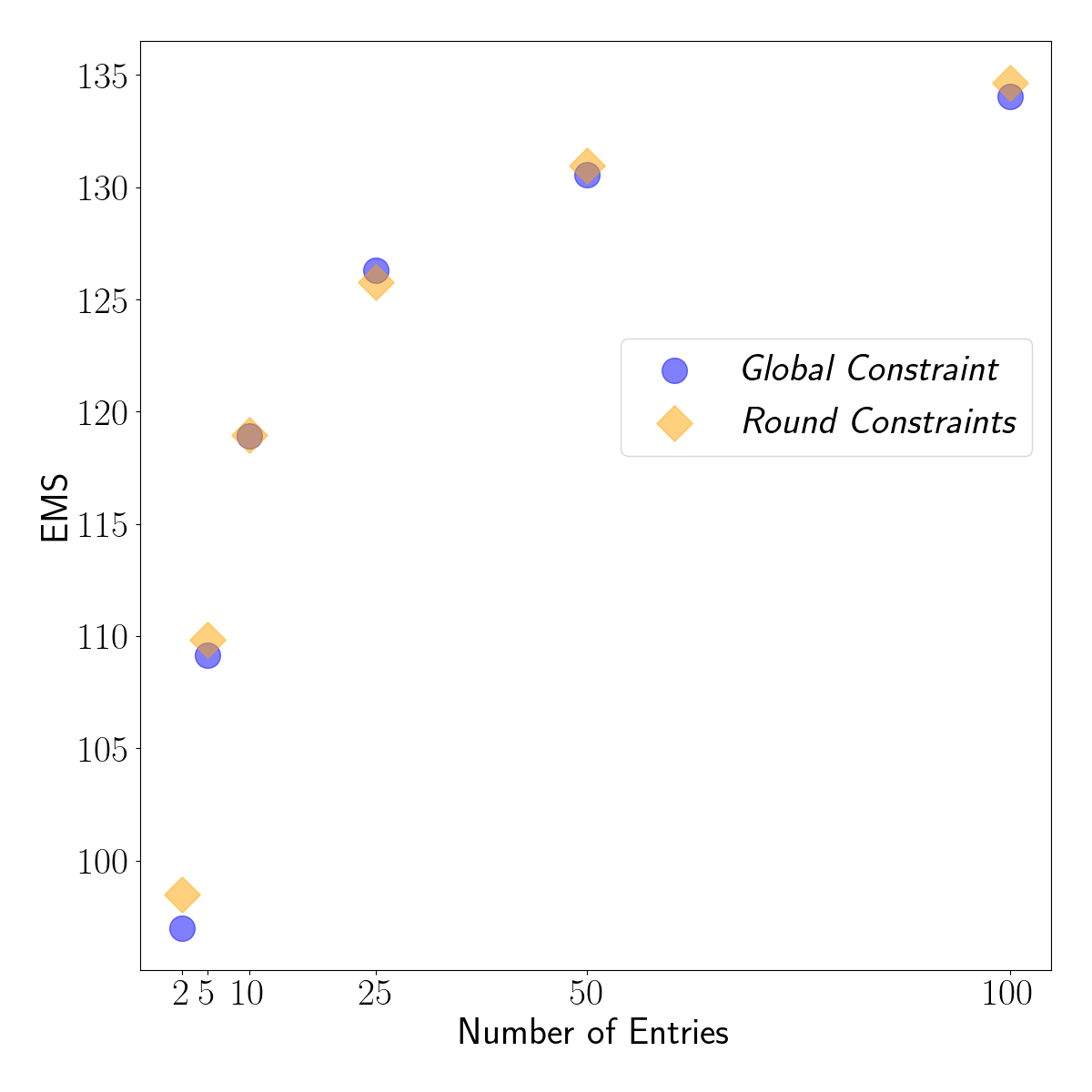}
    \caption{Empirical EMS using the \textit{Global Constraint} or the \textit{Round Constraints}}
    \label{fig:GlobalvsRounds}
\end{figure}

Figure~\ref{fig:GlobalvsRounds} displays the empirical EMS of the best 2-, 5-, 10-, 25-, 50-, and 100-entry solutions found using \SIP with either the \textit{Global Constraint} or the \textit{Round Constraints}. 
Figure~\ref{fig:sigma_tuning} shows that the \textit{Round Constraints} finds better solutions for 2, 5, 10, 50, and 100 entries. However, the \textit{Global constraints} finds a better 25-entry solution; this is in agreement with the results presented in Figure~\ref{fig:sigmaR_tuning_25entries}, as they suggest that the \textit{Round Constraints} are less important as the number of entries becomes larger. 

Based on the observations above, we parameterize $D(\overline{x})$ according to the number of entries~$\nentries$ as follows:
\begin{itemize}
    \item $\nentries\in\{2,\ldots,4\}$: \textit{Champion Constraint}, \textit{Finalist Constraint}, and \textit{Round Constraints} with $\sigma_1=30$, $\sigma_2=11$, $\sigma_3=7$, and $\sigma_4=1$;
    \item $\nentries\in\{5,\ldots,9\}$: \textit{Champion Constraint}, \textit{Finalist Constraint}, and \textit{Round Constraints} with $\sigma_1=31$, $\sigma_2=13$, $\sigma_3=6$, and $\sigma_4=2$;
    \item $\nentries\in\{10,\ldots,25\}$: \textit{Champion Constraint}, \textit{Finalist Constraint}, and \textit{Round Constraints} with $\sigma_1=32, \sigma_2=15, \sigma_3=7,$ and $\sigma_4=4$; and 
    \item $\nentries\geq26$: \textit{Champion Constraint}, \textit{Finalist Constraint}, and \textit{Global Constraint} with $\sigma=54$.
\end{itemize}

\subsection{Calibration of \ref{SAA} and \GreedySAA}

\begin{figure}[ht!]
    \centering
    \subfigure[\ref{SAA} 95\textsuperscript{th} CI]{\includegraphics[width=0.3\textwidth]{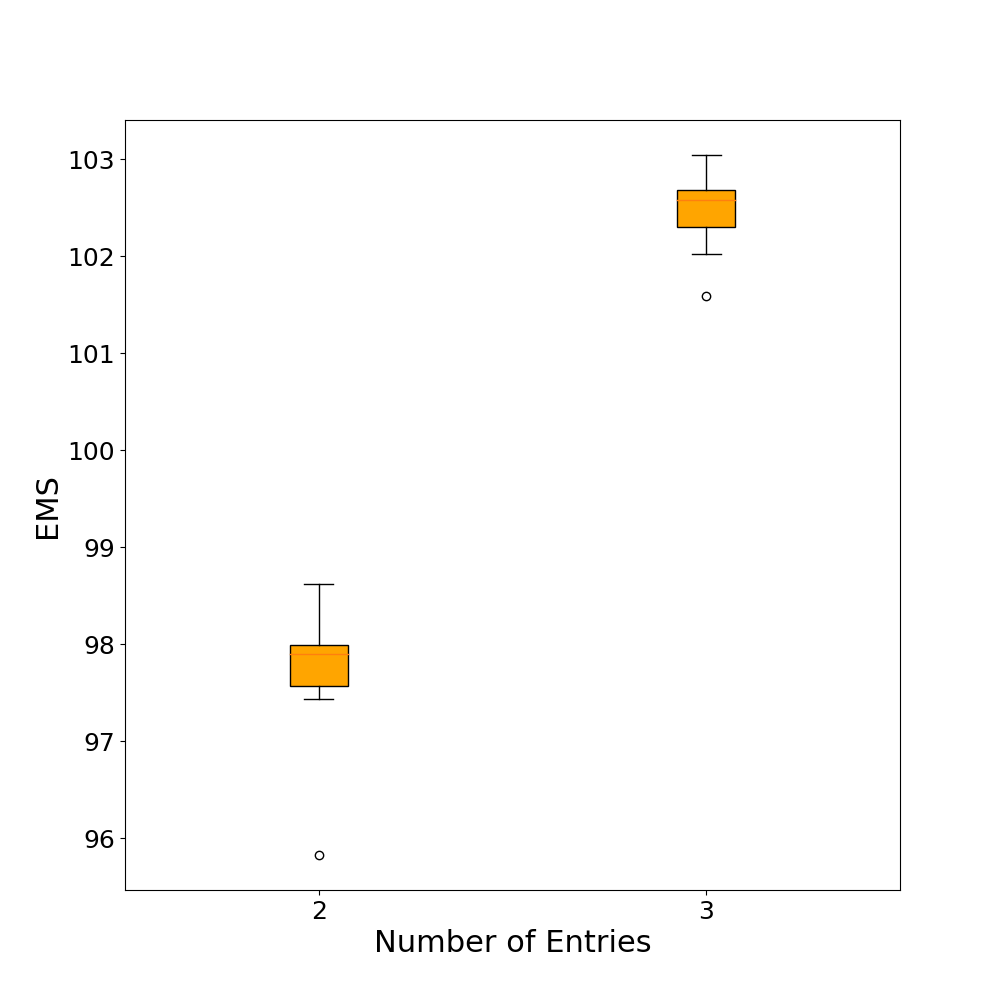}} 
    \subfigure[\GreedySAA 95\textsuperscript{th} CI]{\includegraphics[width=0.3\textwidth]{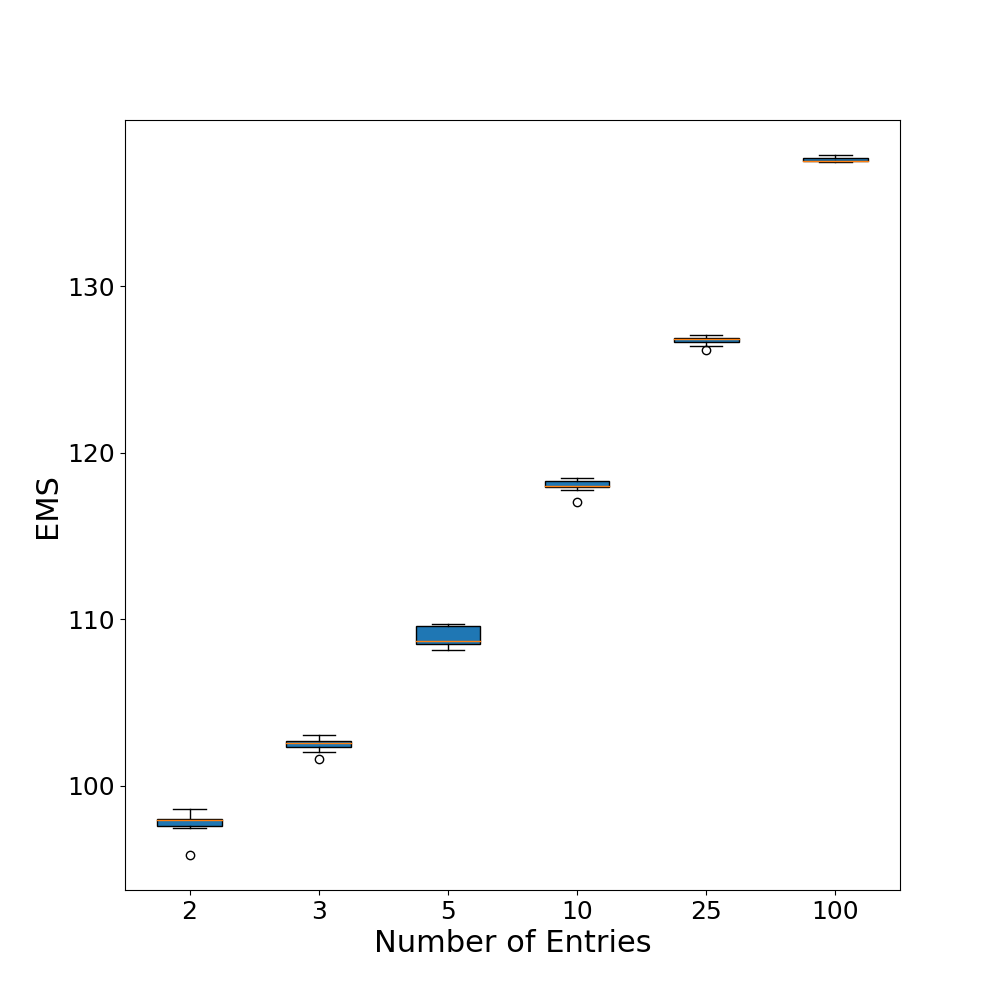}}
    \caption{Comparative Analysis for Multiple Runs of \ref{SAA} and \GreedySAA}
    \label{fig:CI_SAA_GSAA}
\end{figure}

The solutions produced by~\ref{SAA} and~\GreedySAA depend on the set outcome samples used as input. Therefore, we evaluated the accuracy of both algorithms by conducting 10 experiments with each and constructing CIs for the empirical EMS values. Figure~\ref{fig:CI_SAA_GSAA} presents boxplots illustrating our results.
Figure~\ref{fig:CI_SAA_GSAA} shows non-negligible variability in the empirical EMS values obtained through different executions of~\ref{SAA}. Therefore, in our experiments, we conduct 10 independent runs for~\ref{SAA} in each experiment and select the best-performing solution (using another out-of-sample set of randomly generated brackets). In contrast, \GreedySAA is very consistent, so we run the algorithm once per experiment.

\subsubsection{Calibration of \Propplus}

As observed in \ref{sec:EMSCalculation}, adding a lower bound per round to \Prop to create \Propplus transformed the worst proposed algorithm into the best one. We used the Python Package \texttt{Blackbox} \citep{Knysh2016} to find these bounds. Namely, using all tournaments within our dataset, we searched for values of $\widehat{P_{\indexround}}$ for every $\indexround$ in $\setrounds$ that maximized the average empirical EMS obtained over all years in our dataset. We made 800 calls to the \texttt{Blackbox} with $\widehat{P_{1}}\in [0.4,0.5]$, $\widehat{P_{2}}\in [0.35,0.45]$, $\widehat{P_{3}}\in [0.30,0.40]$,  $\widehat{P_{4}}\in [0.15,0.30]$,  $\widehat{P_{5}}\in [0.0,0.10]$, and $\widehat{P_{6}}\in [0.0,0.05]$. The optimal hyperparameters are reported in Table~\ref{tab:LB_Prop}.

\section{Estimating the Team-by-Team Probability Matrices $\Pteam$}
\label{appendix:pteamMatrix}

Our experiments use the procedure presented by \texttt{538} to estimate~$\Pteam$, which rates teams based on the most popular rating systems, such as Kenpom, Elo, and Sagarin, and the pre-season Associated Press poll ranking. Given ratings~$r_A$ and~$r_B$ for teams $\indexteam_A$ and $\indexteam_B$, respectively, \texttt{538} estimates~$\Pteami_{\indexteam_A,\indexteam_B}$ using

\[
    \Pteami_{\indexteam_A,\indexteam_B}=\frac{1}{1+10^{(r_A - r_B)\cdot30.464/400}}.
\]

We compare the accuracy of \texttt{538}'s estimation model with
\texttt{cbbdata} and \texttt{seed-based}. 
Tables~\ref{tab:ProbAccuracy} and~\ref{tab:ProbLogloss} display the accuracy and the \textit{Logloss} of the models, respectively, across all tournaments of our dataset. Accuracy is the percentage of match-ups in which the team with the highest win probability wins. The \textit{Logloss} of a model for a set of~$N$ samples, each with actual outcome~$y_i$ with predicted probability~$p_i$, is given by
\[
Logloss = -\frac{1}{N} \sum_{i=1}^{N} \left[ y_i \cdot \log(p_i) + (1 - y_i) \cdot \log(1 - p_i) \right].
\]
We note that a lower \textit{Logloss} value indicates higher performance. The results suggest that all three models have similar performance, but the 538 model has a slightly better accuracy on average.

\begin{table}[h!]
\footnotesize
\begin{tabular}{|c|c|c|c|c|c|c||c|}
\hline
          & \textbf{2017}  & \textbf{2018}  & \textbf{2019}  & \textbf{2021}  & \textbf{2022}  & \textbf{2023}  & \textbf{Average}\\ \hline\hline
\texttt{538}       & 74.60 & 73.02 & 74.60 & 69.84 & 69.84 & 66.67 & 71.34\\ \hline
%toRvik 
\texttt{cbbdata}     & 68.25 & 65.08 & 71.43 & 68.25 & 69.84 & 68.25 & 68.51 \\ \hline
\texttt{seed-based}  & 74.63 & 67.16 & 65.67 & 68.18 & 65.67 & 69.84 & 68.52 \\ \hline
\end{tabular}
\caption{Probability Models' Accuracy}
\label{tab:ProbAccuracy}
\end{table}

\begin{table}[h!]
\footnotesize
\begin{tabular}{|c|c|c|c|c|c|c||c|}
\hline
          & \textbf{2017} & \textbf{2018} & \textbf{2019} & \textbf{2021} & \textbf{2022} & \textbf{2023} & \textbf{Average}\\ \hline\hline
\texttt{538}       & 0.50 & 0.57 & 0.47 & 0.61 & 0.62 & 0.64 & 0.57\\ \hline
%toRvik
\texttt{cbbdata}& 0.50 & 0.59 & 0.47 & 0.62 & 0.60 & 0.62 & 0.57\\ \hline
\texttt{seed-based} & 0.64 & 0.61 & 0.64 & 0.61 & 0.61 & 0.59 & 0.62\\ \hline
\end{tabular}
\caption{Probability Models' \textit{Logloss}}
\label{tab:ProbLogloss}
\end{table}

Lastly, we report how each of these matrices differs from one another by measuring their distance from the $\nteams\times\nteams$ matrix with 0.5 on every off-diagonal. We measure these distances using the Kullback-Leibler (KL) divergence measures, calculated as
\[
D_{\text{KL}}(P||Q)=\sum_{\substack{(\indexteam,\indexteam')\in\setteams:\Pteami_{\indexteam,\indexteam'}>0.5}}\Pteami_{\indexteam,\indexteam'}\log\left(\frac{\Pteami_{\indexteam,\indexteam'}}{0.5}\right).
\]
Table~\ref{tab:KLTable} reports the KL divergence for all three matrices for all \textit{March Madness} competitions composing our dataset. The results show that, for every year, the \texttt{538} probability matrices are further away from the reference matrix, followed by \texttt{cbbdata} and the \texttt{seed-based} models.
\begin{table}[h!]
\footnotesize
\begin{tabular}{|c|ccccccc|}
\hline
           & \multicolumn{1}{c|}{\textbf{2017}}   & \multicolumn{1}{c|}{\textbf{2018}}   & \multicolumn{1}{c|}{\textbf{2019}}   & \multicolumn{1}{c|}{\textbf{2021}}   & \multicolumn{1}{c|}{\textbf{2022}}   & \multicolumn{1}{c|}{\textbf{2023}}   & \textbf{Average} \\ \hline\hline
\texttt{538}        & \multicolumn{1}{c|}{628.13} & \multicolumn{1}{c|}{604.78} & \multicolumn{1}{c|}{609.63} & \multicolumn{1}{c|}{592.32} & \multicolumn{1}{c|}{579.41} & \multicolumn{1}{c|}{557.26} & 595.26  \\ \hline
\texttt{cbbdata}    & \multicolumn{1}{c|}{580.36} & \multicolumn{1}{c|}{522.45} & \multicolumn{1}{c|}{571.91} & \multicolumn{1}{c|}{517.46} & \multicolumn{1}{c|}{516.92} & \multicolumn{1}{c|}{480.65} & 531.62  \\ \hline
\texttt{seed-based} & \multicolumn{7}{c|}{501.62}                                                                                                                                                             \\ \hline
\end{tabular}
\caption{Kullback-Leibler Divergence Measures per Probability Matrix}
\label{tab:KLTable}
\end{table}

%\end{APPENDICES}

\end{document}